\pdfoutput=1
\documentclass[11pt,a4paper,twoside,openright]{report}
\usepackage[english]{ETHDAsfs}%--> ETHDASA + fancyheadings + ... "umlaute" 
\usepackage{pdfpages}%%to include the confirmation of originality (plagiarism
\usepackage{amsbsy}%% for \boldsymbol and \pmb{.}
\usepackage{amssymb}%% calls  amsfonts...
\usepackage{graphicx}%-- für PostScript-Grafiken (besser als  psfig!)
\usepackage{subcaption} %several figures
\usepackage[longnamesfirst]{natbib}%was {sfsbib}%- Für  Literatur-Referenzen
\usepackage{texab}%- 'tex Abkürzungen' /u/sfs/tex/tex/latex/texab.sty
\usepackage{amsmath}
\usepackage{enumerate}% Fuer s elbstdefinierte Nummerierungen
\usepackage{relsize}%-> \smaller (etc) used here
\usepackage{color} %% to allow cloring in code listings
\usepackage{listings}% Fuer R-code, C-code, ....  and settings for these:
\usepackage{xfrac} % diagonal fraction style
\usepackage[toc,page, titletoc]{appendix}
\usepackage{float} %force figures to lie where you want with [H]
\usepackage{bbm} %for indicator function
\usepackage{mathtools} %for ceiling shorthand

\definecolor{Mygrey}{gray}{0.75}% for linenumbers only!
\definecolor{Cgrey}{gray}{0.4}% for comments
\newtheorem{definition}{Definition}[subsection]

\newtheorem{theorem}[definition]{Theorem}

\theoremstyle{definition} 
\newtheorem{example}[definition]{Example}

\numberwithin{equation}{subsection}

\graphicspath{{./Pictures/}}

\newcommand{\argmaxM}[1]{\underset{#1}{\operatorname{arg}\,\operatorname{max}}\;}
\newcommand{\argminM}[1]{\underset{#1}{\operatorname{arg}\,\operatorname{min}}\;}

\newcommand\mypropto{\mathrel{\overset{\makebox[0pt]{\mbox{\normalfont\tiny\sffamily $\mu$}}}{\propto}}}

\DeclarePairedDelimiter\ceil{\lceil}{\rceil}

\def\hypclass{\mathcal{C}(O^{(n)})}
\def\weights{w_{\beta}(c, X^{(n)})}
\def\ER{\hat{R}(c, X^{(n)})}
\def\ERs{\hat{R}(c, X)}
\def\Zb{Z_{\beta}(X^{(n)})}
\def\ZbOne{Z_{\beta}(X')}
\def\ZbTwo{Z_{\beta}(X'')}
\def\dZb{\Delta Z_{\beta}(X',X'')} 
\def\ZbOnez{Z_{\beta}(\xi_1)}
\def\ZbTwoz{Z_{\beta}(\xi_2)}
\def\dZbz{\Delta Z_{\beta}(\xi_1,\xi_2)} 
\def\Gibbs{P_G(c; \beta, X^{(n)})}
\def\GibbsOne{P_G(c; \beta, X')}
\def\GibbsTwo{P_G(c; \beta, X'')}
\def\wb{w_{\beta}(c,X)}
\def\wbOne{w_{\beta}(c,X')}
\def\wbTwo{w_{\beta}(c,X'')}
\def\IC{I_{\beta}}
\def\SNCCt{Shannon's Noisy-Channel Coding Theorem }
\def\SNCCts{Shannon's Noisy-Channel Coding Theorem}
\def\DGM{data generating mechanism }
\def\dc{(\mathcal{X}, p(y|x), \mathcal{Y})}
\def\rate{\frac{1}{n}logM}
\def\ILP{Idealized Learning Protocol }
\def\LP{Learning Protocol }
\newcommand\ci{\perp\!\!\!\perp}

\def\realnumbers{\mathbb{R}}
\def\binarynumbers{\mathbb{B}}
\def\positivenumbers{\mathbb{Z}^+}

\begin{document}
\bibliographystyle{chicago}% ---> Hampel,F., E.Ronchetti,... W.Stahel(1986) ...
 %was \bibliographystyle{sfsbib}\citationstyle{dcu} %OR DEFAULT : \citationstyle{agsm}

\pagenumbering{roman}%- roman numbering for first few pages

%%%%%%%%%%%%%%%%%%%%%%%%%%%%%%%%%%%%%%%%%%%%%%%%%
%%% Title page                                %%%
%%%%%%%%%%%%%%%%%%%%%%%%%%%%%%%%%%%%%%%%%%%%%%%%%
\period{Autumn 2015}
\dasatype{Semester Project}
\students{Emiliano Díaz}
\mainreaderprefix{Supervisor:}
\mainreader{Prof.\ Dr.\ Markus Kalisch, Prof. Joachim M. Buhmann}
\alternatereaderprefix{Adviser}
\alternatereader{Nico Gorbach, Stefan Bauer}
\submissiondate{April 26th 2016}
\title{Sparse Mean Localization \\ by Information Theory }

\maketitle%- Titelseitex wird abgeschlossen
\cleardoublepage
 %%~~~~~~~~~~~~~~~~~~~~~~~~~~~~~~~~~~~~~~~~

%%%%%%%%%%%%%%%%%%%%%%%%%%%%%%%%%%%%%%%%%%%%%%%%%
%%% Insert here acknowledgements and abstract %%%
%%%%%%%%%%%%%%%%%%%%%%%%%%%%%%%%%%%%%%%%%%%%%%%%%
%% Dedication (optional)
\markright{}
\vspace*{\stretch{1}}
\begin{center}
    I would like to express my gratitude to Stefan Bauer and Nico Gorbach who gave me all the support I needed, to Prof. Dr. Joachim Buhmann for the opportunity to work in such as a fascinating topic, and to Dr. Markus Kalisch for all the support, in this semester project and throughout the Masters' programme. 
\end{center}
\vspace*{\stretch{2}}

% Preface (optional)
%\newpage
%\markboth{Preface}{Preface}
%\include{Preface}

% Abstract should not be longer than one page.
\newpage
\markboth{Abstract}{Abstract}
\chapter*{Abstract}

Sparse feature selection is necessary when we fit statistical models, we have access to a large group of features, don't know which are relevant, but assume that most are not. Alternatively, when the number of features is larger than the available data the model becomes overparametrized and the sparse feature selection task involves selecting the most informative variables for the model. When the model is a simple location model and the number of relevant features does not grow with the total number of features, sparse feature selection corresponds to sparse mean estimation. We deal with a simplified mean estimation problem consisting of an additive model with gaussian noise and mean that is in a restricted, finite hypothesis space (parameter space). This restriction simplifies the mean estimation problem into a selection problem of combinatorial nature. Although the hypothesis space is finite, its size is exponential in the dimension of the mean. In limited data settings and when the size of the hypothesis space depends on the amount of data or on the dimension of the data, choosing an approximation set of hypotheses is a desirable approach.  Choosing a set of hypotheses instead of a single one implies replacing the bias-variance trade off with a resolution-stability trade off. Generalization capacity provides a resolution selection criterion based on allowing the learning algorithm to communicate the largest amount of information in the data to the learner without error. In this work the theory of approximation set coding and generalization capacity is explored in order to understand this approach. We then apply the generalization capacity criterion to the simplified sparse mean estimation problem and detail an importance sampling algorithm which at once solves the difficulty posed by large hypothesis spaces  and the slow convergence of uniform sampling algorithms (caused by the skewed distribution of hypothesis costs). Finally we explore how the generalization capacity criterion can be a applied to a more realistic version of the sparse feature selection problem where the number of relevant features grows with the total number of features.

%\cite{Buhm97} \cite{Buhm10} \cite{Buhm12} \cite{Buhm14} \cite{Buhm13} \cite{Vapnik} \cite{Cover} \cite{Yeung} \cite{Lehmer} \cite{BuhmNotes} \cite{vanDeGeer} \cite{Jaynes1} \cite{Jaynes2} \cite{Jaynes3} \cite{Tikochinsky}

%%% Local Variables: 
%%% mode: latex
%%% TeX-master: "MasterThesisSfS"
%%% End: 

%%%%%%%%%%%%%%%%%%%%%%%%%%%%%%%%%%%%%%%%%%%%%%%%%
%%% Table of contents and list of figures and %%%   
%%% tables (no need to change this usually)   %%%
%%%%%%%%%%%%%%%%%%%%%%%%%%%%%%%%%%%%%%%%%%%%%%%%%
\newpage
\tableofcontents
\newpage
\listoffigures
%\newpage
%\listoftables

%% Notations and glossary (optional)
%\cleardoublepage
%\phantomsection
%\addcontentsline{toc}{chapter}{\protect\numberline{}{Notation}}
%\markboth{Notation}{Notation}
%\include{Notation}

\cleardoublepage
\pagenumbering{arabic}%--- switch back to standard numbering 

%%%%%%%%%%%%%%%%%%%%%%%%%%%%%%%%%%%%%%%%%%%%%%%%%
%%% Your text... Either write here directly,  %%%
%%% or even better: write in separate files   %%%
%%% that you just have to include here.       %%% 
%%%%%%%%%%%%%%%%%%%%%%%%%%%%%%%%%%%%%%%%%%%%%%%%%
\chapter{Introduction}

It is often the case that when fitting statistical models, the majority of available features are not informative in the sense of the underlying learning task. In other cases the limited amount of data available implies that most features can't be used, even if they are all informative, because the model becomes overparametrized. In both instances sparse feature selection must be done prior or simultaneous to model fitting. In this work we deal with the sparse feature selection problem as it applies to a simplified location model. We first assume the number of relevant features is small and fixed and then explore the case where the number of relevant features is small but grows with the total number of features. Although this problem is well known and studied, for example in \cite{vanDeGeer}, we are interested in how we can apply approximation set coding and generalization capacity to localize the hypothesis class to an \emph{optimal} resolution. 

\section{Structure} 

The report is organized as follows. Section \ref{prbStmnt} gives a description of the problem we will focus on: sparse mean estimation and sparse feature selection. We want to solve this problem using the \emph{approximation set coding} and \emph{generalization capacity}  methodology proposed by \citeauthor{Buhm13}, so in Sections \ref{PA}-\ref{Shannon} we give an introduction to the theory involved. Section \ref{PA} introduces the \emph{pattern analysis} framework for learning problems. In section \ref{ApproxSets} we explore how, by defining \emph{approximation sets} of hypotheses instead of proposing a single hypothesis as the solution, we are able to move from the normal bias-variance trade-off of learning problems to a resolution-stability trade-off. In section \ref{GC}, with the help of concepts from Sections \ref{PA} and \ref{ApproxSets}, we define various information theoretic concepts such as Boltzmann weights, Gibbs distributions and partition functions,  culminating in the definition of \emph{generalization capacity}. We try to give an intuitive understanding of each of these concepts except that of generalization capacity itself. In Section \ref{Shannon} we motivate the concept of generalization capacity in analogy to Shannon's noisy channel coding theorem from which it is derived. 

In Chapter \ref{MeanLoc} we estimate the generalization capacity of the squared loss based, empirical risk function for the non-sparse version of the mean localization problem. We concentrate on low-dimensional cases. In Section \ref{MLGC} the information theoretic concepts defined in Section \ref{GC} are applied to the problem at hand culminating in an expression for the generalization capacity that suggests an \emph{exhaustive} simulating algorithm for its estimation. Section \ref{NSEA} includes the pseudo code for implementing this algorithm.  Section \ref{NSSR} includes the results of implementing the exhaustive simulating algorithm to estimating generalization capacity. Section \ref{logSum} is a note on how to avoid \emph{underflow} problems when implementing this algorithm. In Section \ref{VarRed} we explore different ways in which we may incoprorate the use of \emph{common random numbers} into our algorithm as a variance reduction technique. 

In Chapter \ref{chp3} we estimate the generalization capacity for the sparse mean localization problem. Section \ref{SMEA} describes the changes and additional tools necessary to implement the algorithm described in \ref{NSEA} to the sparse version of the problem. Section \ref{SSR} includes the results of implementing this algorithm to estimating generalization capacity. Since the algorithm will be shown to be inadequate in the high dimensional case, in Section \ref{SA} we describe a sampling algorithm based on a re-expression of the generalization capacity. Section \ref{SASR} includes the results of this sampling algorithm. This algorithm will be shown to converge too slowly in the number of simulations and so in section \ref{ISA} we describe an \emph{importance} sampling algorithm for estimating generalization capacity. Section \ref{impSampRes} includes the results of this algorithm. 

Chapter \ref{ch:TowardsSparseFeatureSelection} is a brief exploration into a  more realistic version of  sparse feature selection where the number of relevant features grows with the total number of features. We describe the problem and explore some of the difficulties of estimating generalization capacity with a simulation algorithm in this case. 

Chapter \ref{s:Summary} includes a summary of the report and a list of possible related avenues of future research.

\section{Problem statement}
\label{prbStmnt}

We deal with the statistical model studied in \cite{Buhm14}:

$$
		X_i = \mu^0 + \epsilon_i
$$

where

\begin{itemize}
		\item $\mu^0 \in \binarynumbers^d=\{0,1\}^d$
		\item $X_i,\epsilon_i \in \realnumbers^d$
		\item $\epsilon \sim N(0,\sigma^2 I_d)$ 
		\item observations $X_i$ with $i=1,...,n$ are i.i.d. 
\end{itemize}

In the general case estimating $\mu^0$ corresponds to selecting a hypothesis $\mu$ from the hypothesis space $\binarynumbers^d$ which has cardinality $2^d$. While we first deal with this problem, we will be more interested in a modified version of this problem where:

\begin{enumerate}
	\item $||\mu^0||_1=k$
	\item $\mu^0 \in \binarynumbers^d_k=\{\mu \in \binarynumbers^d: ||\mu||_1=k\}$
	\item $|\binarynumbers^d_k|={d \choose k} $
	\item It is assumed that $k$ is known.
\end{enumerate}

We first deal with the general case where $k,d \in \positivenumbers$, $k \leq d$, and then with a \emph{sparse} case where $k$ is kept constant and $d$ grows toward infinity. In Chapter \ref{ch:TowardsSparseFeatureSelection} we briefly discuss another \emph{sparsity} condition where $k\approx\frac{d}{\log(d)}$.

\section{Pattern analysis} \label{PA}

Although the classical framework of parameter inference, in which estimators $\hat{\theta}(\cdot)$ are maps from a sample space $\mathcal{X}$ to a parameter space $\Theta$, is appropriate for the problem at hand we introduce \emph{Aproximation Set Coding} (ASC) and \emph{Generalization Capacity} (GC) within the framework of \emph{Pattern Analysis} since they are more relevant in this wider context. The rest of this introductory chapter follows \cite{Buhm13} closely.

The problem described in Section \ref{prbStmnt} belongs to the class of problems which are the object of \emph{Pattern Analysis}. The goal of pattern analysis is to map a set of object configurations to a pattern space. Concretely, we want to  choose a hypothesis $c \in \mathcal{C}(O^{(n)})$ where: 

\begin{itemize}
	\item $O_i \in \mathcal{O}$ are objects in an object space.
	\item $O^{(n)}=\{O_1,...,O_n\} \in \mathcal{O}^{(n)}$ are object sets.
	\item $c:\mathcal{O}^{(n)} \rightarrow \mathcal{P}$ is a hypothesis in a hypothesis class $\mathcal{C}(O^{(n)})$ and $\mathcal{P}$ is a pattern space. 
\end{itemize}

A few remarks about this framework:

\begin{enumerate}
	\item The hypothesis class $\mathcal{C}(O^{(n)})$ may or may not depend on the object set. Specifically, the size of the hypothesis class may depend on the object set or not. 
	\item In this exposition the objects in the object set $O_i \in \mathcal{O}$ may be tuples of objects from more fundamental object sets, i.e. $O_i = (o_{i1},...,o_{ir}), o_{ij} \in \mathcal{O}_j$. However, the objects $O_i$ are at the level of the mapping $c$.
	\item The hypothesis map $\mathcal{C}(O^{(n)})$ is actually a composition of the maps $X: \mathcal{O}^{(n)} \rightarrow  \mathcal{X}^{n}$ and $t: \mathcal{X}^{n} \rightarrow  \mathcal{P}$ where $\mathcal{X}^{n}$ is a measurement space. 
	\item The pattern space $\mathcal{P}$ may be related to the data generating process or not. It is an \emph{interpretation} space: a set of abstract, mutually exclusive \emph{properties} which we wish to assign to object configurations. 
\end{enumerate}	

We present some examples to clarify the pattern analysis framework. \\

\begin{example}[Mean estimation] \label{meanEst}
	We want to estimate the population mean height of swiss women given a sample of 100. We make no assumptions regarding the data generating process. 
	\begin{itemize}
			\item $\mathcal{O} = \{$swiss women$\}$
			\item $O^{(100)}$ is the set of sampled women. 
			\item $X(O^{(100)})$ are the heights of the sampled women.
			\item $\mathcal{X}^n \subset (\realnumbers^+)^{100}$ is the set of possible heights for the 100 women. 
			\item $\mathcal{P} \subset \realnumbers^+$ is the set of possible population mean heights. 
			\item $\mathcal{C}(O^{(100)})=\mathcal{O}^{(100)} \times \mathcal{P}$ is the hypothesis class which does not depend on the size $n=100$ of the object set. 
	\end{itemize}	
\end{example}
\hspace{1mm}

	\begin{example}[Clustering - Population] We want to cluster 100 people into 4 groups according to height and weight. We assume the underlying data generating process is a gaussian mixture with parameters $\{(\mu_1, \Sigma_1),...,(\mu_4, \Sigma_4)\}$ with $\mu_i \in \realnumbers^2$ and $\Sigma_i \in \realnumbers^{2 \times 2}$.
	\begin{itemize}
			\item $\mathcal{O} = \{$people$\}$.
			\item $O^{(100)}$ is the set of sampled people. 
			\item $X(O^{(100)})=\{(h_1,w_1),...,(h_{100},w_{100})\}$ are the heights and weights of the sampled people.
			\item $\mathcal{X}^n \subset \realnumbers^{100 \times 2}$ is the set of possible heights and weights for the 100 people. 
			\item $\mathcal{P} \subset \realnumbers^{2 \times 4} \times \realnumbers^{2 \times 2 \times 4}$ is the set of possible population mean and covariances.
			\item $\mathcal{C}(O^{(100)})=\mathcal{O}^{(100)} \times \mathcal{P}$ is the hypothesis class which does not depend on the size $n$ of the object set.  
	\end{itemize}
	\textbf{Remark}: Notice how our assumptions about the data generating process inform our choice of pattern space $\mathcal{P}$. 
	\end{example}
	
	\hspace{1mm}
	
	\begin{example}[Clustering - Sample] \label{clustSmpl} We want to cluster 100 people into 4 groups according to height and weight. We do not assume anything about the underlying data generating process and are just interested in finding a clustering that defines homogenous groups for \emph{this} sample and not the entire population. 
	\begin{itemize}
			\item $\mathcal{O} = \{$people$\}$.
			\item $O^{(100)}$ is the set of sampled people. 
			\item $X(O^{(100)})=\{(h_1,w_1),...,(h_{100},w_{100})\}$ are the heights and weights of the sampled people.
			\item $\mathcal{X}^n \subset \realnumbers^{100 \times 2}$ is the set of possible heights and weights for the 100 people. 
			\item $\mathcal{P} = \{1,2,3,4\}^{100}$ are all the possible ways we can group 100 people into 4 groups. 
			\item $\mathcal{C}(O^{(100)})=\mathcal{O}^{(100)} \times \mathcal{P}$ is the hypothesis class which in this case \textbf{does depend} on the size $n$ of the object set.  
	\end{itemize}
	\end{example}
	
	\hspace{1mm}
	
	\begin{example}[Dyadic data] We are interested in predicting if a user will make a purchase at a given website, based on the age and gender of the person and on the type of website (there are $m$ types). We don't assume anything about the data generating process but have already decided to model the probability of purchase using a logistic regression model. 
	\begin{itemize}
			\item $\mathcal{O} = \mathcal{O}_p \times \mathcal{O}_w = \{$people$\} \times \{$websites$\}$.
			\item $O^{(n)}=\{(o_{1p}, o_{1w}),...,(o_{np}, o_{nw})\}$ is the set of sampled person-website pairs. 
			\item $X(O^{(n)})=\{(a_1,g_1,t_1,p_1),...,(a_n,g_n,t_n,p_n)\}$ are the age, geneder, website-type and purchase outcome of the sampled person-website pairs. 
			\item $\mathcal{X}^n \subset \realnumbers^+ \times \{male, female\} \times \{1,...,m\} \times \{0,1\}$ is the sample space. 
			\item $\mathcal{P} = \{(\beta_0,\beta_1,\beta_2) \in \realnumbers^3\}$ is the parameter space for the logistic model. 
			\item $\mathcal{C}(O^{(n)})=\mathcal{O}^{(n)} \times \mathcal{P}$ is the hypothesis class which does not depend on the size $n$ of the object set.  
	\end{itemize}
\end{example}	

As we can see the pattern analysis framework fits a wide range of problems. Although problems such as mean estimation and regression, in which hypothesis classes with infinite cardinality are involved, can be tackled using the pattern analysis framework, in the rest of this introductory chapter we focus on classes with a finite number of hypotheses. In other words we assume: 

\begin{align}
	|\mathcal{C}(O^{(n)})| < \infty
\end{align}	

\section{Approximation sets} \label{ApproxSets}

In classical statistical learning theory, in order to solve an inference decision problem, we choose a loss function $\rho(c,x)$ with which we construct the risk function $R(c)=\mathbb{E}_X[\rho(c,X)]$. We then choose a single hypothesis $c^*$ that minimizes the empirical risk $\hat{R}(c, X^{(n)})$ for a given data set $X^{(n)}$: 

\begin{align}
	c^*(X^{(n)}) &\in \argminM{c \in \mathcal{C}(O^{(n)})} \hat{R}(c, X^{(n)})\\
	\hat{R}(c, X^{(n)}) &= \frac{1}{n}\sum_{i=1}^n\rho(c,X_i)
\end{align}	

If $n$ is large then the \emph{Emprirical Risk Minimizer} (ERM) will be close to the minimizer of the risk function $R(c,X^{(n)})$. However, in general we know that when $n$ is not large then the ERM will tend to overfit the data. Instead of choosing a single hypothesis we can choose a subset of the hypothesis class which includes \emph{good} hypotheses: hypothesis with low costs.  Qualitatively, we would like this set to be composed of low cost hypotheses which we cannnot (partially) order further because their costs are statistically indistinguishable. The goal is to choose a subset of hypotheses that are stable with respect to fluctuations in the cost measurements. We may code this selection with a weight function $w_{\beta}(c, X^{(n)})$ over the hypothesis class where:

\begin{align}
  w: \mathcal{C} \times \mathcal{X}^n \times \realnumbers^+ \rightarrow [0,1]
\end{align} 

\begin{align}
  (c,X^{(n)},\beta) \mapsto w_\beta(c,X^{(n)})
 \end{align}
 
 \begin{align}
  w_{\beta}(c, X^{(n)})  =
    \begin{cases}
      1, & \hat{R}(c, X^{(n)}) \leq \hat{R}(c^*, X^{(n)}) + \sfrac{1}{\beta} \\
      0, & \text{otherwise}
    \end{cases}
\end{align}

Where $\beta$ can be interpreted as the degree of certainty we have that $c^*$ is the best solution. We may generalize the concept of approximation sets by allowing \emph{fuzzy}, non-binary selection where hypotheses belong to the solution set to varying degrees, i.e. $ w_{\beta}(c, X^{(n)}) \in [0,1]$. In this case valid weight vectors satisfy: 

\begin{align} \label{condApprox}
	 w_{\beta}(c, X^{(n)}) \geq  w_{\beta}(c', X^{(n)}) \Leftrightarrow \hat{R}(c, X^{(n)}) \leq \hat{R}(c', X^{(n)})
\end{align}	

The sum of the weights over the hypothesis class indicates the \emph{equivalent} number of hypotheses selected. The bigger this sum the more unsure we are about $c^*$. We will sometimes say that $w_{\beta}(c, X^{(n)})$ is the \emph{approximation set of hypotheses}, meaning that it encodes the (fuzzy) membership of the hypotheses in the set.  A parametric family of weights which satisfies condition \ref{condApprox} is: 

\begin{equation} \label{eq:parFam}
		w_{\beta}^f(c, X^{(n)}) = \{w_{\beta}(c, X^{(n)})=e^{-\beta f(\hat{R}(c, X^{(n)}))}: \beta \in \realnumbers^+,  f  \textrm{ increasing} \}
\end{equation}	

Notice that if we normalize the weights such that $\sum_{c \in \mathcal{C}(O^{(n)})}w_{\beta}(c, X^{(n)})=1$ we can interpret the weights as a posterior probability distribution over the hypothesis class. 

In the classical statistical setting, when we have limited data, obtaining unbiased estimators often means these estimators have high variance: estimations change dramatically from one data set to the next. Lowering the variance can sometimes be achieved by introducing bias into our estimator. This is the bias-variance trade-off that, when there is limited data, is usually resolved through some sort of regularization. As we shall see in Section \ref{Shannon} the ASC approach leads to a resolution-stability trade-off which replaces the bias-variance trade-off. \emph{Resolution} refers to the equivalent number of hypotheses selected while \emph{stabiity} refers to obtaining similar approximation sets for different $X^{(n)} \in \mathcal{X}^n$. Adopting the ASC approach the trade-off becomes, do we obtain a very stable set of \emph{good} hypotheses that don't change a lot depending on the data set but that is quite large (low resolution) or do we focus in on a small number of \emph{very good} hypotheses but such that they will change from one data set to the next (unstable). 

Notice that the $\beta$ parameter in our weight function $w_{\beta}$ is the resolution parameter that determines how this trade-off is resolved. Adopting the view of our normalized weight vector as a posterior over the hypothesis class, the higher $\beta$ is the more probability is \emph{spread} or \emph{smoothed} among all the hypotheses. In limited data settings, choosing the parameter $\beta$ corresponds to regularizing our empirical risk function. 

In general, the justification for using the ASC approach is: 

\begin{enumerate}[I]
	\item \textbf{Inference}. It allows us to identify hypotheses which are similar in cost but which might be distinct according to other criteria not included in the cost function. This benefit is also common to bayesian inference. 
	\item \textbf{Learnability}. For $\hypclass$ to be \emph{learnable}, ERM theory requires that it should not be too complex. In other words, $\hypclass$ should have a finite \emph{VC-dimension}. For certain problems, such as \ref{clustSmpl}, the size of $\hypclass$ increases too quickly in $n$, meaning that as $n \rightarrow \infty$ the empirical risk minimizer does not converge to the true risk minimizer. For this type of problem it is not even theoretically possible to converge to the true $c \in \hypclass$ as $n \rightarrow \infty$ so an approximation set solution seems more reasonable. 
\end{enumerate}	

\section{Generalization capacity} \label{GC}
\begin{definition}[Boltzmann weights] %[subsection]
If, from the parametric family \ref{eq:parFam}, we choose $f(x)=x$ to construct our weight vector we obtain the so called Boltzmann  weights:
\begin{align}
	\weights := e^{-\beta\ER}
\end{align}	
\end{definition}

\begin{definition}[Partition function]
	The sum over $\hypclass$ of the Boltzmann weights is a function of the data $X^{(n)}$. We call it the partition function with respect to $w_\beta$ and define it as: 
	
	\begin{align}
			\Zb := \sum_{c \in \hypclass} \weights 
	\end{align}	
\end{definition}	

\begin{definition}[Gibbs Distribution]
	The normalized Boltzmann weights define a Gibbs distribution, $P_G(c; \beta, X^{(n)})$ over $\hypclass$ with respect to the cost function $\ER$: 
	
	\begin{align}
			P_G(c; \beta, X^{(n)}) := \frac{\weights}{\Zb}
	\end{align}	
\end{definition}	

This choice of weight vector can be justified from an information theoretic perspective. The Gibbs distribution is the maximum \emph{entropy} distribution among all distributions $p(c)$ over $\hypclass$ such that: 

\begin{align}
	\mathbb{E}_{p(c)}[\ER] = \mu_{\beta}
\end{align}	

where $\mu_{\beta}$ is a non-increasing function of $\beta$. As we increase $\beta$, the resolution parameter, the expected cost with respect to the Gibbs distribution, decreases. In the limit, as $\beta \rightarrow \infty$, the Gibbs distribution becomes a single point mass distribution over $c^*$ and $\mu_{\beta} \rightarrow \hat{R}(c^*, X^{(n)})$. The Gibbs distribution $\Gibbs$ preserves the same (partial) ordering of $\hypclass$ as $-\ER$, but rescales so that differences in cost on the low end of the cost spectrum are \emph{exaggerated} and differences in cost on the high end of the cost spectrum are \emph{smoothed} out. 

\begin{figure}[hbt!]%--- Picture 'H'ere, 'B'ottom or 'T'op; '!' Try to
                    %impose your will to LaTeX
 % \centering 
  
  \begin{subfigure}{.5\textwidth}
    \centering
	\includegraphics[width=1\textwidth]{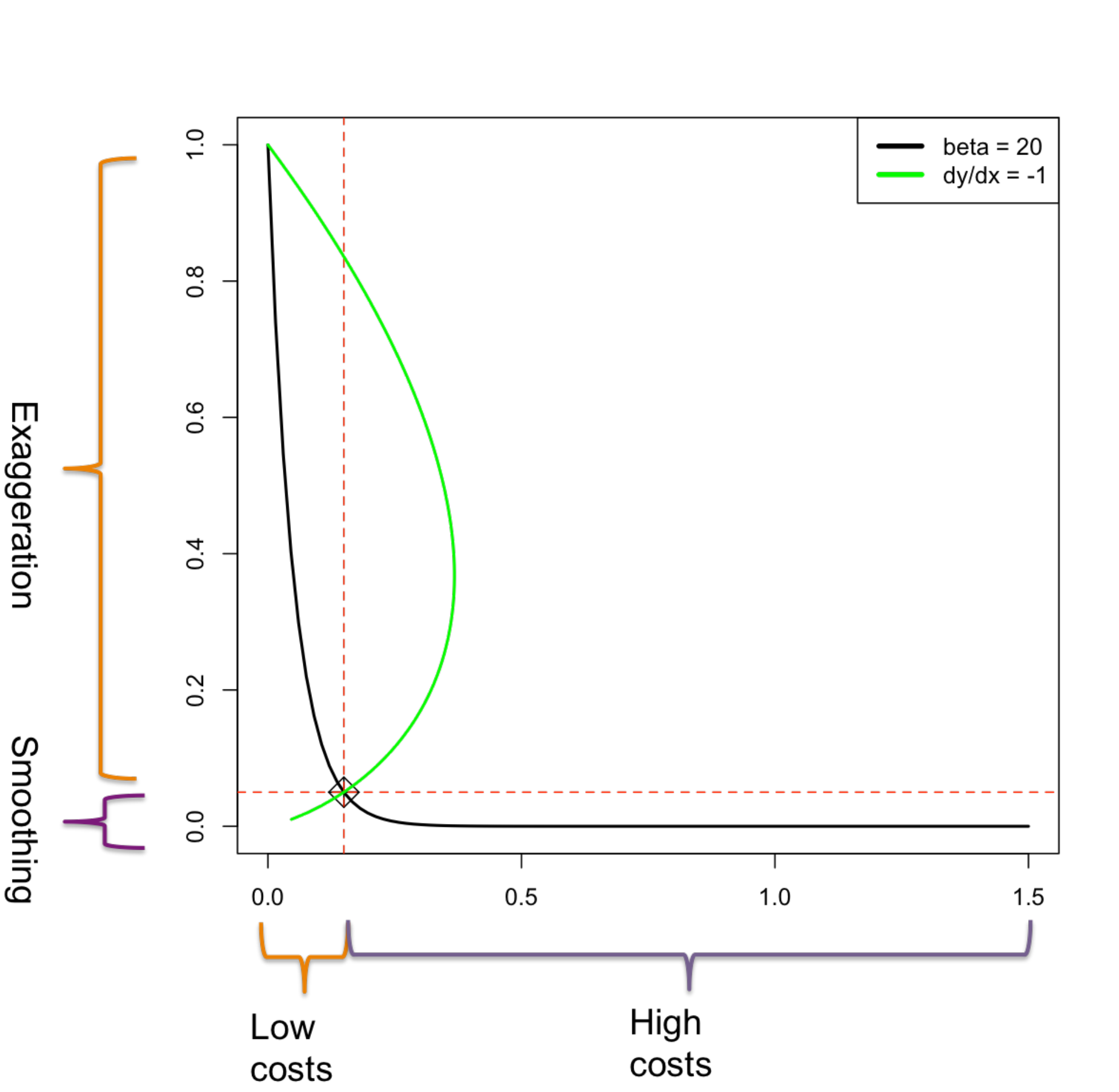} 
    \caption{$\beta = 20$}
    \label{fig:sfig1}
  \end{subfigure}%
  \begin{subfigure}{.5\textwidth}
    \centering
    \includegraphics[width=1\textwidth]{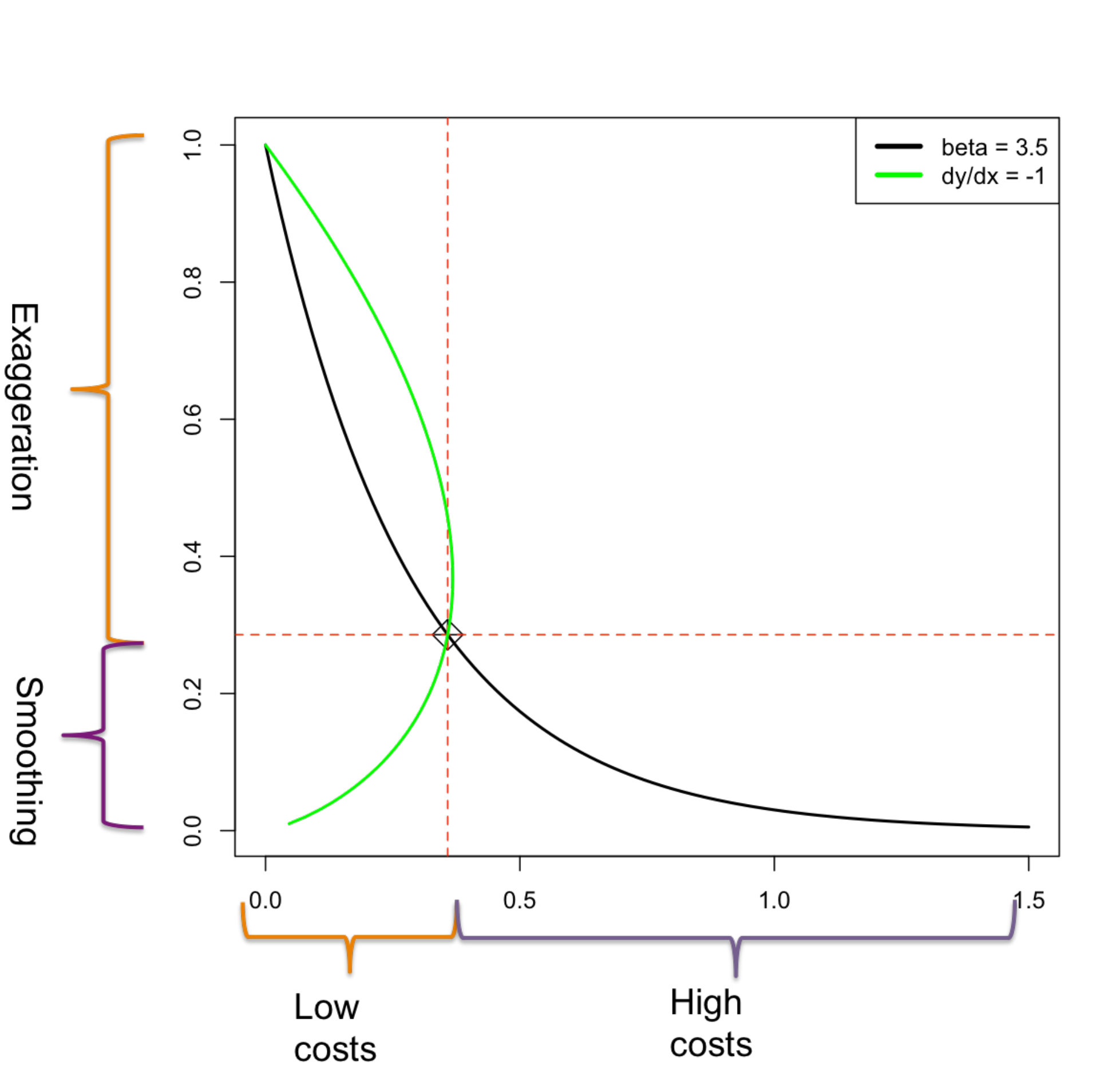}
    \caption{$\beta = 3.5$}
    \label{fig:sfig2}
  \end{subfigure}
  \begin{subfigure}{.5\textwidth}
    \centering
    \includegraphics[width=1\textwidth]{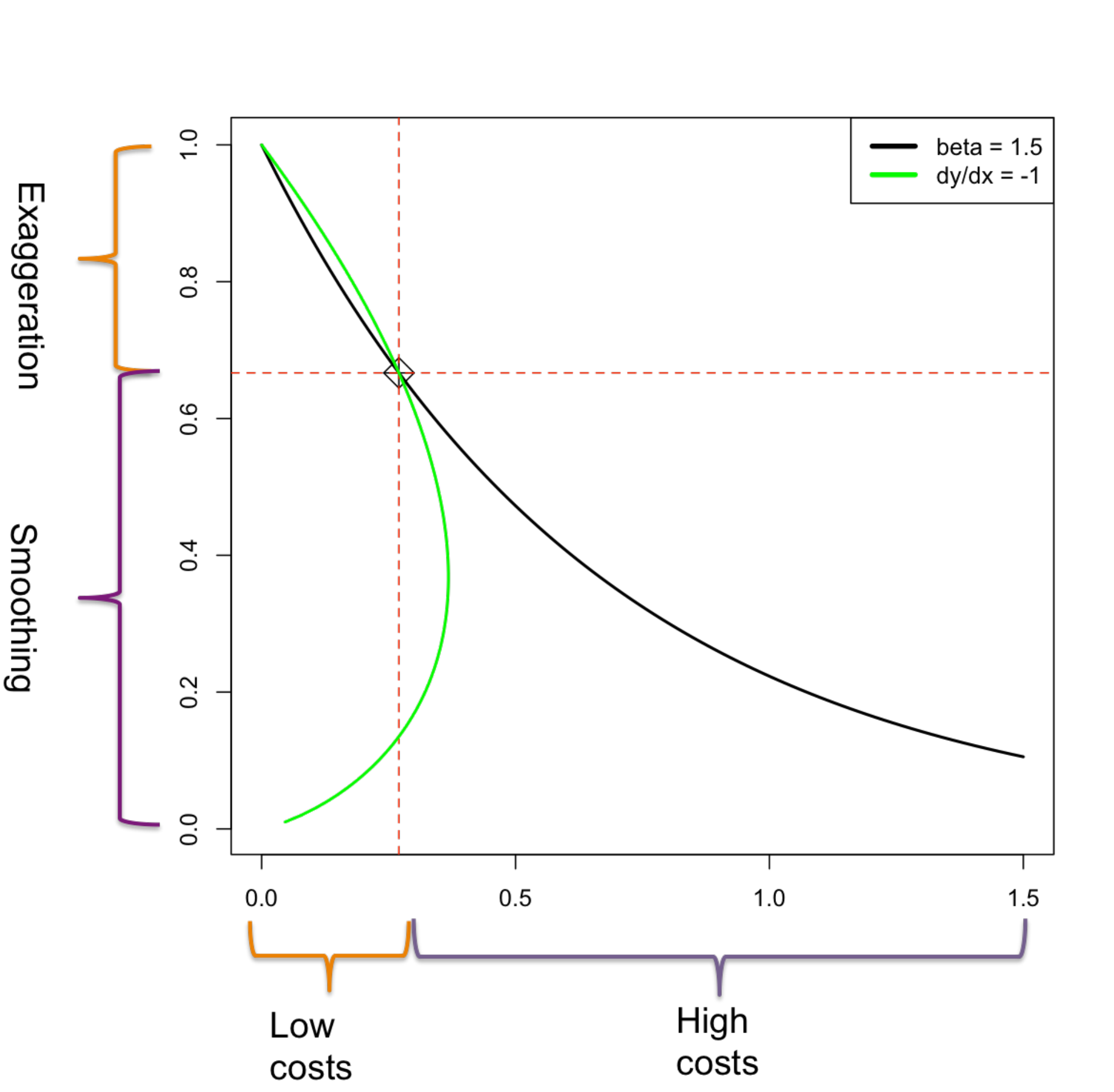}
    \caption{$\beta = 1.5$}
    \label{fig:sfig3}
  \end{subfigure}
  \begin{subfigure}{.5\textwidth}
    \centering
    \includegraphics[width=1\textwidth]{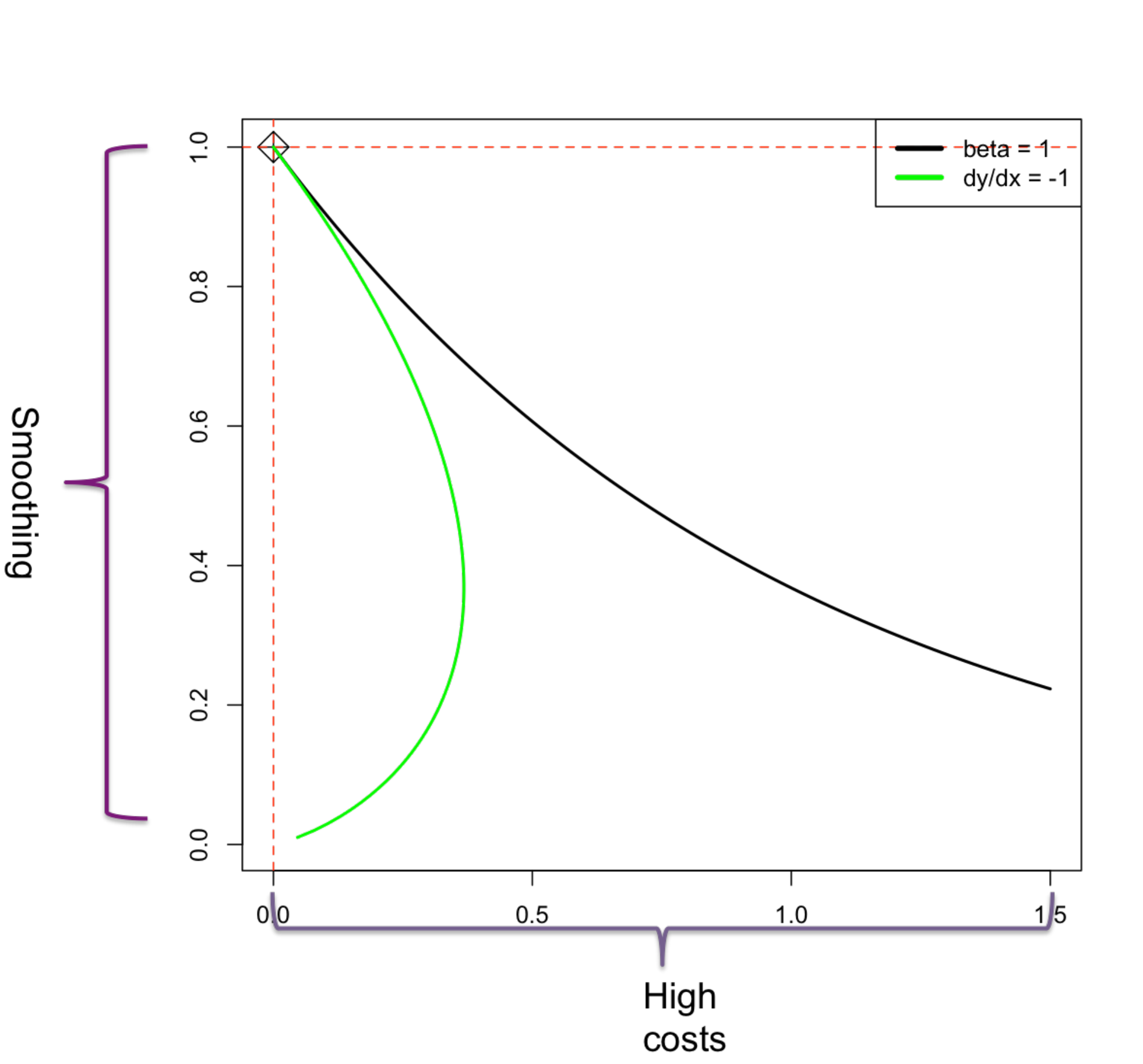}
    \caption{$\beta = 1$}
    \label{fig:sfig4}
  \end{subfigure}
  %%         --- .5\textwidth stands for 50% of text width
  \caption[Boltzmann smoothing: smoothing of costs with Boltzmann weight function]%<<-- Legend for the list of figures at the beginning of you thesis
  {Boltzmann smoothing: smoothing of costs with Boltzmann weight function}% legend displayed below the graph.
  \label{fig:boltzSmooth}
\end{figure}

Figure \ref{fig:boltzSmooth} illustrates the mechanics of the smoothing of costs with the Boltzmann weight function. Costs scaled by the $\beta$ parameter are plotted on the x-axis ($x=\beta\ER$) and the corresponding Boltzmann weights on the y-axis ($y=\weights=e^{-x}$). The black lines show the Boltzmann weights for two different resolution values: $\beta=20,1.5$. The green line represents the points $(\sfrac{\log(\beta)}{\beta},\sfrac{1}{\beta})$ which are the points that satisfy $\frac{dy}{dx}=-1$. Let the point of intersection between a given weight function $\weights$ indexed by $\beta$ and the green line be called the critical point $(x_{\beta}^c,y_{\beta}^c)$ for that $\beta$. For a given $\beta$, scaled costs to the right of the critical point $x_{\beta}^c$ (high costs) are smoothed onto the interval $(0,y_{\beta}^c)$ while scaled costs to the left of $x_{\beta}^c$ (low costs) are exaggerated onto the interval $(y_{\beta}^c,1)$. This is how the Boltzmann weight function $\weights$ and the parameter $\beta$ control the level of smoothing: for high resolution $\beta$ only the lowest cost hypotheses remain relevant, while for low resolution levels most hypotheses retain some measure of relevance. 

The characteristics discussed above, are shared by all functions from the parametric family \ref{eq:parFam}.  These characteristics allow the Boltzmann weight function to be used in global optimization strategies such as \emph{simulated} and \emph{deterministic annealing} where the smoothing out of less important features in the cost surface in early iterations prevent the search algorithm from getting stuck in local minima. As is established in \cite{Jaynes1}, \cite{Jaynes2} and \cite{Jaynes3}, particular to the Gibbs distribution (for which $f(x)=x$), is the fact that for a given level of resolution, manifested as an expectation, $\mathbb{E}_{p(c)}[\ER]=\mu_{\beta}$, that is a certain distance from  $\hat{R}(c^*, X^{(n)})$, it has maximum entropy among distributions with this characteristic  This means that if we use the Gibbs distribution to describe our uncertainty about the true hypothesis,  the only information extracted from $X^{(n)}$ is that obtained using $\ER$. Interpreting the entropy of a distribution as a measure of its uncertainty and supposing we know that $\mathbb{E}_{p(c)}[\ER]=\mu_{\beta}$, then $\Gibbs$ is the maximally non-comittal distribution with respect to information different to that contained in this restriction. \cite{Tikochinsky} established another characteristic that makes the Boltzmann weights and Gibbs distribution an appealing choice as the ASC weighting function: it is maximally stable. If we change our desired resolution level from $\beta_1$ to $\beta_2$, the change in the induced Gibbs distributions is minimal, in the $L_2$ norm sense, among any two distributions $p_1$ and $p_2$ that satisfy $\mathbb{E}_{p_1(c)}[\ER]=\mu_{\beta_1}$ and $\mathbb{E}_{p_2(c)}[\ER]=\mu_{\beta_2}$. 

We have discussed the role of the resolution parameter $\beta$ in the context of the resolution-stability trade-off, so how can we determine the best value of $\beta$? For this purpose \citeauthor{Buhm13} developed the concept of \emph{Generalization Capacity} which we will first define and then describe in analogy to the \emph{Channel Capacity} concept of information theory. \\

\begin{definition}[Joint partition function]
Before  we define Generalization Capacity we define the joint partition function between two data sets which measures the equivalent number of hypotheses selected by a weighting function $\wb$ for two different data sets $X'$ and $X''$:
	\begin{align}
		\dZb = \sum_{c \in \mathcal{C}} \wbOne \wbTwo
	 \end{align}	
	 
\textbf{Remarks:}
\begin{enumerate}
	\item We have dropped the superindex (n) for better readability: $X':=X'^{(n)}$, $X'':=X''^{(n)}$, $O':=O'^{(n)}$ and $O'':=O''^{(n)}$.
	\item As this definition already suggests GC will involve comparing the approximation sets obtained with different data sets of the same size. 
	\item For some pattern analysis problems such as \ref{clustSmpl} the hypothesis class depends on the object set $O^{(n)}$ so that $\mathcal{C}(O') \neq \mathcal{C}(O'')$. In this case we need a mapping $\psi: \mathcal{O}' \rightarrow \mathcal{O}''$ so that $\dZb$ can be properly defined as $\dZb = \sum_{c \in \mathcal{C}(O'')} w_{\beta}(c, \psi (X')) \wbTwo$. Although these types of problems are very important in the context of ASC given the infinite VC dimension of the hypothesis class, for the sparse mean estimation problem described in Section \ref{prbStmnt} this is not the case, so we will simply assume, from now on, that $\mathcal{C} := \mathcal{C}(O') = \mathcal{C}(O'')$. This also means we can dispense with the mappings $\psi$ in this exposition.
\end{enumerate}		 
\end{definition}
\hspace{20 mm}
\begin{definition}[Information Content]
	The information content retrievable from data $(X',X'')$ by a cost function $\ERs$ with resolution $\beta$ is: 
	
	\begin{align}
		\IC := \log \frac{|\mathcal{C}|\dZb}{\ZbOne \ZbTwo}
	\end{align}	
	
	\textbf{Remarks:}\\
	\begin{enumerate}[i.]

	\item $\IC$ is a normalized and rescaled version of $\dZb$ which measures the equivalent number of selected hypotheses with cost function $\ERs$ for both data sets $X'$ and $X''$. 

	\item Since $\lim_{\beta \to 0} \ZbOne  = \lim_{\beta\to 0} \ZbTwo = \lim_{\beta\to 0} \dZb = |\mathcal{C}|$ it holds that $\lim_{\beta\to 0} I_\beta = 0$, which means that for resolution $\beta=0$, where all hypothesis are given a weight of 1, the information content is zero. 

	\item Let $\mathcal{C}=\{c_1,...,c_{|\mathcal{C}}|\}$, $a_i = w_\beta(c_i,X') \geq 0$ and $b_i = w_\beta(c_i,X'') \geq 0$, then:

	\begin{align}
		0 &\leq \frac{\dZb}{\ZbOne \ZbTwo} = \frac{\sum_{i=1}^{|\mathcal{C}|} a_i b_i}{(\sum_{i=1}^{|\mathcal{C}|} a_i)  (\sum_{i=1}^{|\mathcal{C}|} b_i)} \\
		& = \frac{\sum_{i=1}^{|\mathcal{C}|} a_i b_i}{\sum_{i=1}^{|\mathcal{C}|}\sum_{j=1}^{|\mathcal{C}|} a_i b_j} = \frac{\sum_{i=1}^{|\mathcal{C}|} a_i b_i}{\sum_{i=1}^{|\mathcal{C}|} a_i b_i + \sum_{i \neq j} a_i b_j} \leq 1
	\end{align}	

	and we can see that $\lim_{\beta\to \infty} I_\beta \leq \log |\mathcal{C}|$. This means that for maximum resolution the information content can reach up to the log-size of the hypothesis class. 
	
	%\item Since $0 \leq \frac{\dZb}{\ZbOne \ZbTwo} \leq 1$ we have that $I_\beta \leq \log |\mathcal{C}|$ however, depending on how \emph{similar} $X'$ and $X''$ are (i.e. how noisy the data is), $\frac{\dZb}{\ZbOne \ZbTwo}$ is not necessarily monotonous in $\beta$ so that  $I_\beta$ doesn't always reach the maximum of $\log |\mathcal{C}|$.

	\end{enumerate}
\end{definition}	
\hspace{20 mm}

\begin{definition}[Generalization Capacity]
	The generalization capacity of a cost function $\ERs$ defined over a hypothesis class $\mathcal{C}$ and data space $\mathcal{X}^n$ is:
	
	\begin{align}
		I := \max_{\beta \in \realnumbers^+} \mathbb{E}_{(X',X'')}\IC
	\end{align}	
	
\textbf{Remarks:}\\
\begin{enumerate}[i.]

\item Since  $\lim_{\beta\to 0} I_\beta = 0$ we have that $\lim_{\beta\to 0} I = 0$ and,s

\item since  $\lim_{\beta\to \infty} I_\beta \leq \log |\mathcal{C}|$ we have that $\lim_{\beta\to \infty} I \leq \log |\mathcal{C}|$

\end{enumerate}
	
\end{definition}

\section{GC and Shannon's noisy-channel coding theorem}
\label{Shannon}

To motivate the relevance of the Generalization Capacity as an important quantity in itself aswell as a criterion for deciding between cost functions in a pattern analysis problem, we briefly study \SNCCts, the communication protocol suggested therein and the role of \emph{Channel Capacity}. We then move from the communication context to the pattern learning context and study an analagous \emph{learning protocol} suggested by \citeauthor{Buhm13} where the generalization capacity emerges as a natural counterpart to channel capacity. The exposition of \SNCCt is based on \cite{Cover} and \cite{Yeung}.

\SNCCt deals with the rate at which information can be passed through a channel so we first define what information and channels are. \\

\begin{definition}[Shannon Information]
	The Shannon information of an outcome $x$ of a random variable $X \in \mathcal{X}$, where $\mathcal{X}$ is a finite set and $p(x)$ is the probability distribution of $X$ is:
	
	$$
		I(x) = -\log p(x)
	$$	
	
	If we interpret the informativeness of an outcome in terms of the worth of knowing its value the following properties make this a useful measure of information: 
	\begin{enumerate}
		\item $I(x) \geq 0$ $ \forall x$
		\item Assigns 0 to a certain outcome
		\item The rarer an outcome the more informative: $p(x)=p, p(y)=q, p<q \Rightarrow I(x)>I(y)$
		\item It is continuous in $p(x)$: $\forall \epsilon>0$ $\exists$ $\delta>0 : \textrm{ if } |p-q|<\delta \Rightarrow |I(x)-I(y)|<\epsilon$
		\item Additivity: $p_{XY}(x,y)=p_X(x)p_Y(y) \Rightarrow I(x,y) = I(x) + I(y)$ 
	\end{enumerate}
\end{definition}	

\hspace{1mm}

\begin{definition}[Entropy]
	The entropy of a random variable $X \in \mathcal{X}$, where $\mathcal{X}$ is a finite set, is its expected Shannon Information:
	\begin{align}
		H(X):= -\mathbb{E}_X[\log p(X)]=\mathbb{E}_X[I(X)]=\sum_{x \in \mathcal{X}} p(x)\log p(x)
	\end{align}	
\end{definition}	

We can interpret entropy as the average information rate of a random variable. If we want to send \emph{messages} from a finite message set $\mathcal{W}=\{1,...,M\}$, we may define the information rate of the message set by assuming messages will be sent according to the uniform distribution. In this case:

\begin{align}
	H(\mathcal{W}) := -\sum_{i=1}^M \frac{1}{M}\log\frac{1}{M} = \log M
\end{align}	

We now define joint and conditional entropy. \\ 

\begin{definition}[Joint Entropy]
	For random variables $X \in \mathcal{X}$ and $Y \in \mathcal{Y}$, with $\mathcal{X}$ and $\mathcal{Y}$ finite sets, the joint entropy of $X$ and $Y$ is defined as:
	
	\begin{align}
		H(X,Y) := -\sum_{(x,y) \in \mathcal{X} \times \mathcal{Y}} p(x,y)\log p(x,y) = -\mathbb{E}_{(X,Y)}\log p(X,Y)
	\end{align}	
	
\end{definition}	

\begin{definition}[Conditional Entropy]
	For random variables $X \in \mathcal{X}$ and $Y \in \mathcal{Y}$, with $\mathcal{X}$ and $\mathcal{Y}$ finite sets, the joint entropy of $Y$ given $X$ is defined as:
	
	\begin{align}
		H(Y|X) := -\sum_{(x,y) \in \mathcal{X} \times \mathcal{Y}} p(x,y)\log p(y|x) = -\mathbb{E}_{(X,Y)}\log p(Y|X)
	\end{align}	
\end{definition}

	Conditional entropy is a measure of the mean information left in $Y$ once we know the outcome of $X$. It turns out that $H(X,Y) = H(X) + H(Y|X)$ so the joint entropy can be interpreted as the mean amount of information in $X$ plus the mean amount of information left in $Y$ once the outcome of $X$ is known. \\
	
	\begin{definition}[Mutual Information]
		For random variables $X \in \mathcal{X}$ and $Y \in \mathcal{Y}$, with $\mathcal{X}$ and $\mathcal{Y}$ finite sets, the mutual information between $X$ and $Y$ is defined as:
		\begin{align}
		I(X;Y):= \sum_{(x,y) \in \mathcal{X} \times \mathcal{Y}} p(x,y)\log \frac{p(x,y)}{p(x)p(y)} = \mathbb{E}_{(X,Y)}\log \frac{p(X,Y)}{p(X)p(Y)}
		\end{align}
	\end{definition}	
	
	Using that $I(X;Y) = H(X) - H(X|Y)$ we can interpret the mutual information as the reduction in information left in $X$ once $Y$ is known (or vice versa). Alternatively, we can interpret $I(X;Y)$ as the information that is common to $X$ and $Y$. If $X$ and $Y$ are independent then they have no information in common and if $X$ depends deterministically on $Y$ then they have the same information. \\
	
	\begin{definition}[Discrete Channel]
		Let $\mathcal{X}$ and $\mathcal{Y}$ be discrete sets and $p(y|x)$ be a transition matrix from $\mathcal{X}$ to $\mathcal{Y}$ that is a valid distribution for all $x$. Then the tuple $\dc$ is a discrete channel where $X \in \mathcal{X}$ and $Y \in \mathcal{Y}$ are the input and output respectively. 
		\textbf{Remark}: We sometimes refer to the discrete channel simply as $p(y|x)$. 
	\end{definition}	
	
	\begin{figure}[hbt!]%--- Picture 'H'ere, 'B'ottom or 'T'op; '!' Try to
	                    %impose your will to LaTeX
	  \centering
	  \includegraphics[width=.5\textwidth]{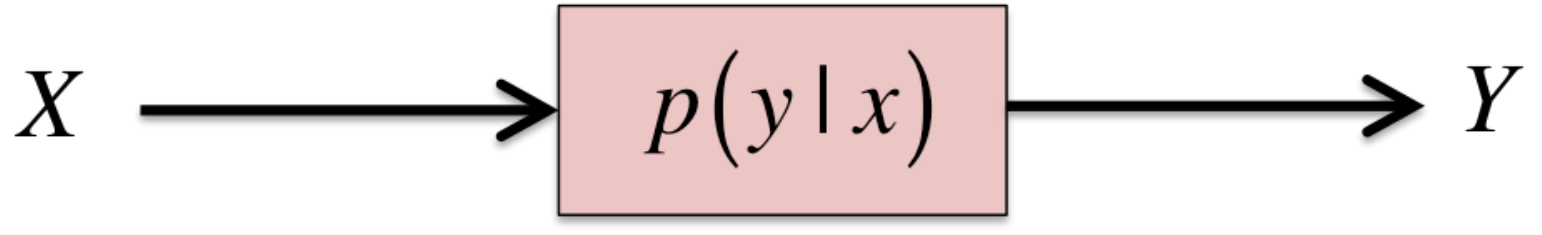} %<< no file extension
	  %%         --- .5\textwidth stands for 50% of text width
	  \caption[Discrete channel: schematic of input and output of a discrete channel]%<<-- Legend for the list of figures at the beginning of you thesis
	  {Discrete channel: schematic of input and output of a discrete channel}% legend displayed below the graph.
	  \label{fig:channel}
	\end{figure}
	
	If Input $X=x$ is \emph{sent} through the channel then the output $Y$ is distributed according to $p(y|x)$.
	
	\hspace{1mm}
	
	\begin{definition}[Discrete Memoryless Channel (DMC)]
		A discrete memoryless channel is a discrete channel $\dc$ such that if a sequence of inputs $X_1, X_2,...$ are sent through the channel then:
		
		\begin{align}
			Y_t \ci \{X_1,Y_1,X_2,Y_2,...,X_{t-1},Y_{t-1}\} | X_t
		\end{align}
		
	\end{definition}	
	
	Intuitively, the channel \emph{forgets} all previous communication such that the ouptut of $Y_t$ only depends on the input $X_t$ and on the distribution $p(y|x)$.\\
	
	\begin{definition}[Capacity of a DMC]
		The capacity of a DMC $\dc$ is defined as:
		\begin{align}
			C := \max_{p(x)} I(X;Y)
		\end{align}	
		Where $X$ and $Y$ are the input and output of the channel. 
	\end{definition}	
	
	The capacity of a DMC is a measure of the amount of common information between the input and output in the most optimistic scenario. As we will see later in this section, \SNCCt shows why the capacity of a DMC is an important quantity. 
	
	Since a given channel $\dc$ only takes as input $X \in \mathcal{X}$ we need an encoder function to transform our message $w \in \mathcal{W}$ into an acceptable input. If $|\mathcal{X}|=M$ then we may simply assign each message an element of the input set $\mathcal{X}$, however this doesn't help us avoid errors in communication. If the channel transforms the message such that the output is not the same as the input then an error will occur. 
	
	If $|\mathcal{X}|>M$ we have some \emph{slack} in our input set $\mathcal{X}$ which may help us to avoid errors. Suppose that $\mathcal{X}=\{x_1, x_2, x_3\}$ and $\mathcal{W}=\{0,1\}$ then we can assign 0 to $x_1$ and 1 to $x_2$ and $x_3$. If an $x_2$ is sent through and the channel distorts it into an $x_3$ we still avoid error. 
	
	If $|\mathcal{X}|\leq M$ we may add \emph{slack} to our coding scheme by encoding each message $w \in \mathcal{W}$ with a sequence of $n$ symbols $x_i \in \mathcal{X}$. In this case we have $|\mathcal{X}|^n$ sequences to encode $M$ messages. If we let $n$ grow then we increase the slack in our code and so reduce the probability of error, especially if we assign sets of sequences to each message in a \emph{smart} way. To prove the Noisy-Channel Coding theorem Shannon constructed such a smart assignment procedure using ideas of \emph{typicality} which we explore somewhat further on. 
	
	We can already touch on how the pattern analysis problem bares some resemblance to the problem of sending a message through a noisy channel: in the former there is some \emph{truth} or property in nature which is a hypothesis $c \in \mathcal{C}$ and it is encoded in a slack way by a \DGM such that for each hypothesis there correspond many possible data sets $X^{(n)} \in \mathcal{X}^n$. 

Having broached the idea of slack codes we now define an $(n,M)$ code and give a schematic description of Shannon's communication scenario.\\ 

\begin{definition}[$(n,M)$ code]
An $(n,M)$ code for a DMC $\dc$ is defined by an encoding function $f$ and a decoding function $g$:

\begin{tabular}{p{5cm} p{5cm}}
	\centering
	$f:\mathcal{W} \rightarrow \mathcal{X}^n$ & $g: \mathcal{Y}^n \rightarrow \mathcal{W}$ \\
\end{tabular}	

Where: 
\begin{itemize}
	\item $\mathcal{W}=\{1,...,M\}$ is called the message set,
	\item $f(1),...,f(M) \in \mathcal{X}^n$ are the codewords and
	\item $\{f(1),...,f(M)\}$ is the codebook.
\end{itemize}	
\end{definition}

	\begin{figure}[hbt!]%--- Picture 'H'ere, 'B'ottom or 'T'op; '!' Try to
	                    %impose your will to LaTeX
	  \centering
		  \includegraphics[width=1\textwidth]{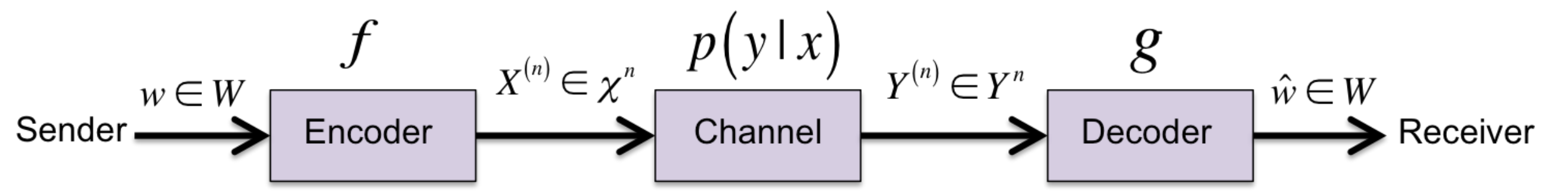} %<< no file extension
	  %%         --- .5\textwidth stands for 50% of text width
	  \caption[Communication channel: schematic of communication channel]%<<-- Legend for the list of figures at the beginning of you thesis
	  {Communication channel: schematic of communication channel}% legend displayed below the graph.
	  \label{fig:commChannel}
	\end{figure}

		The rate of an $(n,M)$ code is defined as $\frac{1}{n}\log M$.  This corresponds to the rate of a uniform random variable over the message set $\mathcal{W}$ divided by $n$ so that it is in the units of bits (or nats depending on the base of the logarithm) per symbol $X \in \mathcal{X}$ and not bits per sequence $X^{(n)} \in \mathcal{X}^n$. 

A rate $R$ for a DMC is said to be \emph{asymptotically achievable} if there exists an $(n,M)$ code that for a sufficiently large $n$ can transmit at a rate arbitrarily close to $R$ with arbitrary precision.\\ 

\begin{theorem}[\SNCCt]
	A rate $R$ is asymptotically achievable for a DMC $\iff R \leq C$ 
\end{theorem}	

This theorem justifies the \emph{Capacity} of a channel as an interesting quantity: it implies that we can achieve error-free communication at a rate equal to the capacity of the DMC. The proof of this part of the theorem involves proposing the $(n,M)$ code shown below and then proving that for an $n$ such that the rate $R=\rate$ of the code is close to the capacity of the DMC, the probability of error $P(w\neq \hat{w})$ is small. In this work we are especially interested in the $(n,M)$ code proposed in Shannon's proof since it forms the basis of a similar coding scheme and communication protocol in which generalization capacity plays an analogous role to that of channel capacity. \\

\begin{definition}[Shannon's (n,M) code]
The $(n,M)$ code proposed is the following:
\begin{enumerate}[1]
	\item Sample $M$ sequences $X^{(n)}$ uniformly at random from $\mathcal{X}^n$ and randomly assign each sequence sampled to one of the messages. This establishes the encoding function $f$. Both sender and receiver have the codebook $\{f(1),...,f(M)\}$. 
	
	\item Compare the joint entropy of  $H(X,Y)$, to the empirical entropy of the the pairs of sequences $(f(1), Y^{(n)}),...,(f(M), Y^{(n)})$. Choose message $i$ such that the empirical entropy of $(f(i), Y^{(n)})$ is close to the entropy $H(X,Y)$. If there is more than one pair of sequences that satisfies this condition decode to $\hat{w}=1$. This establishes the decoding function $g$. 
\end{enumerate}	
\end{definition}

The proof that this $(n,M)$ code can asymptotically achieve a rate $R=C$ involves the concept of \emph{typicality} which is an application of the \emph{Law of Large Numbers}. Although we do not give the formal proof we give a sequential illustration of the ideas. 

\begin{enumerate}
	\item \textbf{Start with a large $n$ for a slack code.} First we choose $n$ large so that we have a lot of slackness in our code, more than we will need, and uniformly at random choose $M$ sequences $X^{(n)} \in \mathcal{X}^n$ as our codewords. 
	
	\begin{figure}[H]%--- Picture 'H'ere, 'B'ottom or 'T'op; '!' Try to
	                    %impose your will to LaTeX
	  \centering
		  \includegraphics[width=0.2\textwidth]{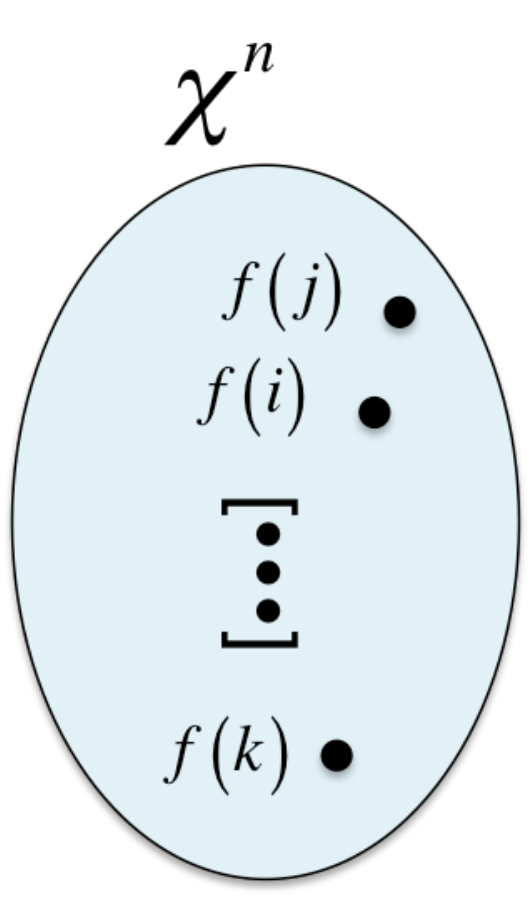} %<< no file extension
	  %%         --- .5\textwidth stands for 50% of text width
	  \caption[Shannon code: creation of codebook]%<<-- Legend for the list of figures at the beginning of you thesis
	  {Shannon code: creation of codebook}% legend displayed below the graph.
	  \label{fig:typ1}
	\end{figure}
	
	\item \textbf{Channel sends messages to non-overlapping regions}. We have chosen $n$ so large that even with a lot of noise, when $f(1),...,f(M) \in \mathcal{X}^n$ are transformed into $Y_1^{(n)},..., Y_M^{(n)} \in \mathcal{Y}^n$ by the channel $\dc$ the probability that $Y_i$ and $Y_j$ are \emph{close} for any $i \neq j$ is essentially zero. The i-th region $\mathcal{R}_i$ represents the sequences $Y^{(n)} \in \mathcal{Y}^n$ which are jointly typical with $f(i)$. Given that the sequence passed over the channel is $f(i)$, the probability that the sequence received by the decoder is outside this region is essentially zero. If $Y_j^{(n)} \in \mathcal{R}_i$ and $\nexists k \neq i : Y_j^{(n)} \in \mathcal{R}_k$ then $g(Y_j^{(n)})=i$. 
	
	\begin{figure}[H]%--- Picture 'H'ere, 'B'ottom or 'T'op; '!' Try to
	                    %impose your will to LaTeX
	  \centering
		  \includegraphics[width=0.5\textwidth]{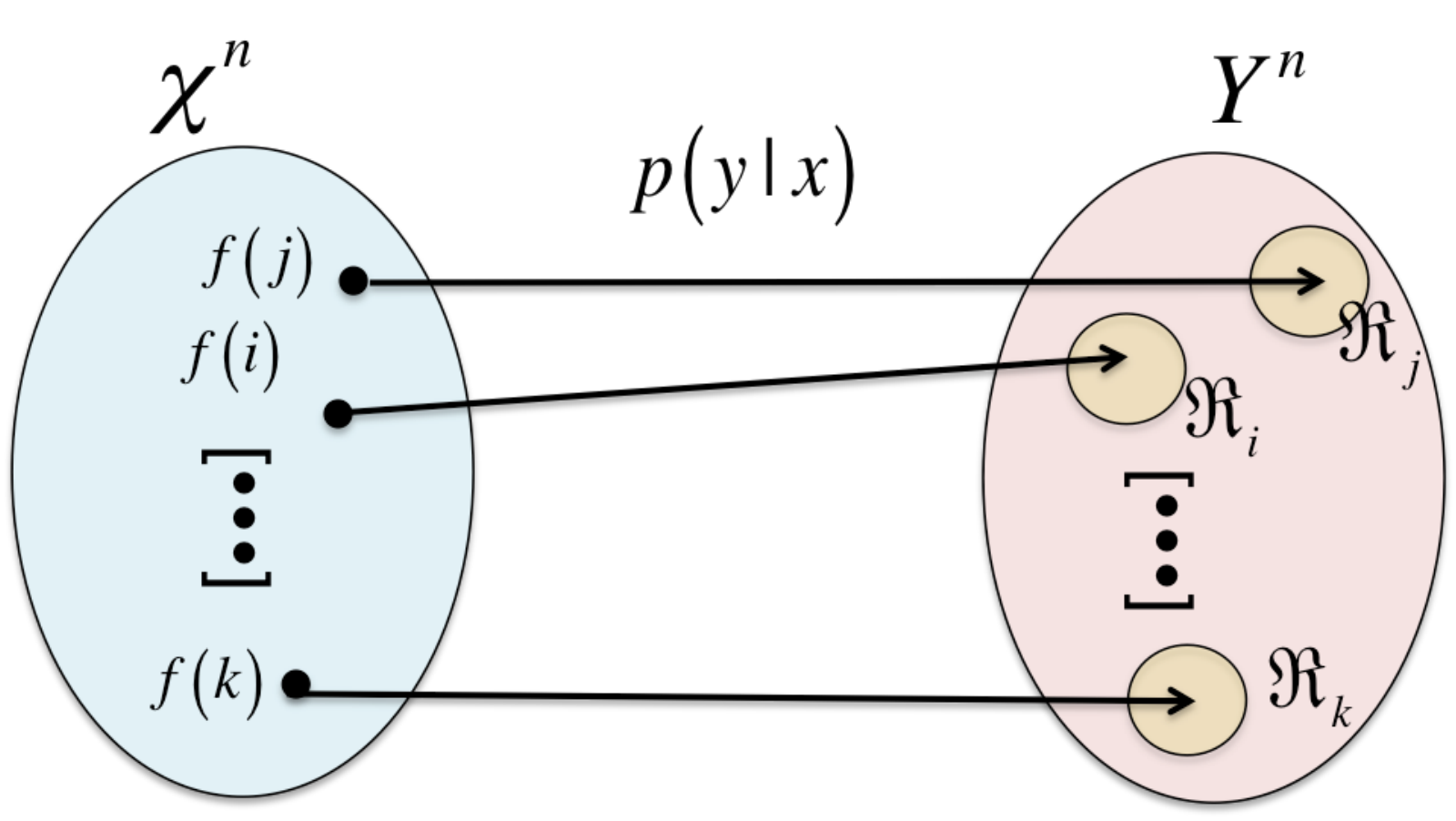} %<< no file extension
	  %%         --- .5\textwidth stands for 50% of text width
	  \caption[Shannon code: decoding]%<<-- Legend for the list of figures at the beginning of you thesis
	  {Shannon code: decoding}% legend displayed below the graph.
	  \label{fig:typ2}
	\end{figure}
	
	\item \textbf{Decrease $n$ until $\mathcal{Y}^n$ is tight around regions}. Since $n$ is large the rate $R=\rate$ of the code is low. \SNCCt says that we can decrease $n$ so that the rate increases to close to $C$ and the error stays very small. By decreasing $n$ we decrease the size of $\mathcal{Y}^n$ so that all the regions $\mathcal{R}_i$ are tightly crowded within. If we are at capacity, the overlap between the regions is still essentially zero, but if we make $\mathcal{Y}^n$ any smaller by decreasing $n$ further, the overlap will start to grow, meaning the probability of error grows. 
	\begin{figure}[H]%--- Picture 'H'ere, 'B'ottom or 'T'op; '!' Try to
	                    %impose your will to LaTeX
	  \centering
		  \includegraphics[width=0.5\textwidth]{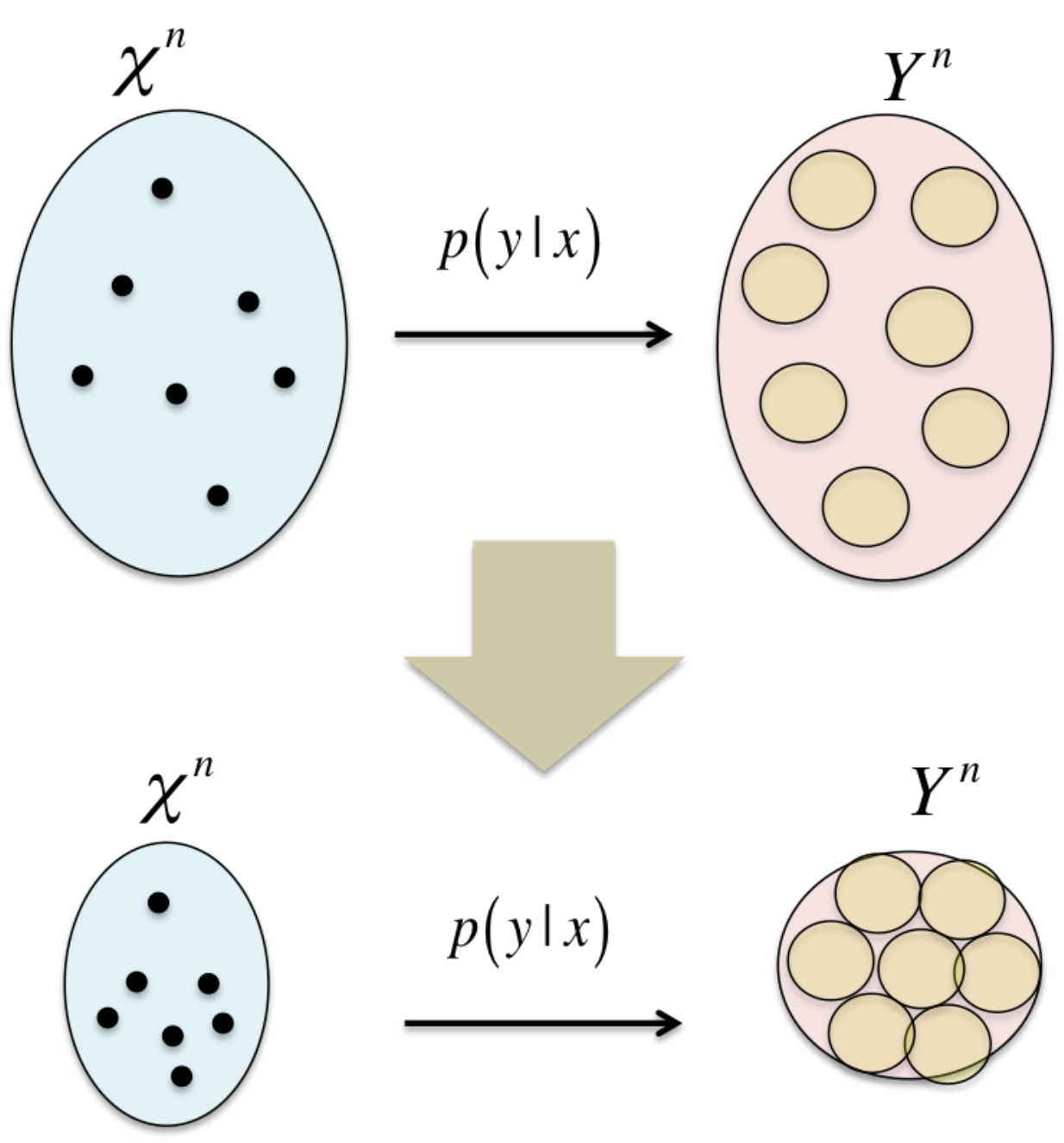} %<< no file extension
	  %%         --- .5\textwidth stands for 50% of text width
	  \caption[Shannon code: towards capacity]%<<-- Legend for the list of figures at the beginning of you thesis
	  {Shannon code: towards capacity}% legend displayed below the graph.
	  \label{fig:typ3}
	\end{figure}
\end{enumerate}	

\hspace{.5mm}
As we have already hinted at, the pattern analysis problem can be seen as a special case of the communication problem. We explore this further by proposing the following \emph{\ILP}:

	\begin{figure}[H]%--- Picture 'H'ere, 'B'ottom or 'T'op; '!' Try to
	                    %impose your will to LaTeX
	  \centering
		  \includegraphics[width=1\textwidth]{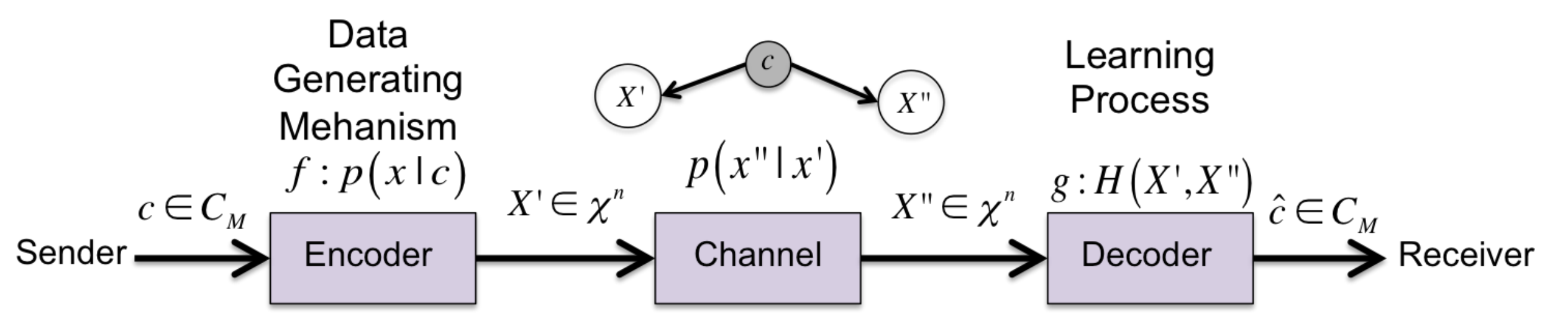} %<< no file extension
	  %%         --- .5\textwidth stands for 50% of text width
	  \caption[Idealized learning protocol]%<<-- Legend for the list of figures at the beginning of you thesis
	  {Idealized learning protocol}% legend displayed below the graph.
	  \label{fig:IdeLearn}
	\end{figure}

The communication protocol has the following characteristics:

\begin{enumerate}[1]
	\item The sender picks $M$   uniformly at random to construct $\mathcal{C}_M =\{c_1,...,c_M\} \subseteq \mathcal{C}$: $c_i \sim p(c)=\frac{1}{|\mathcal{C}|}$.
	\item The sender selects a message $\widetilde{c}_s \in \mathcal{C}_M$ uniformly at random: $\widetilde{c}_s \sim p_s(c)=\frac{1}{M}$. \textbf{Remark:} we use the tilde to separate $c_i \sim p(c)=\frac{1}{|\mathcal{C}|}$, $i \in \{1,...,M\}$ from $\widetilde{c}_s \sim p_s(c)=\frac{1}{M}$.
	\item The sender and receiver have access to the data generating mechanism $p(x|c)$ which they use to set up the following $(n,M)$ code:
		\begin{enumerate}[a]
				\item The encoder function $f$ is constructed by randomly sampling $M$ times a data set of size $n$ from the data generating mechanism $p(x|c)$ to obtain the codebook $\{f(c_1),...,f(c_M)\}$ where $f(c_i) \in \mathcal{X}^n$. Notice that in many problems such as clustering and regression $\mathcal{X}=\realnumbers^d$ so that $\mathcal{X}^n = \realnumbers^{n \times d}$, i.e. we code each message using a data matrix. 
				\item Our decoding function $g$ works as in Shannon's $(n,M)$ code except that $\mathcal{X}$ is not necessarily a finite set so we might need to use joint \emph{differential entropy} instead of joint entropy. 
		\end{enumerate}
	\item We set up a channel $(\mathcal{X}^n,p(x''|x'),\mathcal{X}^n)$ with our knowledge of the hypothesis and data generating mechanisms $p_s(c)$ and $p(x|c)$ respectively:
	
	\begin{align} \label{channDist}
		p(x''|x') \propto p(x',x'') = \sum_{c \in \mathcal{C}} p(c,x',x'') &= \sum_{c \in \mathcal{C}} p_s(c)p(x',x''|c) \\
		&= \frac{1}{M}\sum_{c \in \mathcal{C}} p(x'|c)p(x''|c)
	\end{align}
	
	Notice that we are assuming that successive sample sets of size $n$ from the data generating mechanism are independent given $c$. 
				
\end{enumerate}	

With the exception that we are using what may be a set $\mathcal{X}$ with infinite cardinality to code the message set $\mathcal{C}_M=\{c_1,...,c_M\}$ the above \ILP corresponds to the previous communication protocol. The channel $p(x''|x')$ characterizes the noisiness of the pattern analysis problem since in a noise-free scenario we would obtain the same data set for each realization of the data generating mechanism, i.e. $X'=X''$. Since messages are always selected uniformly at random according to $p_s(x')=p_s(c)=\frac{1}{M}$ we may consider the capacity of the channel to be $I(X';X'')$. The capacity of the channel is a measure of the noisiness (the higher the capacity the less is the noise) and is an upper bound on the rate at which any learning algorithm can extract information from specific realizations $X \in \mathcal{X}^n$, that generalizes accross realizations, i.e. information about $c$ and not about the noise. In the above \ILP we can achieve the capacity rate, as before, by choosing a suitable $n$. Since this scenario is highly idealized we make successive changes to it until we arrive at the more useful \LP proposed by \citeauthor{Buhm13} and from which generalized capacity is derived:

\begin{enumerate}[I]
	\item \textbf{Change \emph{expressiveness} of codebook instead of size of code sequences}. Suppose we can no longer change $n$, the size of our sequence $X \in \mathcal{X}^n$, i.e. it is fixed. Instead we are allowed to change $M$ the number of   selected to form $\mathcal{C}_M$. We can now achieve a rate close to capacity by increasing $M$ instead of decreasing $n$. Our message set $\mathcal{C}_M$ and codebook $\{f(c_1),...,f(c_M)\}$ become more \emph{expressive} as we increase $M$. 
	
	\item \textbf{Coding based on transformation set}. Additionally, suppose we can only use the data generating mechanism $p(x|c)$ and channel $p(x''|x')$ once. We still know the form of $p(x|c)$ and $p(x''|x')$ (and so can calculate entropies for decoding) but can only generate with it once. Furthermore, suppose that we don't know what  hypothesis $c_D$  is selected and passed to the data generating mechanism $p(x|c)$. Since we don't have any information about $c_D$ other than the data $X'$ and $X''$, we use a uniform prior $c_D \sim p(c)=\frac{1}{|\mathcal{C}|}$ to describe our uncertainty regarding the true $c_D$. All this means we can only generate two data sets $X',X'' \in \mathcal{X}^n$: we generate $X'$ using $p(x|c)$ and then send it through the channel $p(x'|x'')$ to get $X''$. Without access to the data generating mechanism and the true $c_D$ we need an alternative way to construct a codebook (i.e. another $f$ for our $(n,M)$ code). Consider the set of unique maps:
	
	\begin{align} \label{TransfSet}
		\mathcal{T}^h = \{t^h \neq u^h \in \mathcal{T}: \forall c \neq d \in \mathcal{C}  \Rightarrow t^h(c) \neq t^h(d) \textrm{ and } t^h(c) \neq u^h(c) \}
	\end{align}

	Where $\mathcal{T}= \{t:\mathcal{C} \rightarrow \mathcal{C} \}$.
	
	Notice the following properties about $\mathcal{T}^h$:
	\begin{enumerate}
		\item $|\mathcal{T}^h|=|\mathcal{C}|$
		\item If you apply a fixed $t^h\in\mathcal{T}^h$ on all $c \in \mathcal{C}$ you get $C$ again. 
		\item If you apply all $t^h\in\mathcal{T}^h$ on a fixed $c \in \mathcal{C}$ you get $C$ again. 
	\end{enumerate}	
	
	Additionally, consider the set of maps: 
	
	\begin{align}
		\mathcal{T}^D = \{t^D:\mathcal{X}^n \rightarrow \mathcal{X}^n\} %:  t^D \in \mathcal{T}^D \Rightarrow  \exists t^c \in \mathcal{T}^c : \forall c \in \mathcal{C} \textrm{ and } X \in \mathcal{X}^n \textrm{ } p(c|x)=p(t^h(c)|t^D(x)) \}
	\end{align}
	
	We assume that we have a mapping $\phi:\mathcal{T}^h \rightarrow \mathcal{T}^D$ such that for a given $t^h \in \mathcal{T}^h$, $\phi(t^h)=t^D \in \mathcal{T}^D$ and: 
	
	\begin{align} \label{transfAssumption}
		p(c|x)=p(t^h(c)|t^D(x))
	\end{align}	
	
	Where the posteror $p(c|x)$ is obtained from the data generating mechanism $p(x|c)$ and the prior $p(c)$: 
	
	\begin{align}
		p(c|x) \propto p(x|c)p(c)
	\end{align}	
	
	This means that if data set $X$ is generated under hypothesis $c$ with the data generating mechanism $p(x|c)$ then, for a given \emph{move} $t^h$ within the hypothesis space $\mathcal{C}$ we know how to make a corresponding \emph{move} $t^D$ in the coding/data space $\mathcal{X}^n$.  The assumption that we can obtain a mapping $\phi$ is reasonable in some contexts such as in the sparse mean estimation problem that is the main topic of this work. In other pattern analysis problems such as in the mean estimation problem \ref{meanEst} where the hypothesis class $\mathcal{C}$ has infinite cardinality, the validity of this assumption is not clear. 
	
	With the above assumption we will be able to encode $M$   $c_i \in \mathcal{C}_M$ into a codebook $\{f(c_1),...,f(c_M)\} \subseteq \mathcal{X}^n$ However we can only pass one data set through the channel $p(x''|x')$ and arbitrarily choose to pass $f(c_D)=X'=x'_0$ which gives the random output $X''=x''_0$. Observe that: 
	
	\begin{align}
		p(t^D_i(x'),t^D_i(x'')) &= \sum_{c \in \mathcal{C}} p(c, t^D_i(x'),t^D_i(x''))\\
								&= \sum_{c \in \mathcal{C}} p(c)p(t^D_i(x'),t^D_i(x'')|c) \\
								&= \sum_{c \in \mathcal{C}} p(c)p(t^D_i(x')|c)p(t^D_i(x'')|c) \\
								&= \sum_{c \in \mathcal{C}} p(t^h_i(c))p(t^D_i(x')|t^h_i(c))p(t^D_i(x'')|t^h_i(c)) \\
								&= \sum_{c \in \mathcal{C}} p(c)p(x'|c)p(x''|c) \\
								&= \sum_{c \in \mathcal{C}} p(c)p(x',x''|c) \\
								&= \sum_{c \in \mathcal{C}} p(c,x',x'')=p(x',x'') 
	\end{align}	
	
	Where we have used assumption \ref{transfAssumption} and the properties of \ref{TransfSet}. This implies that:
	
	\begin{align}
		p(t^D_i(x'')|t^D_i(x'))=p(x''|x')
	\end{align}

	We wish to \emph{mimic} the channel $p(x''|x')$ by mimicking the noise process that \emph{contaminated} $x'_0$ to produce $x''_0$. We may think of the output of the channel $p(x''|x')$ as a function $\alpha$ of the input $x'$  and a noise realization of some $N$. We then have that 
	
	\begin{align}
		X'' &= \alpha(x'_0,N) \sim p(x''|x'_0) \\
		t^D_i(X'') &= \alpha(t^D_i(x'_0),N) \sim p(x''|x'_0)
	\end{align}	
	
	Since we have passed $x'_0$ through the channel and have observed $X''=x''_0$ we implicitly have a noise observation $N=n_0$. Although, we can't pass $t^D_i(x'_0)$ through the channel to get $t^D_i(X'')$ (recall we are only allowed to use the channel once, and we have already used it to pass $x_0'$ through), we can mimic the output with $t_i^D(x''_0)$ which is the result of evaluating $\alpha$ on $n_0$ instead of on a new realization of $N$:
	
	\begin{align}
		t^D_i(X'') &= \alpha(t^D_i(x'_0),N)\\
		t^D_i(x''_0) &= \alpha(t^D_i(x'_0),n_0)
	\end{align}	
	
	With the set $\mathcal{T}^c$ and the mapping $\phi$ we have the necessary elements to replace our encoding function $f$. Incorporating the changes to the encoding function and channel, the learning protocol, thus far, consists of:
	
	\begin{enumerate}[1]
		\item The sender picks $M$ transformations uniformly at random to construct $\mathcal{T}_M^h =\{t^h_1,...,t^h_M\} \subseteq \mathcal{T}^h$: $t^h_i \sim p(t^h)=\frac{1}{|\mathcal{T}^h|}=\frac{1}{|\mathcal{C}|} $. Even though we don't know what $c_D$ is we can set up our message set $C_M$ using $\mathcal{T}_M^h$: $C_M= \{c_1,...,c_M\}=\{t^h_1(c_D),...,t^h_M(c_D)\}$. In fact, we may now say that $\mathcal{T}^h_M$ \emph{is} the message set. 
		\item The sender selects a message $\widetilde{t}^h_s \in \mathcal{T}^h_M$ uniformly at random: $\widetilde{t}_s^h \sim p_s(t^h)=\frac{1}{M}$. \textbf{Remark:} we use the tilde to separate $t_i^h \sim p(t^h)=\frac{1}{|\mathcal{C}|}$, $i \in \{1,...,M\}$ from $\widetilde{t}_s^h \sim p_s(t^h)=\frac{1}{M}$.
		\item The sender and receiver have access to the transformation set $\mathcal{T}^h_M$ and the mapping $\phi$ which they use to set up the following $(n,M)$ code:
			\begin{enumerate}[a]
					\item The encoder function $f$ is constructed by applying $\phi$ to each $t^h_i \in \mathcal{T}_M^h$ to construct  $\mathcal{T}_M^D=\{t^D_1,...,t^D_M\}=\{\phi(t^h_1),...,\phi(t^h_M)\}$. Our codebook vector is then $\{f(c_1),...,f(c_M)\}=\{f(t^h_1(c_D)),...,f(t^h_M(c_D))\}=\{t_1^D(X'),...,t_M^D(X')\}$.
					\item Since we still know the distribution $p(x|c)$ we may use it to calculate $p(x',x'')= \frac{1}{M}\sum_{c \in \mathcal{C}} p(x''|c)p(x''|c)$. Notice that since we know how $\widetilde{c}_s=\widetilde{t}_s^h(c_D)$ is selected we use $p_s(c)=\frac{1}{M}$. With $p(x',x'')$ we can calculate the joint entropy $H(X',X'')$ and use it for decoding as before. 
				\end{enumerate}	
				
		\item We can only send one data set $X'=x'_0$ through the channel $(\mathcal{X}^n,p(x''|x'),\mathcal{X}^n)$ and so can only observe one output $X''=x''_0$. However, we may mimic the behavior of the channel for other input data sets $t_i^D(X')$ by using, as derived above, that:
	
		\begin{align} \label{channDist2}
			p(t^D_i(x'')|t^D_i(x')) = p(x''|x')
		\end{align}		
				
		This means that we may mimic the channel $p(x''|x')$ by outputing $t^D_i(x''_0)$ for a given input $t^D_i(x'_0)$. This output corresponds to the output the actual channel $(\mathcal{X}^n,p(x''|x'),\mathcal{X}^n)$ would have given for an input $t^D_i(x'_0)$, assuming the same realization of the noise process as ocurred when $X'$ passed through the channel. Since the noise realization for one data set, is made up of $n$ components, the hope is that the realization observed summarizes the noisiness of the channel. In other words we hope that applying this noise realization to any input data set $t^D_i(x'_0)$, the output data set $t^D_i(x''_0)$ is similar to that we would get by passing $t^D_i(X')$ through the real channel (to which we no longer have access).  	
				
	\end{enumerate}			
	
	The assumption that we still know the distribution $p(x|c)$ is the last \emph{idealized}, unrealistic element of our learning protocol. The last change to the protocol involves dispensing with this assumption. To do so we will use the Boltzmann approximation sets discussed in sections \ref{ApproxSets} and \ref{GC}.\\

	\item \textbf{Decoding based on approximation sets}. Finally suppose we don't know the distribution of the data generating mechanism $p(x|c)$ or that of the channel $p(x''|x')$. This means we need a new decoding function $g$ since we don't know $p(x',x'')$ and so cannot use the joint entropy $H(X',X'')$ for decoding. This is where our learning algorithm comes to the fore in the form of the cost function $\ERs$ and the Gibbs distributions corresponding to both data sets: $\GibbsOne$ and $\GibbsTwo$. We first describe the new decoding function $g$ and then discuss the ideas behind it and its relationship to genealization capacity. 
	
	Let 
	
	\begin{align}
		\Delta Z_{\beta}^j := \sum_{c \in \mathcal{C}} w_{\beta}(c, t^D_j(X'))w_{\beta}(c, t^D_s(X''))
	\end{align}	
	
	Then the decoding rule is 
	
	\begin{align}
	 g_{\beta}(X'') \in \argmaxM{j \in \{1,...,M\}}  \Delta Z_{\beta}^j
	\end{align}
	
	Where ties are resolved by taking the minimum $j$. Before discussing how to choose the resolution parameter $\beta$ we can show the final \LP schematic. 
	
	\begin{figure}[H]%--- Picture 'H'ere, 'B'ottom or 'T'op; '!' Try to
	                    %impose your will to LaTeX
	  \centering
		  \includegraphics[width=1\textwidth]{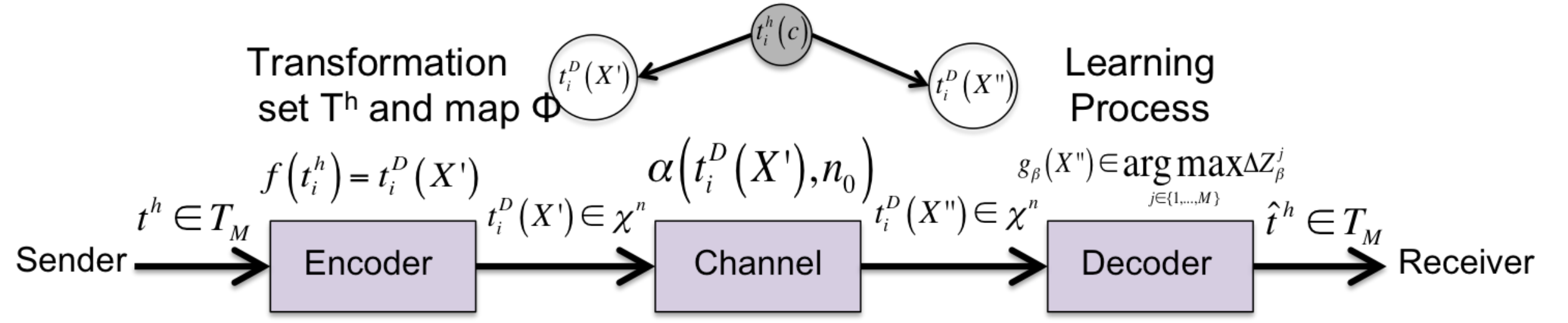} %<< no file extension
	  %%         --- .5\textwidth stands for 50% of text width
	  \caption[Learning protocol]%<<-- Legend for the list of figures at the beginning of you thesis
	  {Learning protocol}% legend displayed below the graph.
	  \label{fig:Learn}
	\end{figure}

What $\beta$ should we use? In general we can use any $\beta$, however to find the channel capacity we must choose it so that for a given $M$ (which determines the rate $R=\frac{1}{n}\log M$, given $n$ is fixed) we can achieve error-free communication. We then increase $M$ to $M^*$ such that if we increase it any further there exists no $\beta$ that allows error-free communication. This $M^*$ determines the maximum  achievable rate of our code. We illustrate the process of finding the maximum achievable rate of the above $(n,M)$ code.

\begin{enumerate}[i]
	\item \textbf{Start with low expressiveness $M$ for slack code}. We uniformly sample $M$   $c_i \in \mathcal{C}_M$ by sampling $M$ transformations $t_i^h \in \mathcal{T}^h_M$ and use the corresponding set $\mathcal{T}^D_M$ to build our codebook. We select a transformation $\widetilde{t}^h_s$ and pass the econcoded message $\widetilde{t}^D_s(X')$ through the channel to decoder that receives $\widetilde{t}^D_s(X'')$. 
	
	\begin{figure}[H]%--- Picture 'H'ere, 'B'ottom or 'T'op; '!' Try to
	                    %impose your will to LaTeX
	  \centering
		  \includegraphics[width=0.5\textwidth]{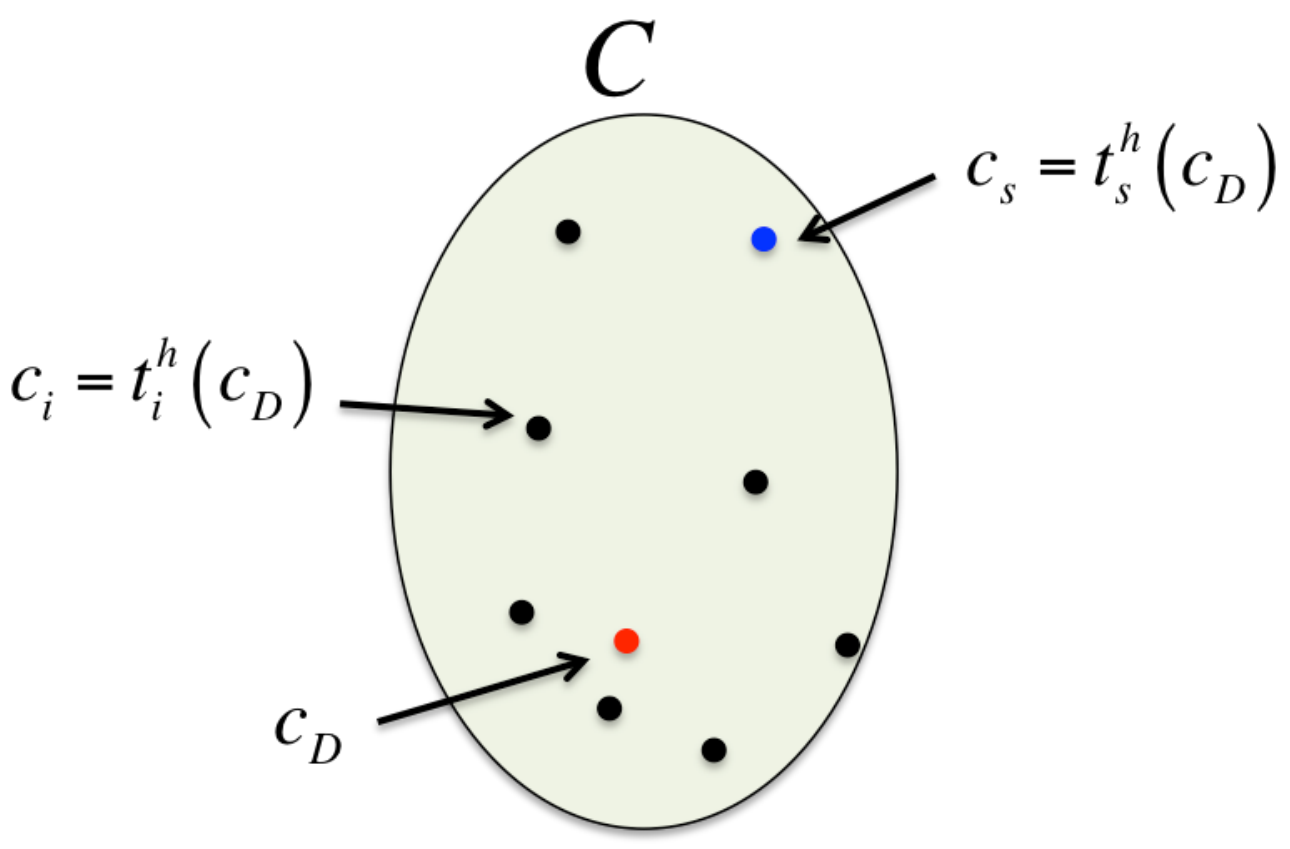} %<< no file extension
	  %%         --- .5\textwidth stands for 50% of text width
	  \caption[Learning protocol: creation of codebook]%<<-- Legend for the list of figures at the beginning of you thesis
	  {Learning protocol: creation of codebook}% legend displayed below the graph.
	  \label{fig:Learn1}
	\end{figure}

		\item \textbf{Create high resolution $\beta$ approximation sets}. Using high resolution $\beta$, we calculate $M$ approximation sets $w_{\beta}(c, t_j^D(X'))$, one for each codeword in the codebook. Using the received data set, we calculate an additional approximation set $w_{\beta}(c, \widetilde{t}_s^D(X''))$, and apply decoding rule. In the case illustrated in figure \ref{fig:Learn2}  $\Delta Z_{\beta}^j=0$ $\forall j$ so by default we decode message to $t_1^h(c_D)$. The red circles represent the approximation sets $w_{\beta}(c, t_j^D(X'))$ while the blue circle represents the approximation set $w_{\beta}(c, \widetilde{t}_s^D(X''))$. Recall from Section \ref{ApproxSets} that although, strictly speaking, $w_{\beta}(c, X)$ is a weight vector over the entire hypothesis class $\mathcal{C}$, the circles represent the subset of $\mathcal{C}$ where the \emph{majority} of the weight is supported.
	
	\begin{figure}[H]%--- Picture 'H'ere, 'B'ottom or 'T'op; '!' Try to
	                    %impose your will to LaTeX
	  \centering
		  \includegraphics[width=0.5\textwidth]{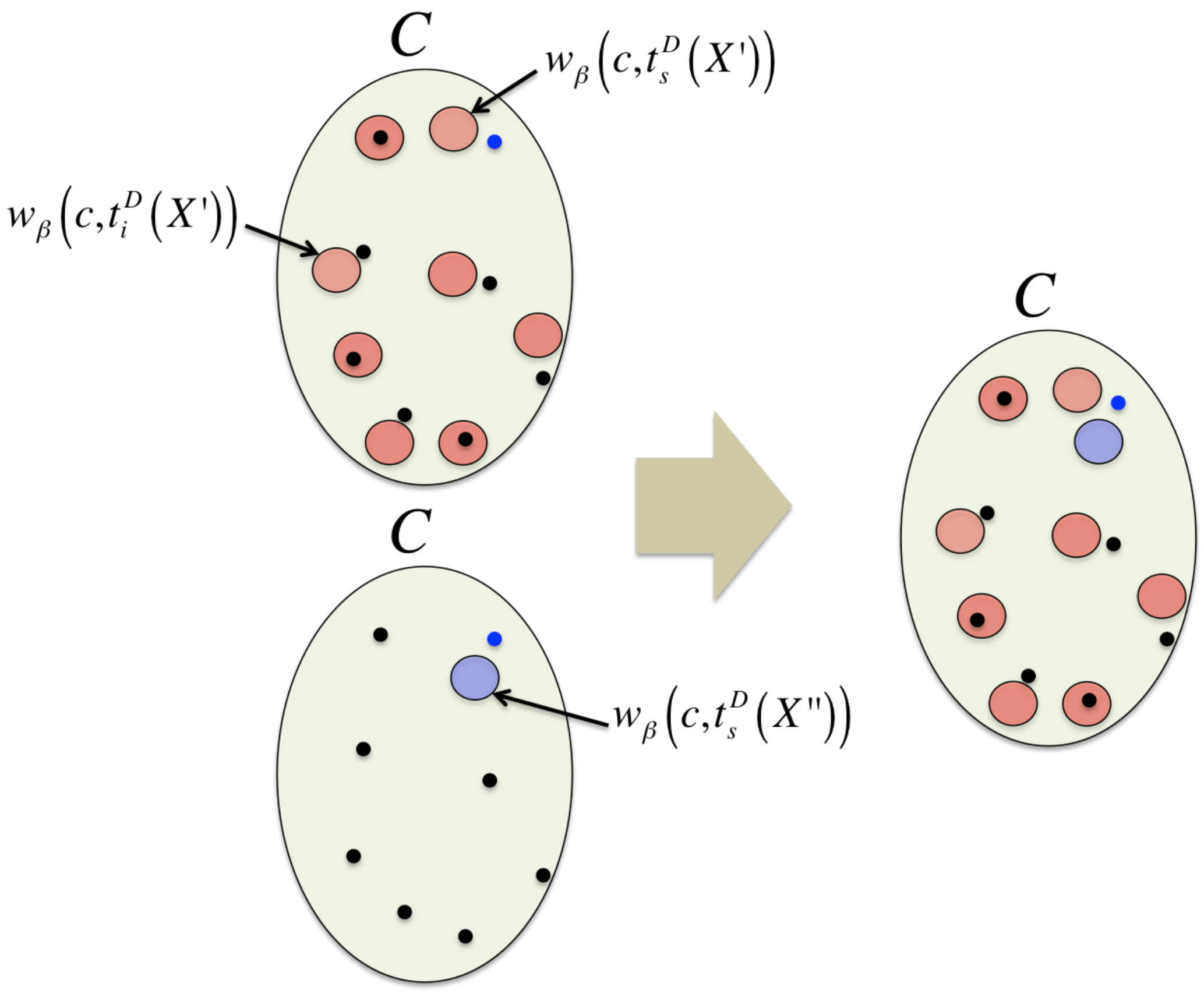} %<< no file extension
	  %%         --- .5\textwidth stands for 50% of text width
	  \caption[Learning protocol: decoding with high resolution]%<<-- Legend for the list of figures at the beginning of you thesis
	  {Learning protocol: decoding with high resolution}% legend displayed below the graph.
	  \label{fig:Learn2}
	\end{figure}
	
	\item \textbf{Lower resolution $\beta$ to increase intersection}. Since the current resolution level doesn't allow significant intersection between $w_{\beta}(c, \widetilde{t}_s^D(X''))$ and any of the $w_{\beta}(c, t_j^D(X'))$ approximation sets we lower the resolution level until there is some intersection meaning we can decode the message without error. Since the hypothesis class $\mathcal{C}$ is not \emph{cluttered} with approximation sets, because we are not near capacity, the intersection need not represent a large percentage of the respective approximation sets. 
	
	\begin{figure}[H]%--- Picture 'H'ere, 'B'ottom or 'T'op; '!' Try to
	                    %impose your will to LaTeX
	  \centering
		  \includegraphics[width=0.5\textwidth]{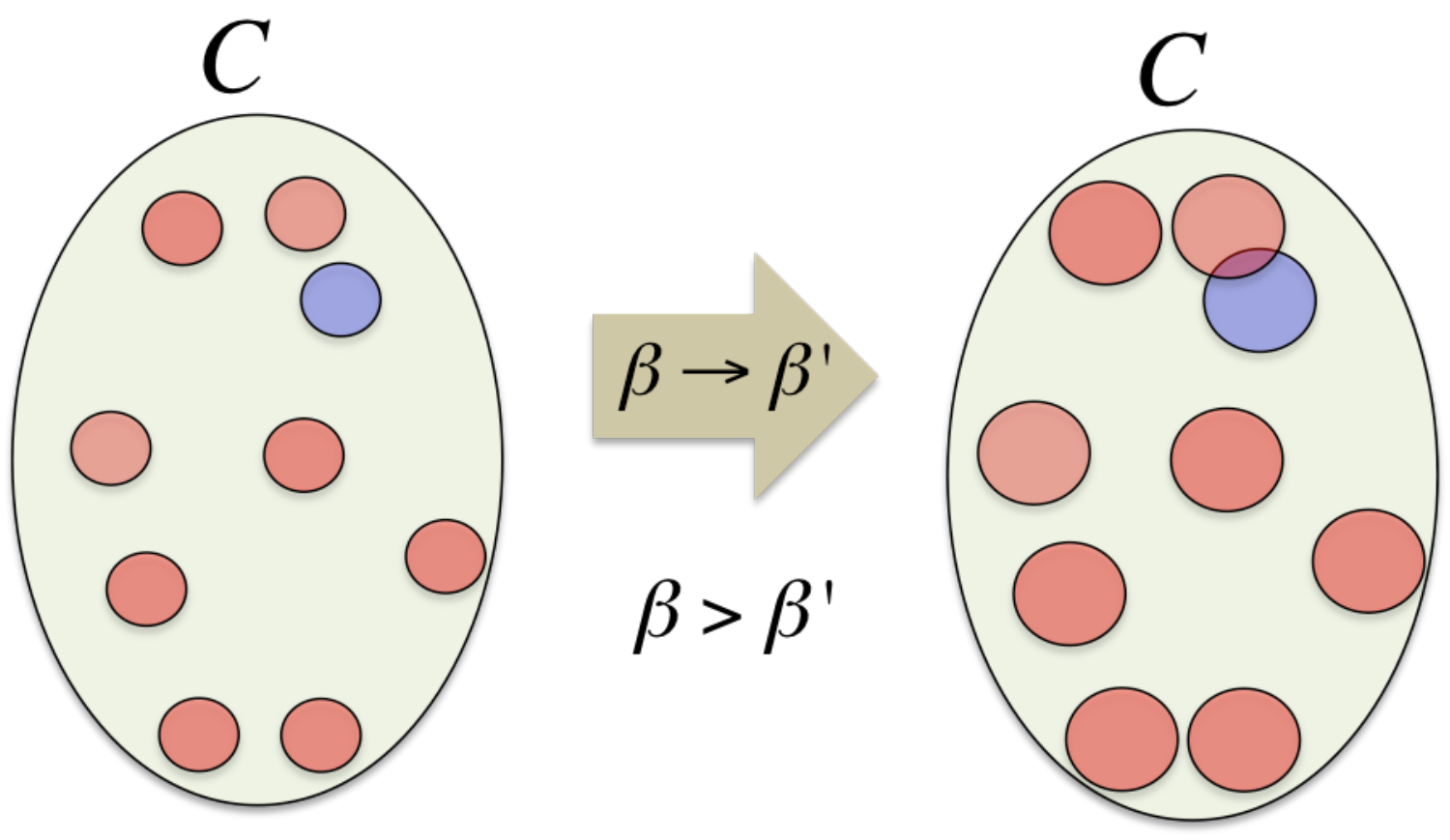} %<< no file extension
	  %%         --- .5\textwidth stands for 50% of text width
	  \caption[Learning protocol: decoding with low resolution]%<<-- Legend for the list of figures at the beginning of you thesis
	  {Learning protocol: decoding with low resolution}% legend displayed below the graph.
	  \label{fig:Learn3}
	\end{figure}
	
	\item \textbf{Increase expressiveness $M$, while adjusting resolution $\beta$}. In finding the maximum rate of our code we increase the size of the message set which means more approximation sets over $\mathcal{C}$. However, as happens in part (b) of figure \ref{fig:Learn4}, for a given resolution $\beta$ the approximation set $w_{\beta}(c, \widetilde{t}_s^D(X''))$ intersects with more than one approximation set $w_{\beta}(c, t_j^D(X'))$ meaning the probability of decoding the wrong message increases. We can, as is shown on part (c), fix this by decreasing the resolution but this increases the size of the approximation sets $w_{\beta}(c, t_j^D(X'))$ so that they intersect between themselves. This also increases the probability of error since the set $w_{\beta}(c, \widetilde{t}_s^D(X''))$ intersects with several $w_{\beta}(c, t_j^D(X'))$ sets simultaneously. Finally in part (d) we obtain obtain the maximum rate of our code: if we add any more approximation sets $w_{\beta}(c, t_j^D(X'))$, they will become cluttered between themselves (we go back to the situation in part c) and if we then fix this by increasing resolution, more than one of these sets will intersect significantly with $w_{\beta}(c, \widetilde{t}_s^D(X''))$ (we go back to the situation in part b). 
	
	\begin{figure}[H]%--- Picture 'H'ere, 'B'ottom or 'T'op; '!' Try to
	                    %impose your will to LaTeX
	  \centering
		  \includegraphics[width=1\textwidth]{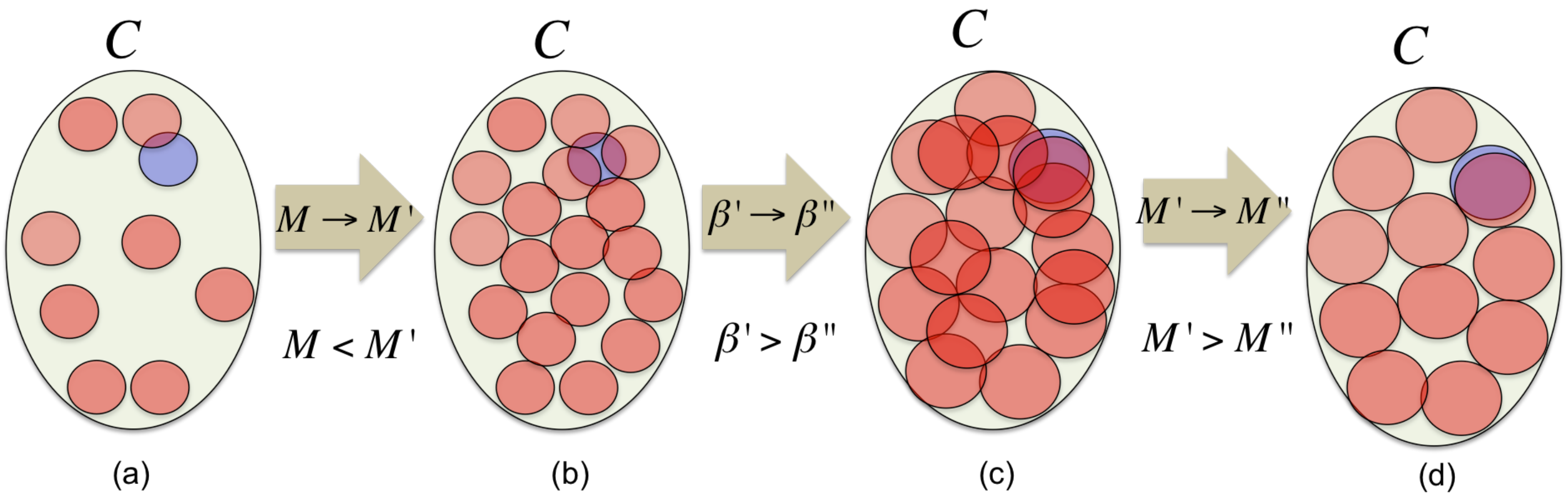} %<< no file extension
	  %%         --- .5\textwidth stands for 50% of text width
	  \caption[Learning protocol: maximizing learning rate]%<<-- Legend for the list of figures at the beginning of you thesis
	  {Learning protocol: maximizing learning rate}% legend displayed below the graph.
	  \label{fig:Learn4}
	\end{figure}
	
\end{enumerate}	

We have arrived at a fundamental trade-off between, on the one hand, the expressiveness of our code (the cardinality of $\mathcal{T}^h_M$) and the resolution of our decoding mechanism (the parameter $\beta$) and on the other, the stability of the learning protocol measured by the probability of error, itself a function of the \emph{quality} of the overlap $\max_{j \in \{1,...,M\}} \Delta Z_{\beta}^j$. 

\end{enumerate}	

Suppose the sender sends the hypothesis $c_j=t_j^h(c_D)$ corresponding to the dataset $t_j^D(X')$. We want to compare the approximation sets $w_{\beta}(c, t_j^D(X'))$ and $w_{\beta}(c, t_j^D(X''))$ to get an idea of what needs to happen to be able to achieve the maximum learning rate. For a given resolution $\beta$:

\begin{enumerate}[1]
	\item We want to be able to increase the size of the message set, taking care that the approximation sets $w_{\beta}(c, t_j^D(X'))$ don't become cluttered. We can achieve this by maximizing:
	\begin{align}
		\frac{|\mathcal{C}|}{Z_{\beta}(t_j^D(X'))}
	\end{align}	
	
	This criterion will tend to make $Z_{\beta}(t_j^D(X'))$ small. 
	\item We want to make sure that the \emph{quality} of the overlap $\Delta Z_{\beta}(t^D_j(X'),t^D_j(X''))$ is good. However we need a relative measure since low resolution communication will in general lead to a bigger $\Delta Z_{\beta}(t^D_j(X'),t^D_j(X''))$. We can achieve this by maximizing: 
	
	\begin{align}
		\frac{\Delta Z_{\beta}(t^D_j(X'),t^D_j(X''))}{Z_{\beta}(t_j^D(X''))}
	\end{align}	
	
	This criterion will tend to make $Z_{\beta}(t_j^D(X''))$ small and $\Delta Z_{\beta}(t^D_j(X'),t^D_j(X''))$ large. 
\end{enumerate}	

Now notice that for a given $X'$, $X''$ and $\beta$ these two quantities don't depend on the transformation $t^D_j$ due to  assumption \ref{transfAssumption} so that we may assume without loss of generality that $t_j^h(c)=c$ and $t_j^D(X)=X$. So for a given $X'$ and $X''$ the bigger $\frac{|C|}{Z_{\beta}(X')}$ and $\frac{\Delta Z_{\beta}(X',X'')}{Z_{\beta}(X'')}$ the larger the learning rate will be. This gives us a qualitative notion of why generalization capacity defines the maximum learning rate of our learning algorithm. 

We now analyze the probability of error of the above \LP to more formally understand the role of generalization capacity. 
	
Recall what the random quantities in the \LP are:

\begin{itemize}
	\item $X'$,$X'' \sim p(x|c)$
	\item $\mathcal{T}^h_M=\{t^h_1,...,t^h_M\}$, where $t^h_j \sim p(t^h)=\frac{1}{|\mathcal{C}|}$
	\item $\widetilde{t}^h_s \sim p_s(t^h)=\frac{1}{M}$ 
\end{itemize}	

Now let $J_{s^-} := \{1,...,M\} \setminus \{s\}$. We have that: 

\begin{align}
	\mathbb{P}(\hat{t}^h \neq \widetilde{t}^h_s | \widetilde{t}^h_s) &= \mathbb{P}(\max_{j \in J_{s^-}} \Delta Z_{\beta}^j \geq \Delta Z_{\beta}^s | \widetilde{t}^h_s) \\
											 &= \mathbb{P}(OR_{j \in J_{s^-}} (\Delta Z_{\beta}^j \geq \Delta Z_{\beta}^s) | \widetilde{t}^h_s) \\
											 &\leq \sum_{j \in J_{s^-}}\mathbb{P}(\Delta Z_{\beta}^j \geq \Delta Z_{\beta}^s | \widetilde{t}^h_s) \\
											 &= \sum_{j \in J_{s^-}}\mathbb{P}\bigg(\frac{\Delta Z_{\beta}^j}{\Delta Z_{\beta}^s} \geq 1 | \widetilde{t}^h_s\bigg)
\end{align}	

Where we have used the union bound to establish the inequality. Note that $\frac{\Delta Z_{\beta}^j}{\Delta Z_{\beta}^s}$ is a random variable depending on $X'$, $X''$, $\mathcal{T}^h_M$ and $\widetilde{t}^h_s$. Using the Markov inequality and the independence of $(X',X'')$ and $t^h_j$ we see the following: 

\begin{align}
	\mathbb{P}(\hat{t}^h \neq \widetilde{t}^h_s | \widetilde{t}^h_s) &  \leq \sum_{j \in J_{s^-}}\mathbb{P}\bigg(\frac{\Delta Z_{\beta}^j}{\Delta Z_{\beta}^s} \geq 1 | \widetilde{t}^h_s\bigg) \\
	&\overset{(1)}{\leq} \sum_{j \in J_{s^-}}\mathbb{E}_{(X',X'',\mathcal{T}^h_M, \widetilde{t}^h_s)}\bigg[\frac{\Delta Z_{\beta}^j}{\Delta Z_{\beta}^s}  | \widetilde{t}^h_s\bigg] \\
	&\overset{(2)}{=} \sum_{j \in J_{s^-}}\mathbb{E}_{(X',X'',\mathcal{T}^h_M)}\bigg[\frac{\Delta Z_{\beta}^j}{\Delta Z_{\beta}^s}  \bigg] \\
	&\overset{(3)}{=} \sum_{j \in J_{s^-}}\mathbb{E}_{(X',X'')}\bigg\{\mathbb{E}_{\mathcal{T}^h_M}\bigg[\frac{\Delta Z_{\beta}^j}{\Delta Z_{\beta}^s}\bigg] | \mathcal{T}^h_M  \bigg\} \\
	&\overset{(4)}{=} \sum_{j \in J_{s^-}}\mathbb{E}_{(X',X'')}\bigg\{\mathbb{E}_{t^h_j}\bigg[\frac{\Delta Z_{\beta}^j}{\Delta Z_{\beta}^s}\bigg] | t^h_j  \bigg\} \\
	&\overset{(5)}{=} \sum_{j \in J_{s^-}}\mathbb{E}_{(X',X'')}\bigg\{\mathbb{E}_{t^h_j}\bigg[\frac{\Delta Z_{\beta}^j}{\Delta Z_{\beta}^s}\bigg]\bigg\} \\
	&\overset{(6)}{=} \sum_{j \in J_{s^-}}\mathbb{E}_{(X',X'')}\bigg\{\frac{1}{\Delta Z_{\beta}^s} \mathbb{E}_{t^h_j}\bigg[\Delta Z_{\beta}^j\bigg]\bigg\} \\
	&\overset{(7)}{=} \sum_{j \in J_{s^-}}\mathbb{E}_{(X',X'')}\bigg\{\frac{1}{\Delta Z_{\beta}^s} \mathbb{E}_{t^h_j}\bigg[\Delta Z_{\beta}^j | X',X''\bigg]\bigg\} \\
	&\overset{(8)}{=} \mathbb{E}_{(X',X'')}\bigg\{\frac{1}{\Delta Z_{\beta}^s} \sum_{j \in J_{s^-}}\mathbb{E}_{t^h_j}\bigg[\Delta Z_{\beta}^j | X',X''\bigg]\bigg\} \\
	&\overset{(9)}{=} (M-1)\mathbb{E}_{(X',X'')}\bigg\{\frac{1}{\Delta Z_{\beta}^s} \mathbb{E}_{t^h_j}\bigg[\Delta Z_{\beta}^j | X',X''\bigg]\bigg\}
\end{align}	

Where:
\begin{itemize}
	\item $(1)$ is due to the Markov inequality,
	\item $(2)$ is due to the fact that $\widetilde{t}_s^h$ is given so there is no need to integrate over it, 
	\item $(3)$ is due to the chain rule of probability,
	\item $(4)$ is due to the fact that $\Delta Z_{\beta}^j$ is independent of $t^h_i$ for $i \neq j$,
	\item $(5)$ and $(7)$ are due to the fact that $(X',X'')$ is independent of $t_j^h$,
	\item $(6)$ is due to the fact that $\Delta Z_{\beta}^s$ is independent of $t_j^h$ for $j \neq s$,
	\item $(8)$ is due to the linearity of expectations, and
	\item $(9)$ is due to the fact that $t_j^h$ are identically distributed for $j \in \{1,...,M\}$.
\end{itemize}	

Now 

\begin{align}
\mathbb{E}_{t^h_j}\bigg[\Delta Z_{\beta}^j | X',X''\bigg] &= \sum_{j=1}^{|\mathcal{T}^h_M|}\frac{1}{|\mathcal{T}^h_M|} \Delta Z_{\beta}^j\\
														  &= \frac{1}{|\mathcal{T}^h_M|} \sum_{j=1}^{|\mathcal{T}^h_M|} \sum_{c \in \mathcal{C}}w_{\beta}(c,t^D_j(X')) w_{\beta}(c,\widetilde{t}^D_s(X''))\\
														  &= \frac{1}{|\mathcal{T}^h_M|} \sum_{c \in \mathcal{C}} w_{\beta}(c,\widetilde{t}^D_s(X'')) \sum_{j=1}^{|\mathcal{T}^h_M|} w_{\beta}(c,t^D_j(X')) \\
														  &= \frac{1}{|\mathcal{T}^h_M|} \sum_{c \in \mathcal{C}} w_{\beta}((\widetilde{t}^h_s)^{-1}(c),X'') \sum_{j=1}^{|\mathcal{T}^h_M|} w_{\beta}((t^h_j)^{-1}(c),X')
\end{align}

Where for the last equivalence we have used assumption \ref{transfAssumption}. Using the properties of \ref{TransfSet} we can establish the following:

\begin{align}
	\mathbb{E}_{t^h_j}\bigg[\Delta Z_{\beta}^j | X',X''\bigg] &= \frac{1}{|\mathcal{T}^h_M|} \sum_{c \in \mathcal{C}} w_{\beta}((\widetilde{t}^h_s)^{-1}(c),X'') \sum_{j=1}^{|\mathcal{T}^h_M|} w_{\beta}((t^h_j)^{-1}(c),X')\\
		 													  &= \frac{1}{|\mathcal{T}^h_M|} \sum_{c \in \mathcal{C}} w_{\beta}((\widetilde{t}^h_s)^{-1}(c),X'') \sum_{c \in \mathcal{C}} w_{\beta}(c,X')\\
															  &= \frac{Z_{\beta}(X')}{|\mathcal{T}^h_M|} \sum_{c \in \mathcal{C}} w_{\beta}((\widetilde{t}^h_s)^{-1}(c),X'') \\
															  &= \frac{Z_{\beta}(X')}{|\mathcal{T}^h_M|} \sum_{c \in \mathcal{C}} w_{\beta}(c,X'') \\
															  &= \frac{Z_{\beta}(X')Z_{\beta}(X'')}{|\mathcal{T}^h_M|}
\end{align}	

So we have that 

\begin{align}
	\mathbb{P}(\hat{t}^h \neq \widetilde{t}^h_s | \widetilde{t}^h_s) &\leq (M-1)\mathbb{E}_{(X',X'')}\bigg\{\frac{1}{\Delta Z_{\beta}^s} \mathbb{E}_{t^h_j}\bigg[\Delta Z_{\beta}^j | X',X''\bigg]\bigg\} \\
											&= (M-1)\mathbb{E}_{(X',X'')}\bigg\{\frac{Z_{\beta}(X')Z_{\beta}(X'')}{|\mathcal{T}^h_M|\Delta Z_{\beta}^s} \bigg\} 
\end{align}	

Again using the properties of \ref{TransfSet} we have that:

\begin{align}
	\Delta Z_{\beta}^s &= \sum_{c \in \mathcal{C}} w_{\beta}(c,\widetilde{t}^D_s(X')) w_{\beta}(c,\widetilde{t}^D_s(X'')) \\
					   &= \sum_{c \in \mathcal{C}} w_{\beta}((\widetilde{t}^h_s)^{-1}(c),X') w_{\beta}((\widetilde{t}^h_s)^{-1}(c),X'') \\
					   &= \sum_{c \in \mathcal{C}} w_{\beta}(c,X') w_{\beta}(c,X'') \\
					   &= \Delta Z_{\beta}(X',X'')
\end{align}	

with which

\begin{align}
	\mathbb{P}(\hat{t}^h \neq \widetilde{t}^h_s | \widetilde{t}^h_s) &\leq  (M-1)\mathbb{E}_{(X',X'')}\bigg\{\frac{Z_{\beta}(X')Z_{\beta}(X'')}{\Delta Z_{\beta}^s} \bigg\} \\
											&=  (M-1)\mathbb{E}_{(X',X'')}\bigg\{\frac{Z_{\beta}(X')Z_{\beta}(X'')}{|\mathcal{T}^h_M|\Delta Z_{\beta}(X',X'')} \bigg\} \\
											&=  (M-1)\mathbb{E}_{(X',X'')}\bigg\{\frac{Z_{\beta}(X')Z_{\beta}(X'')}{|\mathcal{C}|\Delta Z_{\beta}(X',X'')} \bigg\} \\
											&=  \mathbb{E}_{(X',X'')}[e^{\log(M-1)-I_{\beta}}] 
\end{align}	

So error free learning is possible as long as, \emph{on average}, for data sets $X'$ and $X''$ it holds that $\log(M-1) < I_{\beta}$.

Finally we mention that since the generalization capacity defines the maximum learning rate of a cost function $\ERs$ it can be used as a criterion for deciding which cost function to use: simply use the cost function with highest generalization capacity. We will see an application of this in Section \ref{impSampRes}.

\chapter{Mean localization} \label{MeanLoc}

\section{Generalization capacity} \label{MLGC}

Recall from Section \ref{ApproxSets} that the empirical risk function $\hat{R}(\mu, X)$ is defined with respect to a loss function $\rho(\mu,X)$. We will mostly deal with the \emph{square loss} function:

\begin{align}
	\rho_{\mu}(x) = ||x-\mu||_2^2 = \sum_{j=1}^d (x_j - \mu)^2
\end{align}	

We calculate the empirical risk, which is our cost function in the context of ASC, with respect to this loss function. 

\begin{align}
	\hat{R}(\mu, X) &= \frac{1}{n} \sum_{i=1}^n \rho_{\mu}(x_i) = \frac{1}{n} \sum_{i=1}^n \sum_{j=1}^d (x_{ij}-\mu_j)^2 = \sum_{j=1}^d \frac{1}{n} \sum_{i=1}^n  x_{ij}^2-2\mu_jx_{ij}+\mu_j^2 \\
	                &= \sum_{j=1}^d \bigg\{ \frac{1}{n} \sum_{i=1}^n x_{ij}^2 - \frac{1}{n} \sum_{i=1}^n 2\mu_jx_{ij} + \frac{1}{n} \sum_{i=1}^n \mu_{j}^2 \bigg\}\\
					&= \sum_{j=1}^d \bigg\{ \frac{1}{n} \sum_{i=1}^n x_{ij}^2 -   2\mu_j\bar{X}_{j} + \mu_{j}^2 \bigg\}\\
					&\mypropto \sum_{j=1}^d \bigg\{ \mu_{j}^2 - 2\mu_j\bar{X}_{j} \bigg\}\\
					&\mypropto \sum_{j=1}^d \bigg\{ \mu_{j}^2 - 2\mu_j\bar{X}_{j} + \bar{X}_j^2 \bigg\}\\
					&= \sum_{j=1}^d(\mu_j - \bar{X}_j)^2 = ||\mu - \bar{X}||^2_2
\end{align}	

With which we can define the cost function as:

\begin{align} \label{sqrLoss}
	\hat{R}(\mu, X) = ||\mu - \bar{X}||^2_2
\end{align}	

Recall that in the problem at hand $\mathcal{C}=\binarynumbers^d$ so that:

\begin{align}
	I &= \max_{\beta \in \realnumbers^+} \mathbb{E}_{(X',X'')}\IC = \max_{\beta \in \realnumbers^+} \mathbb{E}_{(X',X'')}\log \frac{|\binarynumbers^d|\dZb}{\ZbOne \ZbTwo}\\
	  &= \log 2^d +  \mathbb{E}_{(X',X'')} \log \dZb - \mathbb{E}_{X'} \log \ZbOne - \mathbb{E}_{X''} \log \ZbTwo
\end{align}	

Where: 

\begin{itemize}
	\item $Z_{\beta}(X)=\sum_{\mu \in \binarynumbers^d} e^{-\beta\hat{R}(\mu, X)} = \sum_{\mu \in \binarynumbers^d}e^{-\beta||\mu - \bar{X}||^2_2}$ and
	\item $\dZb = \sum_{\mu \in \binarynumbers^d} e^{-\beta\{\hat{R}(\mu, X')+\hat{R}(\mu, X'')\}} = \sum_{\mu \in \binarynumbers^d} e^{-\beta\{||\mu - \bar{X}_1||^2_2+||\mu - \bar{X}_2||^2_2\}}$
\end{itemize}	

Since we don't know the distribution of  $Z_{\beta}(X)$ or $\dZb$ we will use simulation to estimate $I$. Before describing the simulation algorithm notice that $I$ is a function of the two data set means $\bar{X}_1$ and $\bar{X}_2$ and since $X$ has a normal distribution:

\begin{align}
	X \sim N \bigg(\mu^0,\sigma^2 I_d \bigg) \Rightarrow \bar{X} \sim N \bigg(\mu^0, \frac{\sigma^2}{n} I_d \bigg)
\end{align}	

If we let $\xi = \frac{\bar{X}-\mu^0}{\sfrac{\sigma}{\sqrt{n}}} \sim N(0,I_d)$ then $\bar{X}= \mu^0 + \frac{\sigma}{\sqrt{n}}\xi$ and we have that:

\begin{align}
	I &= \max_{\beta \in \realnumbers^+} \mathbb{E}_{(\xi_1,\xi_2)} \log \frac{|\binarynumbers^d|\dZbz}{\ZbOnez \ZbTwoz}
\end{align}	

Where: 
\begin{itemize}
	\item $\hat{R}(\mu, \xi) = ||\mu - \mu^0 - \frac{\sigma}{\sqrt{n}}\xi||^2_2$,
	\item $w_{\beta}(\mu,\xi) = e^{-\beta\hat{R}(\mu, \xi)}$
	\item $P_G(\mu; \beta, \xi) = \frac{w_{\beta}(\mu,\xi)}{\sum_{\mu \in \binarynumbers^d} w_{\beta}(\mu,\xi)}$
	\item $Z_{\beta}(\xi)= \sum_{\mu \in \binarynumbers^d}e^{-\beta||\mu - \mu^0 - \frac{\sigma}{\sqrt{n}}\xi||^2_2}$ and
	\item $\dZbz  = \sum_{\mu \in \binarynumbers^d} e^{-\beta\{||\mu - \mu^0 - \frac{\sigma}{\sqrt{n}}\xi_1||^2_2+||\mu - \mu^0 - \frac{\sigma}{\sqrt{n}}\xi_2||^2_2\}}$
\end{itemize}	

Using this alternate expression we can reduce the number of simulations by a factor of $n$. Notice that varying both $n$ and $\sigma$ doesn't make sense since $\frac{\sigma^2}{n}$ characterizes the variance of the data set. For this reason we leave $n=100$ fixed and only vary $\sigma$ in our simulation experiments. 

\section{Exhaustive algorithm} \label{NSEA}

To estimate the generalization capacity for the mean localization problem, for a given $\mu^0$ and $\sigma$, take the following steps:

\begin{enumerate}[1.]
	\item Choose a grid of relevant $\beta$ values: $\underline{\beta}=(\beta_1,...,\beta_l)$
	\item For $i=1$ to $m$
		\begin{enumerate}[a.]
			\item Simulate $\xi_1^i,\xi_2^i \sim N(0,I_d)$ 
			\item For $k=1$ to $l$
				\begin{itemize}
					\item Calculate information content:
						\begin{align}
							\hat{I}^i_{\beta_k}=\log \frac{2^d \Delta Z_{\beta_k}(\xi_1^i,\xi_2^i)}{Z_{\beta_k}(\xi_1^i)Z_{\beta_k}(\xi_2^i)}
						\end{align}	
				\end{itemize}	
		\end{enumerate}	
	\item For $k=1$ to $l$
			\begin{itemize}
				\item Estimate mean information content:
				\begin{align}
					\bar{I}_{\beta_k}= \frac{1}{m} \sum_{i=1}^m \hat{I}^i_{\beta_k}
				\end{align}
				\item Estimate generalization capacity:
				\begin{align}
					\hat{I}= \max_{k \in \{1,...,l\}} \bar{I}_{\beta_k}
				\end{align}
			\end{itemize}	
\end{enumerate}	

\textbf{Remark.} When dealing with real data we don't know what $\mu^0$ is. This doesn't matter since GC is independent of any particular hypothesis, rather it depends on the data generating mechanism and on the cost function $\hat{R}(\mu,X)$. Provided we can simulate from the model that we assume generated the data we will always be able to estimate GC by simulating from the model for an arbitrary set of parameters $\mu^0$. By calculating GC we obtain a $\beta^*$ that resolves the resolution-stability trade off. We may then use $\beta^*$ on the real data to obtain an appropriate approximation set of hypotheses. Alternatively we may compare the GC associated to different cost functions and choose the cost function with the highest GC. 

\section{Simulation results} \label{NSSR}

We used the following parameters for the simulation experiments:

\begin{itemize}
	\item $d=8$,
	\item $m=200$,
	\item 100 different $\beta$ values from 0.01 to 20, and
	\item 30 different noise levels $\sigma$ from 0.1 to 10. 
\end{itemize}	

To check simulation results made sense we first calculated the Gibbs distribution $P_G(c; \beta^*, \xi)$ distribution over $\mathcal{C}=\binarynumbers^d$, where $\beta^*$ is the resolution parameter that allows generalization capacity to be reached. Since $d=8$ $P_G(c; \beta^*, \xi)$ corresponds to a vector with $2^8=258$ entries. To display the results in an easy to read fashion we aggregated this vector to produce a component-wise Gibbs distribution:

\begin{align}
	\mathbb{P}_G(\mu^0_j = 1 | \beta^*) = \sum_{\mu \in \binarynumbers^d}P_G(\mu; \beta^*, \xi) \mathbbm{1}_{\{\mu_j = 1\}}
\end{align}	

The following graph is an illustration of the $\mathbb{P}_G(\mu^0_j = 1 | \beta^*)$ estimate for different noise levels. 

	\begin{figure}[H]%--- Picture 'H'ere, 'B'ottom or 'T'op; '!' Try to
	                    %impose your will to LaTeX
	  \centering
		  \includegraphics[width=0.6\textwidth]{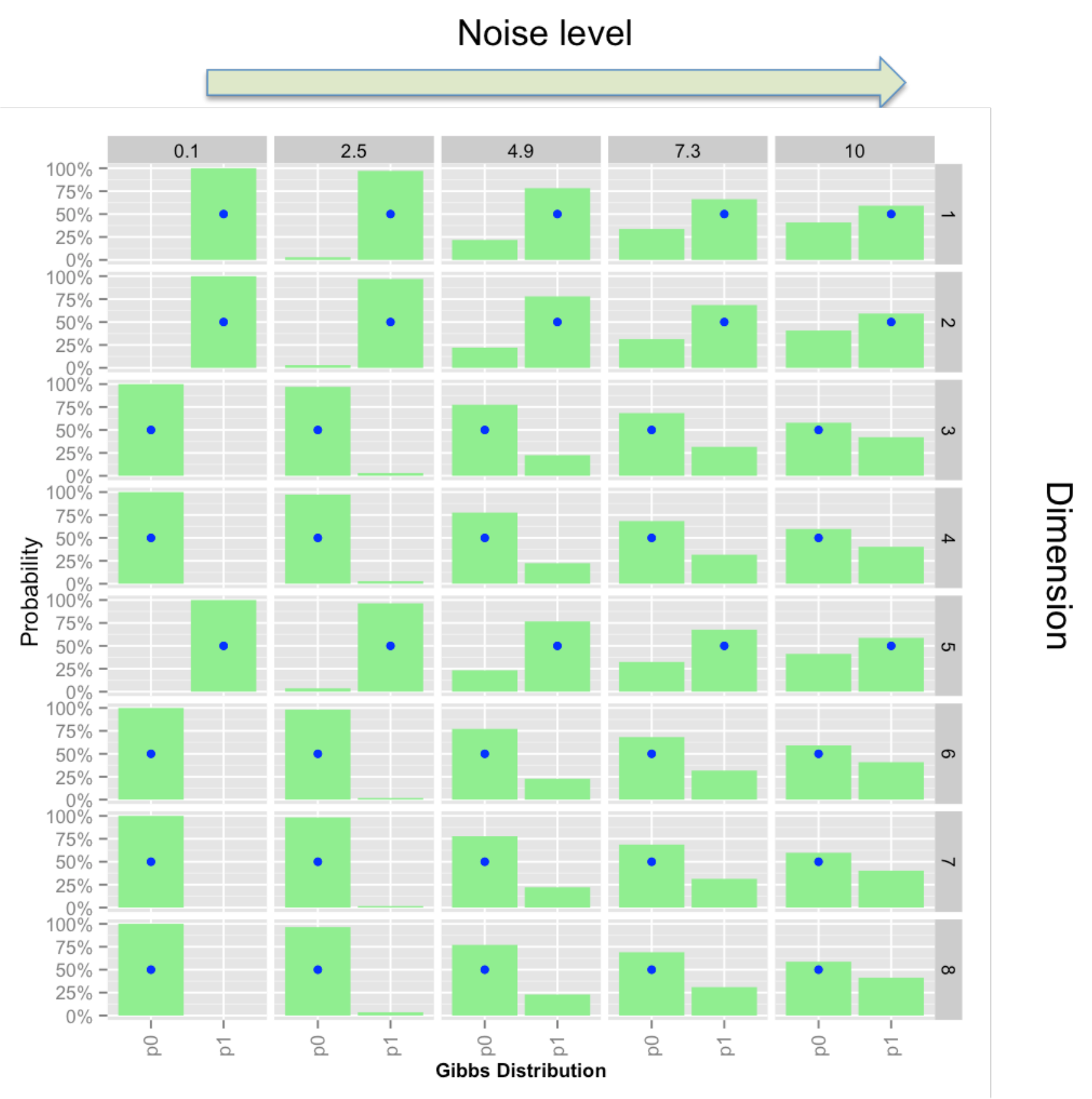} %<< no file extension
	  %%         --- .5\textwidth stands for 50% of text width
	  \caption[Component-wise Gibbs distribution]%<<-- Legend for the list of figures at the beginning of you thesis
	  {Component-wise Gibbs distribution}% legend displayed below the graph.
	  \label{fig:Gibbs}
	\end{figure}
	
The blue dots show the true value of $\mu^0$ for each of its components. The lower the noise $\sigma$ the less uncertainty about the value of the $\mu^0$ we have. We next show the average information content for different resolutions $\beta$ and noise levels $\sigma$. 

	\begin{figure}[H]%--- Picture 'H'ere, 'B'ottom or 'T'op; '!' Try to
	                    %impose your will to LaTeX
	  \centering
		  \includegraphics[width=0.6\textwidth]{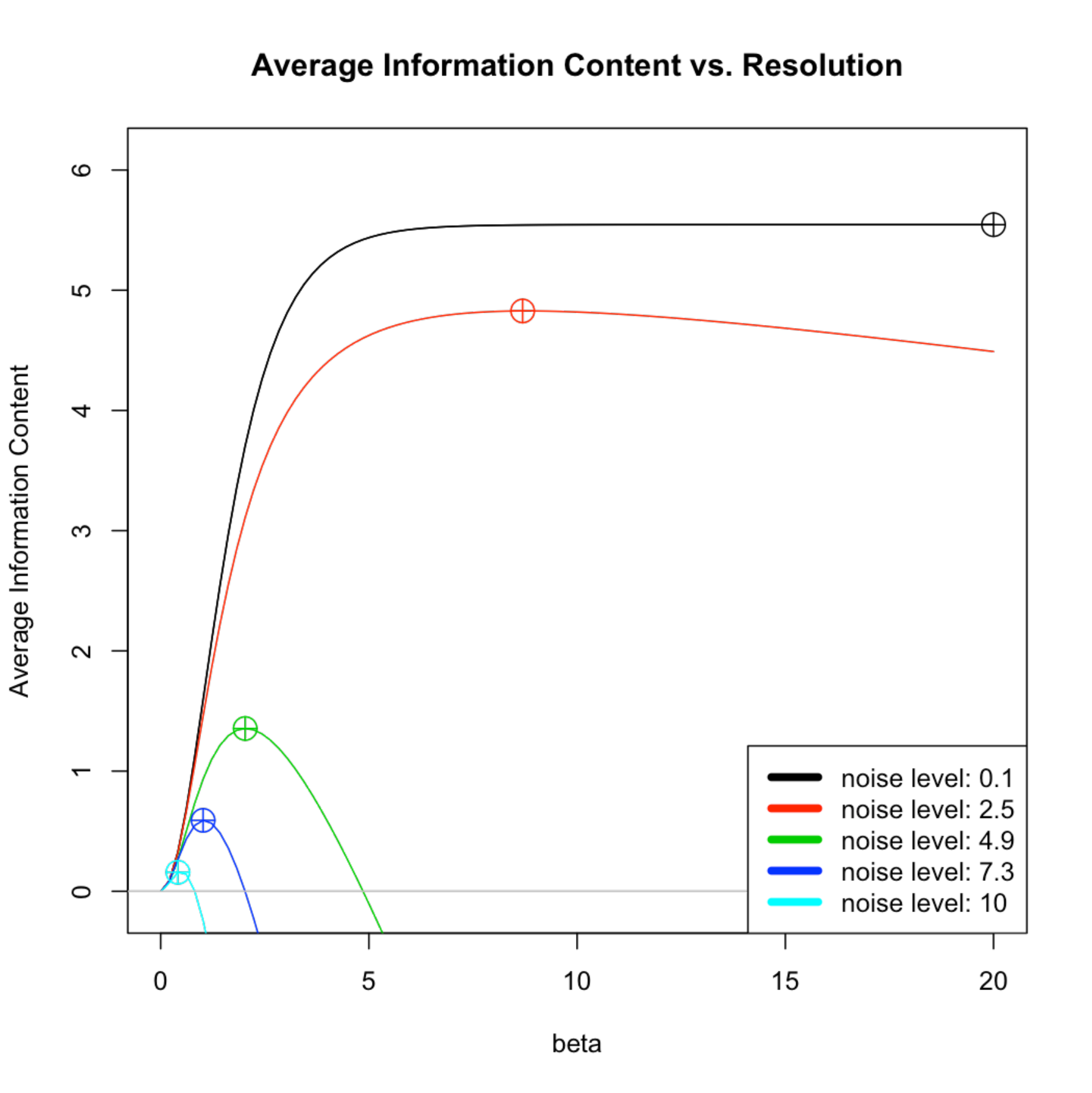} %<< no file extension
	  %%         --- .5\textwidth stands for 50% of text width
	  \caption[Average information content]%<<-- Legend for the list of figures at the beginning of you thesis
	  {Average information content}% legend displayed below the graph.
	  \label{fig:AvgIC}
	\end{figure}
	
The crossed circles represent the pairs $(\beta^*, I)$ where average information content is maximized. The lower the noise level the higher the generalization capacity is. For \emph{low} noise levels, sucha as $\sigma = 0.1$, we can obtain gains in average information content the higher the resolution $\beta$ (albeit at a diminishing rate) i.e. generalization capacity is basically an increasing function of resolution for these noise levels. This means that for such low noise we can let $\beta \rightarrow \infty$ and obtain the empirical risk minimizer.  For \emph{medium} range noise levels, such as $\sigma = 4.9$, once we go past the resolution threshold $\beta^*$, the average information content decreases dramatically i.e. generalization capacity is a convex function of resolution for these noise levels. Here we see the resolution-stability trade-off clearly. For this level of noise we can only decrease our approximation sets to a certain size parametrized by $\beta^*$ before we start to get very unstable sets with little information. 

We now show the generalization capacity for different noise levels. 

	\begin{figure}[H]%--- Picture 'H'ere, 'B'ottom or 'T'op; '!' Try to
	                    %impose your will to LaTeX
	  \centering
		  \includegraphics[width=0.6\textwidth]{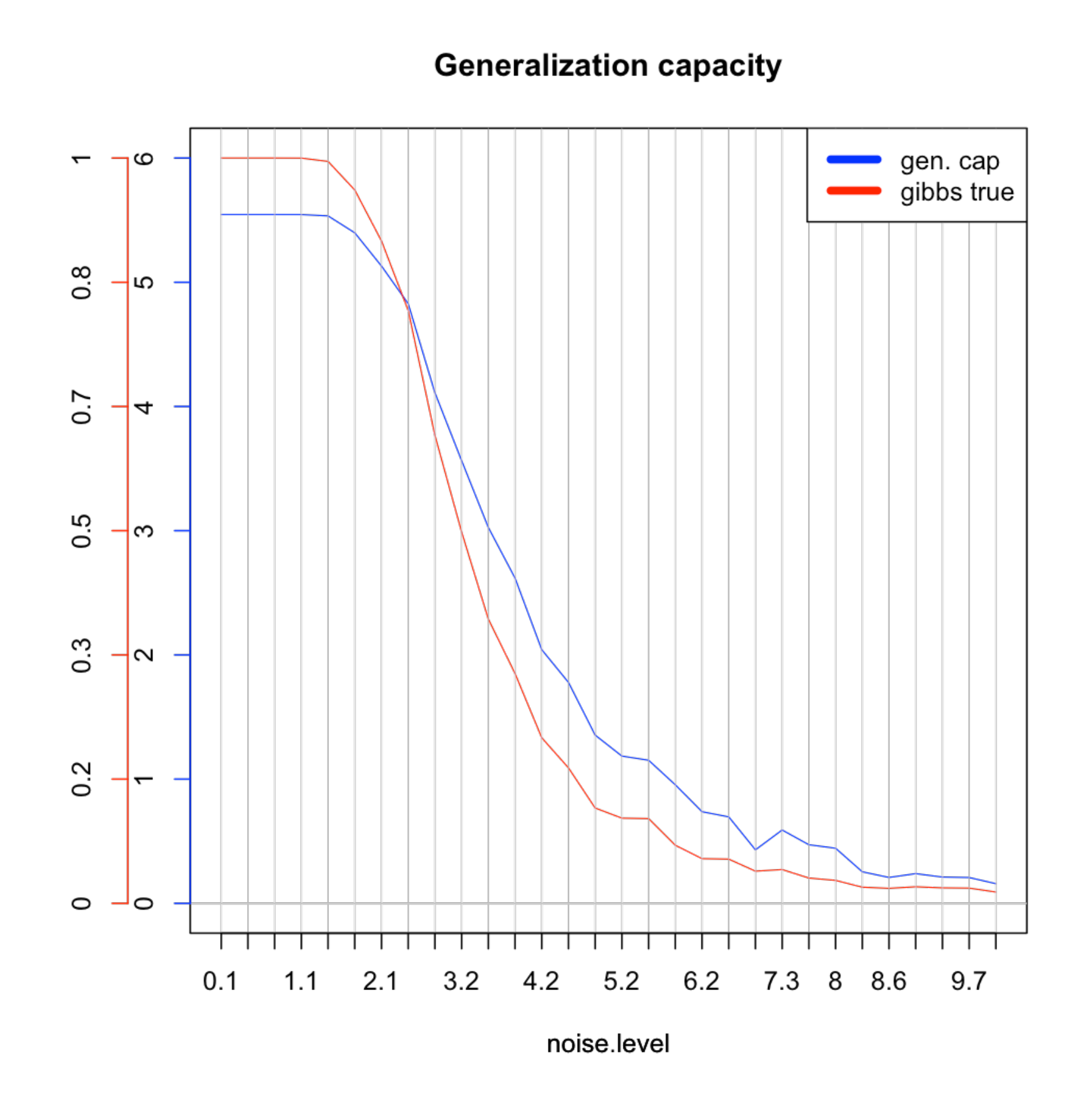} %<< no file extension
	  %%         --- .5\textwidth stands for 50% of text width
	  \caption[Generalization capacity]%<<-- Legend for the list of figures at the beginning of you thesis
	  {Generalization Capacity}% legend displayed below the graph.
	  \label{fig:GC}
	\end{figure}

The blue line shows the generalization capacity while the red line shows the \emph{true} Gibbs probabilty, i.e. $P_G(\mu^0; \beta^*, \xi)$. As expected the generalization capacity decreases toward zero as the noise level becomes so big as to completely drown out the signal $\mu^0$. Notice that for noise levels $\sigma<1.3$, $P_G(\mu^0; \beta^*, \xi)=1$ and $I = d \log 2=8 \log 2 \approx 5.55$ indicating that we can completely recover the signal $\mu^0$ with the empirical risk minimizer ( $I = d \log 2 \Rightarrow \ZbOnez\ZbTwoz=\dZbz$!).

\section{Log-sum-exp trick} \label{logSum}

We can express the information content as 

\begin{align}
	\IC = \log |\mathcal{C}| + \log \dZbz - \log \ZbOnez - \log \ZbTwoz. 
\end{align}	

In turn, we can express the log partition functions as: 

\begin{align}
	\log Z_{\beta}(\xi) &= \log \sum_{i=1}^{\mathcal{C}} e^{a_i} \\
	\log \dZbz &= \log \sum_{i=1}^{\mathcal{C}} e^{b_i} 
\end{align}	

where:
\begin{itemize}
	\item $a_i = -\beta \hat{R}(\xi, \mu_i) < 0 $,
	\item $b_i = -\beta (\hat{R}(\xi_1, \mu_i)+\hat{R}(\xi_2, \mu_i)) < 0 $ and 
	\item $\mathcal{C}=\{ \mu_1,...,\mu_{|\mathcal{C}|} \}$
\end{itemize}	

When implementing the algorithm of Section \ref{NSEA} we may run into the problem that for very high resoluton values $\beta$, $\sum_{i=1}^{\mathcal{C}} e^{a_i}$ and $\sum_{i=1}^{\mathcal{C}} e^{b_i}$ are so small that they are represented as 0 using limited-precision, floating point numbers. This means $\log Z_{\beta}(\xi)$ and $\log \dZbz$ are represented as $-\infty$ with which our calculation of $\IC$ breaks down. To solve this \emph{underflow} problem we use the log-sum-exp trick. Let:

\begin{align}
	A = \max_{i \in \{1,...,|\mathcal{C}| \}} a_i
\end{align}	

Then, 

\begin{align}
	\log Z_{\beta}(\xi)= \log \sum_{i=1}^{\mathcal{C}} e^{a_i} = \log e^A \sum_{i=1}^{\mathcal{C}} e^{a_i-A}= A + \log \sum_{i=1}^{\mathcal{C}} e^{a_i-A}
\end{align}	

If we calculate $\log Z_{\beta}(\xi)$ using the last expression, this solves the underflow problem since $\sum_{i=1}^{\mathcal{C}} e^{a_i-A}>1$ so that $\log \sum_{i=1}^{\mathcal{C}} e^{a_i-A}>0$. 

\section{Variance reduction with common random numbers} \label{VarRed}

If we want to estimate by simulation: 

\begin{align}
	\mu_{f-g} = \mathbb{E}_X[f(X)-g(X)] = E_X[f(X)] - E_X[g(X)]
\end{align}	

We can either: 

\begin{enumerate}[a.]
	\item Generate $2m$ i.i.d. realizations of $X$: $X_1=\{x_{11},...,x_{1m}\}$,$X_2=\{x_{21},...,x_{2m}\}$. Then:
	
		\begin{align}
			&\hat{\mu}_f^1 = \hat{\mathbb{E}}_X^1[f(X)]=\frac{1}{m}\sum_{i=1}^m f(x_{1i})\\ 			
			&\hat{\mu}_g^2 = \hat{\mathbb{E}}_X^2[g(X)]=\frac{1}{m}\sum_{i=1}^m g(x_{2i})\\ 			
			&\hat{\mathbb{E}}_X^{12}[f(X)-g(X)] = \hat{\mathbb{E}}_X^1[f(X)]-\hat{\mathbb{E}}_X^2[g(X)]=\hat{\mu}_f^1-\hat{\mu}_g^2
		\end{align}	
		or we can, 
	\item Generate $m$ i.i.d. realizations of $X$: $X_1=\{x_{11},...,x_{1m}\}$. Then:
		\begin{align}
			&\hat{\mu}_f = \hat{\mathbb{E}}_X[f(X)]=\frac{1}{m}\sum_{i=1}^m f(x_{1i})\\ 			
			&\hat{\mu}_g = \hat{\mathbb{E}}_X[g(X)]=\frac{1}{m}\sum_{i=1}^m g(x_{1i})\\ 			
			&\hat{\mathbb{E}}_X[f(X)-g(X)] = \hat{\mathbb{E}}_X[f(X)]-\hat{\mathbb{E}}_X[g(X)]=\hat{\mu}_f-\hat{\mu}_g
		\end{align}
\end{enumerate}	

The second method is an example of \emph{Common Random Numbers} (CRNs) since we use the same pseudo-random numbers to estimate $\mathbb{E}_X[f(X)]$ and $\mathbb{E}_X[g(X)]$. 

CRNs is a \emph{variance reduction} technique, although, strictly speaking, it does not always succeed in reducing the variance of an estimator. To see when it might succeed we compare the variance of $\hat{\mu}_f^1-\hat{\mu}_g^2$ and $\hat{\mu}_f-\hat{\mu}_g$:

\begin{align}
	\mathbb{V}_X[\hat{\mu}_f^1-\hat{\mu}_g^2] &= \mathbb{V}_X\bigg[\frac{1}{m}\sum_{i=1}^m f(x_{1i})-\frac{1}{m}\sum_{i=1}^m g(x_{2i})\bigg]\\
											  &= \mathbb{V}_X\bigg[\frac{1}{m}\sum_{i=1}^m (f(x_{1i})- g(x_{2i}))\bigg]\\
											  &= \frac{1}{m^2}\sum_{i=1}^m \mathbb{V}_X(f(x_{1i})- g(x_{2i}))\\
											  &= \frac{1}{m^2}\sum_{i=1}^m(\mathbb{V}_X[f(X)] + \mathbb{V}_X[g(X)])\\
											  &= \frac{1}{m}(\mathbb{V}_X[f(X)]+\mathbb{V}_X[g(X)])
\end{align}	

\begin{align}
	\mathbb{V}_X[\hat{\mu}_f-\hat{\mu}_g] &= \mathbb{V}_X\bigg[\frac{1}{m}\sum_{i=1}^m f(x_{1i})-\frac{1}{m}\sum_{i=1}^m g(x_{1i})\bigg]\\
											  &= \mathbb{V}_X\bigg[\frac{1}{m}\sum_{i=1}^m (f(x_{1i})- g(x_{1i}))\bigg]\\
											  &= \frac{1}{m^2}\sum_{i=1}^m \mathbb{V}_X[f(x_{1i})- g(x_{1i})]\\
											  &= \frac{1}{m} \mathbb{V}_X[f(X)- g(X)]\\
											  &= \frac{1}{m} \mathbb{V}_X[f(X)]+ \mathbb{V}_X[g(X)] -2Cov(f(X),g(X))\\
\end{align}	

If $f$ and $g$ are both either monotonically non-decreasing or mononotonically non-increasing then:

\begin{align}
	Cov(X,X)=\mathbb{V}_X[X]>0 &\Rightarrow Cov(f(X), g(X)) \geq 0 \\
		 					&\Rightarrow \mathbb{V}_X[\hat{\mu}_f-\hat{\mu}_g] \leq \mathbb{V}_X[\hat{\mu}_f^1-\hat{\mu}_g^2]
\end{align}	

We now obtain a different expression for $Z_{\beta}(\xi)$ and $\dZbz$ in order to use the CRN technique for the estimation of $\mathbb{E}_{(\xi_1,\xi_2)}[\IC]$. We start with an alternative, but equivalent (proportional in $\mu$) cost function $\hat{R}(\mu, X)$. Using that $\mu_j \in \{0,1\}$ we have that: 

\begin{align}
	\hat{R}(\mu, X) = \frac{1}{n}\sum_{i=1}^n \sum_{j=1}^d (x_{ij}-\mu_j)^2 &=  \frac{1}{n}\sum_{i=1}^n \sum_{j=1}^d \mu_j(x_{ij}-1)^2 + (1- \mu_j)x_{ij}^2\\
						&=  \frac{1}{n}\sum_{i=1}^n \sum_{j=1}^d \mu_j(1-2x_{ij}) + x_{ij}^2\\
						&\mypropto  \frac{1}{n}\sum_{i=1}^n \sum_{j=1}^d \mu_j(1-2x_{ij})\\
						&=  \sum_{j=1}^d \mu_j\frac{1}{n}\sum_{i=1}^n (1-2x_{ij})\\
						&=  \sum_{j=1}^d \mu_j (1-2\bar{x}_j) = \mu^T(1-2\bar{x})
\end{align}	

So that $\mu^T(1-2\bar{x}) \mypropto ||\mu - \bar{x}||^2_2$. 

Again, using that $\bar{X} = \frac{\sigma}{\sqrt{n}}\xi + \mu^0$ we have that:

\begin{align} \label{risk2}
	&\hat{R}(\mu, \xi) = \mu^T(1-2\frac{\sigma}{\sqrt{n}}\xi-2\mu^0)\\
	&\hat{R}(\mu, \xi_1)+\hat{R}(\mu, \xi_2) = 2\mu^T(1-\frac{\sigma}{\sqrt{n}}(\xi_1+\xi_2)-2\mu^0)
\end{align}	

Since $\frac{1}{\sqrt{2}}(\xi_1+\xi_2) \sim N(0,I_d)$ we have that:

\begin{align}
	Z_{\beta}(\xi)        & = \sum_{\mu \in \binarynumbers^d} e^{-\beta\hat{R}(\mu, \xi)}\\
	\Delta Z_{\beta}(\xi_1, \xi_2) & = \Delta Z_{\beta}(\xi) = \sum_{\mu \in \binarynumbers^d} e^{-\beta\hat{R}'(\mu, \xi)}
\end{align}	

Where:
\begin{itemize}
	\item $\hat{R}(\mu, \xi):=\mu^T(1-2\frac{\sigma}{\sqrt{n}}\xi-2\mu^0)$
	\item $\hat{R}'(\mu, \xi):=2\mu^T(1-\frac{\sigma}{\sqrt{n}}\sqrt{2}\xi-2\mu^0)$
\end{itemize}	

Using these expressions we implemented 3 different CRN algorithms to estimate $\IC$ and $I$, and compared the variance estimate $\hat{\mathbb{V}}_{(\xi_1,\xi_2)}[\hat{I}_{\beta^*}]$ of each, where $\beta^* = \argmaxM{\beta \in \{\beta_1,...,\beta_l\}} \hat{\mathbb{E}}_{(\xi_1, \xi_2)}[\hat{I}_{\beta}]$. 

All three algorithms generate $2m$ realizations of $\xi \sim N(0,I_d)$ but differ in wether they use them for the estimation of $\mathbb{E}_{\xi}[\log Z_{\beta}(\xi)]$, $\mathbb{E}_{(\xi_1,\xi_2)}[\log \Delta Z_{\beta}(\xi_1, \xi_2)]=\mathbb{E}_{\xi}[\log \Delta Z_{\beta}(\xi)]$ or both:

\begin{enumerate}[1.]
	\item \textbf{CRN-1} Generate $2m$ realizations of $\xi \sim N(0,I_d)$. Use first $m$ to calculate $\hat{\mathbb{E}}_{\xi_1}[\log Z_{\beta}(\xi_1)]$ and $\hat{\mathbb{E}}_{\xi_2}[\log Z_{\beta}(\xi_2)]$ and second $m$ to calculate $\hat{\mathbb{E}}_{\xi}[\log \Delta Z_{\beta}(\xi)]$. 

	\item \textbf{CRN-2} Generate $2m$ realizations of $\xi \sim N(0,I_d)$. Use all $2m$ to calculate $\hat{\mathbb{E}}_{\xi_1}[\log Z_{\beta}(\xi_1)]$, $\hat{\mathbb{E}}_{\xi_2}[\log Z_{\beta}(\xi_2)]$ and  $\hat{\mathbb{E}}_{(\xi_1,\xi_2)}[\log \Delta Z_{\beta}(\xi_1, \xi_2)]$. 
	
	\item \textbf{CRN-3} Generate $2m$ realizations of $\xi \sim N(0,I_d)$. Use first $m$ to calculate $\hat{\mathbb{E}}_{\xi_1}[\log Z_{\beta}(\xi_1)]$, second $m$ to calculate  $\hat{\mathbb{E}}_{\xi_2}[\log Z_{\beta}(\xi_2)]$ and all $2m$ to calculate $\hat{\mathbb{E}}_{(\xi_1, \xi_2)}[\log \Delta Z_{\beta}(\xi_1, \xi_2)]$. This CRN algorithm is actually the algorithm proposed in Section \ref{NSEA}. We simulate $(\xi_1, \xi_2)$ $m$ times and use it to estimate both $\mathbb{E}_{(\xi_1,\xi_2)}[\log Z_{\beta}(\xi_1)Z_{\beta}(\xi_2)]$ and $\mathbb{E}_{(\xi_1, \xi_2)}[\log \Delta Z_{\beta}(\xi_1, \xi_2)]$.
\end{enumerate}	

We used the following parameters for the simulation experiments:

\begin{itemize}
	\item $d=8$,
	\item $m=100$,
	\item 100 different $\beta$ values from 0.01 to 20, and
	\item 30 different noise levels $\sigma$ from 0.1 to 4. 
\end{itemize}

\begin{figure}[H]%--- Picture 'H'ere, 'B'ottom or 'T'op; '!' Try to
                    %impose your will to LaTeX
 % \centering 
  
  \begin{subfigure}{.5\textwidth}
    \centering
	\includegraphics[width=1\textwidth]{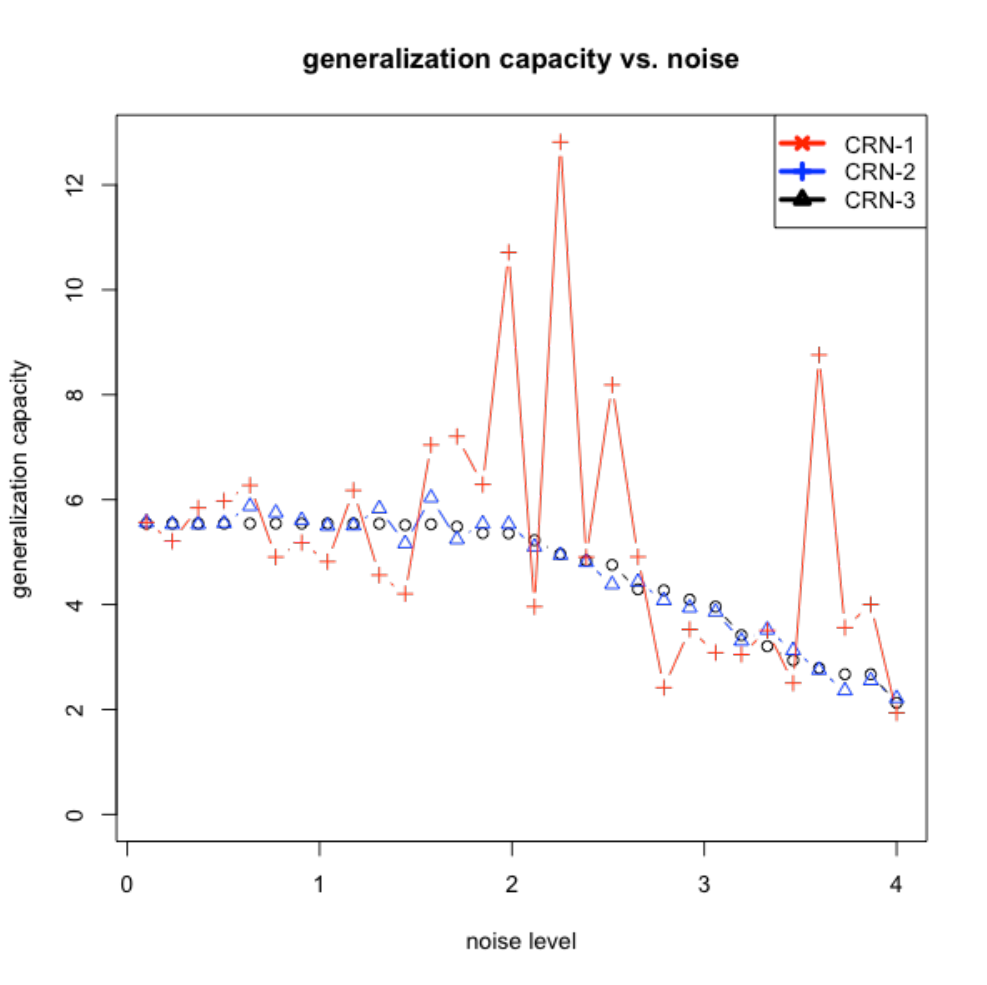} 
    \caption{$\hat{I}$}
    \label{fig:crnGC}
  \end{subfigure}%
  \begin{subfigure}{.5\textwidth}
    \centering
    \includegraphics[width=1\textwidth]{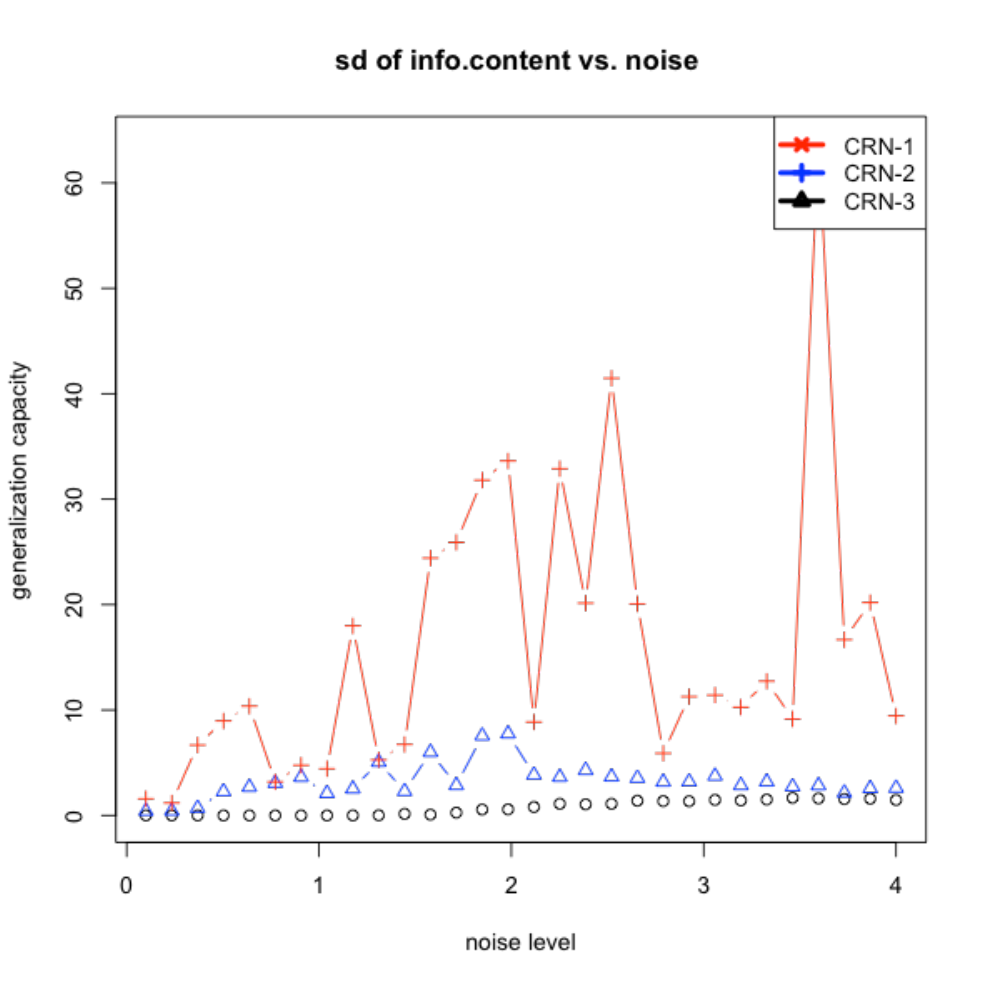}
    \caption{$\sqrt{\hat{\mathbb{V}}_{(\xi_1,\xi_2)}[\hat{I}_{\beta^*}]}$}
    \label{fig:crnVAR}
  \end{subfigure}
  %%         --- .5\textwidth stands for 50% of text width
  \caption[Calculation of generalization capacity with CRNs]%<<-- Legend for the list of figures at the beginning of you thesis
  {Calculation of generalization capacity with CRNs}% legend displayed below the graph.
  \label{fig:CRNs}
\end{figure}

Figure \ref{fig:crnVAR} shows the estimation of the generalization capacity $\hat{I}= \max_{\beta \in \realnumbers^+} \hat{\mathbb{E}}_{(\xi_1,\xi_2)}\log \hat{I}_{\beta}$ and the square root of variance $\sqrt{\hat{\mathbb{V}}_{(\xi_1,\xi_2)}[\hat{I}_{\beta^*}]}$ using all 3 methods. 
	
Clearly method CRN-3, the method described in Section \ref{NSEA}, has the least variance. This makes sense because if $\xi_1, \xi_2, \xi_3$ are i.i.d.: 

\begin{enumerate}[a.]
	\item $Cov(\log Z_{\beta}(\xi_1),\log Z_{\beta}(\xi_1))>0 \Rightarrow \mathbb{V}[\log Z_{\beta}(\xi_1) + \log Z_{\beta}(\xi_1)] > \mathbb{V}[\log Z_{\beta}(\xi_1) + \log Z_{\beta}(\xi_2)]$, so using the same random number to estimate $Z_{\beta}(\xi_1)$ and $Z_{\beta}(\xi_2)$ actually increases the variance, while, 
	\item $Cov(\log Z_{\beta}(\xi_1), \log \Delta Z_{\beta}(\xi_1, \xi_2))>0 \Rightarrow \mathbb{V}[\log \Delta Z_{\beta}(\xi_1, \xi_2) - \log Z_{\beta}(\xi_1)] < \mathbb{V}[\log \Delta Z_{\beta}(\xi_1, \xi_2) - \log Z_{\beta}(\xi_3)]$, so using the same random number to estimate $\Delta Z_{\beta}(\xi_1, \xi_2)$ and $Z_{\beta}(\xi_1)$ decreases the variance. 
\end{enumerate}	
	
\chapter{Sparse mean localization} 
\label{chp3}

\section{Exhaustive algorithm} \label{SMEA}

For the sparse mean localization problem the hypothesis space is restricted to binary vectors with $k$ entries equal to one:

\begin{align}
	\mathcal{C}^k := \{\mu \in \binarynumbers^d: ||\mu||_1 = k\}
\end{align}	

Notice that $|\mathcal{C}^k|={d \choose k}$. In terms of the algorithm in Section \ref{NSEA}, the only thing that changes is that the sums involved in $Z_{\beta}(\xi)$ and $\dZbz$ are over $\mathcal{C}^k$ instead of the entire $\mathcal{C}=\binarynumbers^d$. 

In the non-sparse case, to generate the vectors $\mu_i$ such that $\mathcal{C}=\{\mu_1,...,\mu_{2^d}\}$ we can simply use the mapping $\kappa:\{0,...,2^d-1\} \rightarrow \mathcal{C}$ where $\kappa$ maps a positive integer $0\leq i \leq 2^d-1$ to its $d$ length binary representation. 

In the sparse case, to generate the vecors $\mu_i$ such that $\mathcal{C}^k=\{\mu_1,...,\mu_{{d \choose k}}\}$ we use the mapping $\gamma:\{1,...,{d \choose k}\} \rightarrow \mathcal{C}^k$ where $\gamma$ maps a positive integer $1 \leq i \leq {d \choose k}$ to the i-th element of $\mathcal{C}^k$, assuming that $\mathcal{C}^k$ is in \emph{lexicographical order}. 

The algorithm shown below, adapted from \cite{Lehmer} (pp. 27-29), can be used to produce the mapping $\gamma$. It is based on the fact that if $\mathcal{C}^k$ is in reverse lexicographical order then we can represent the number $1 \leq i \leq {d \choose k}$ as:

\begin{align}
	i = {p_k \choose k} + ... + {p_2 \choose 2} + {p_1 \choose 1}
\end{align}	

Where $p_j \in \{1,...,d\}$ gives the position of the j-th entry of $\mu_i$ that is equal  to one. i.e. $\mu_{ip_j}=1$ $ \forall j \in \{1,...,k\}$.

\hspace{1mm}

\textbf{Algorithm to obtain $\mu_i$}
\begin{enumerate}[1.]
	\item Initalize:
		\begin{itemize}
			\item Set $\mu_{ij} \leftarrow 0$ $\forall j \in \{1,...,d\}$.
			\item Set $m \leftarrow i-1$.
		\end{itemize}
	\item For $j$ from $k$ to $1$:
	 	\begin{enumerate}[a.]
			\item $p_j \leftarrow \argmaxM{l \in \{t: {t \choose j} \leq m\}} {l \choose j}$
			\item $m \leftarrow m - {p_j \choose j}$
			\item $\mu_{ip_j} \leftarrow 1$
		\end{enumerate}	
\end{enumerate}	

\section{Simulation results: exhaustive algorithm} \label{SSR}

We used the following parameters for the simulation experiments:

\begin{itemize}
	\item $d=10$,
	\item $k=1,2,3,4,5$
	\item $m=100$,
	\item 100 different $\beta$ values from 0.01 to 20, and
	\item 30 different noise levels $\sigma$ from 0.1 to 10. 
\end{itemize}	

First we show the information content for different values of $k$, $\sigma$ and $\beta$. Since $d=10$ we look at $k \in \{1,2,...,5\}$ only. For $k \in \{6,7,...,10\}$ the behavior will be equivalent and the only difference is that the role of 0 and 1 ($\mu_j \in \{0,1\}$) is reversed. 

\begin{figure}[H]%--- Picture 'H'ere, 'B'ottom or 'T'op; '!' Try to
                    %impose your will to LaTeX
 % \centering 
  
  \begin{subfigure}{.5\textwidth}
    \centering
	\includegraphics[width=1\textwidth]{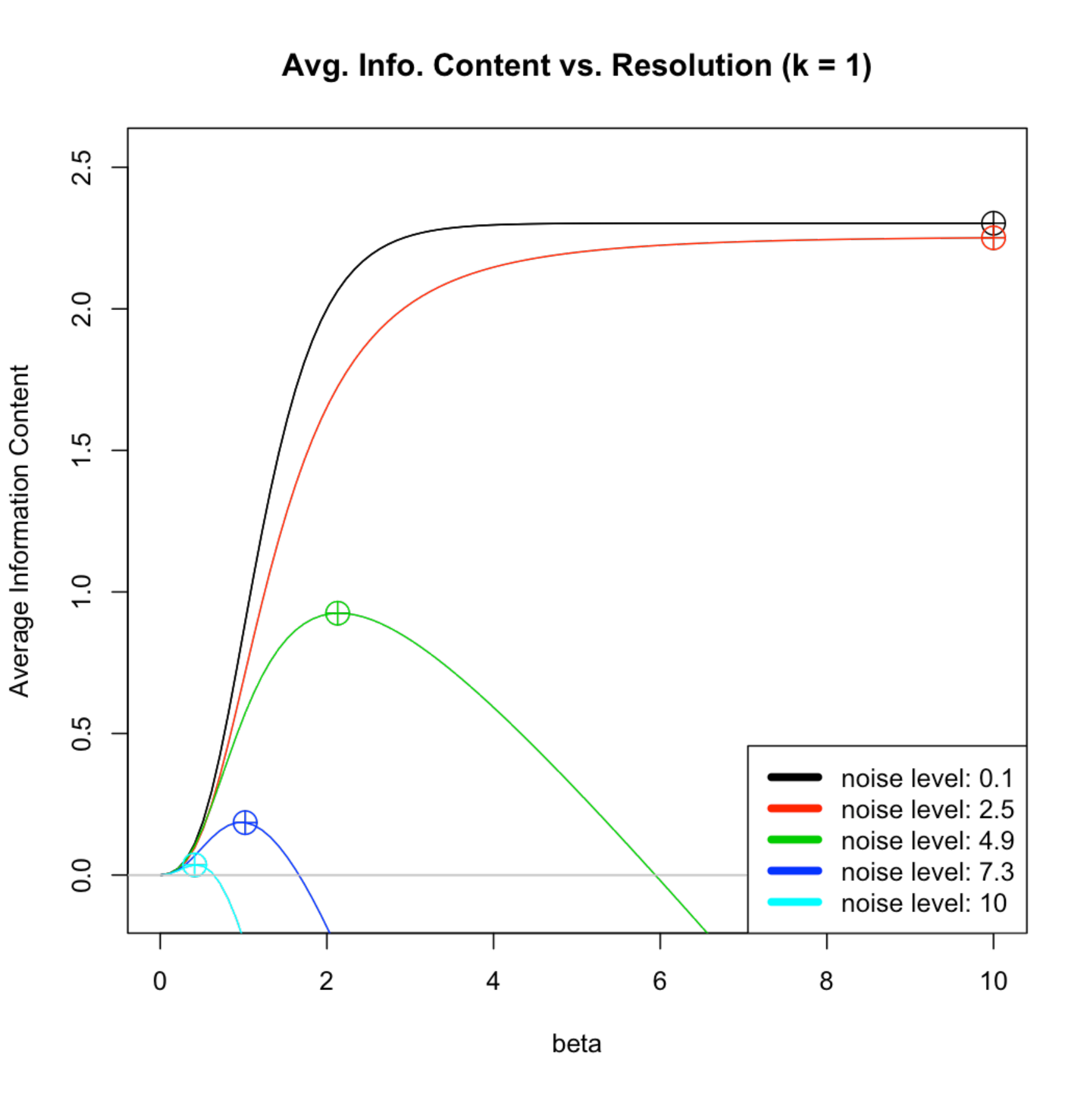} 
    \caption{$k=1$}
    \label{fig:avgICk1}
  \end{subfigure}%
  \begin{subfigure}{.5\textwidth}
    \centering
    \includegraphics[width=1\textwidth]{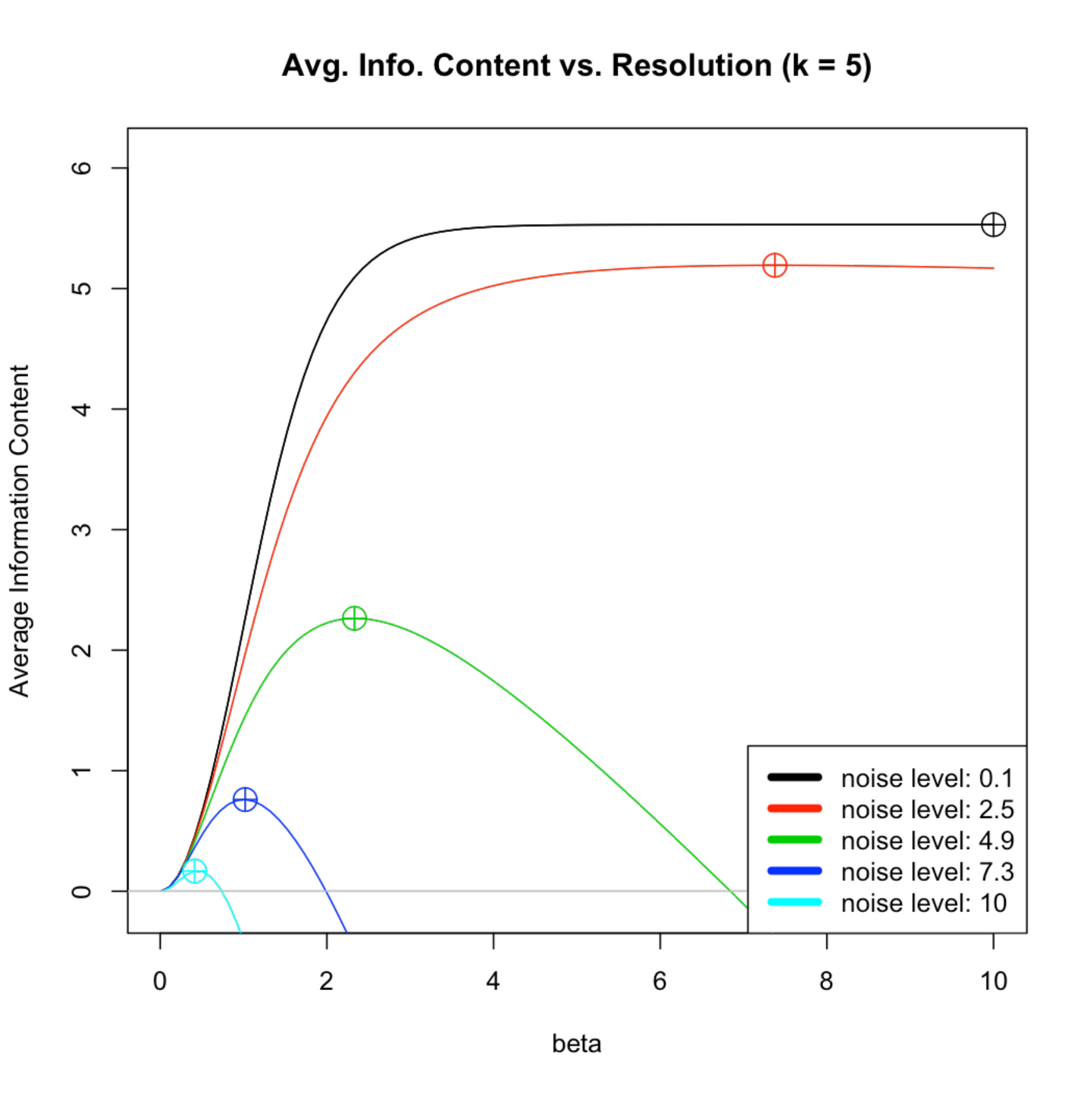}
    \caption{$k=5$}
    \label{fig:avgICk5}
  \end{subfigure}
  %%         --- .5\textwidth stands for 50% of text width
  \caption[Average information content for different $k$]%<<-- Legend for the list of figures at the beginning of you thesis
  {Average information content for different $k$}% legend displayed below the graph.
  \label{fig:avgICk}
\end{figure}	

\begin{figure}[H]%--- Picture 'H'ere, 'B'ottom or 'T'op; '!' Try to
                    %impose your will to LaTeX
 % \centering 
  
  \begin{subfigure}{.5\textwidth}
    \centering
	\includegraphics[width=1\textwidth]{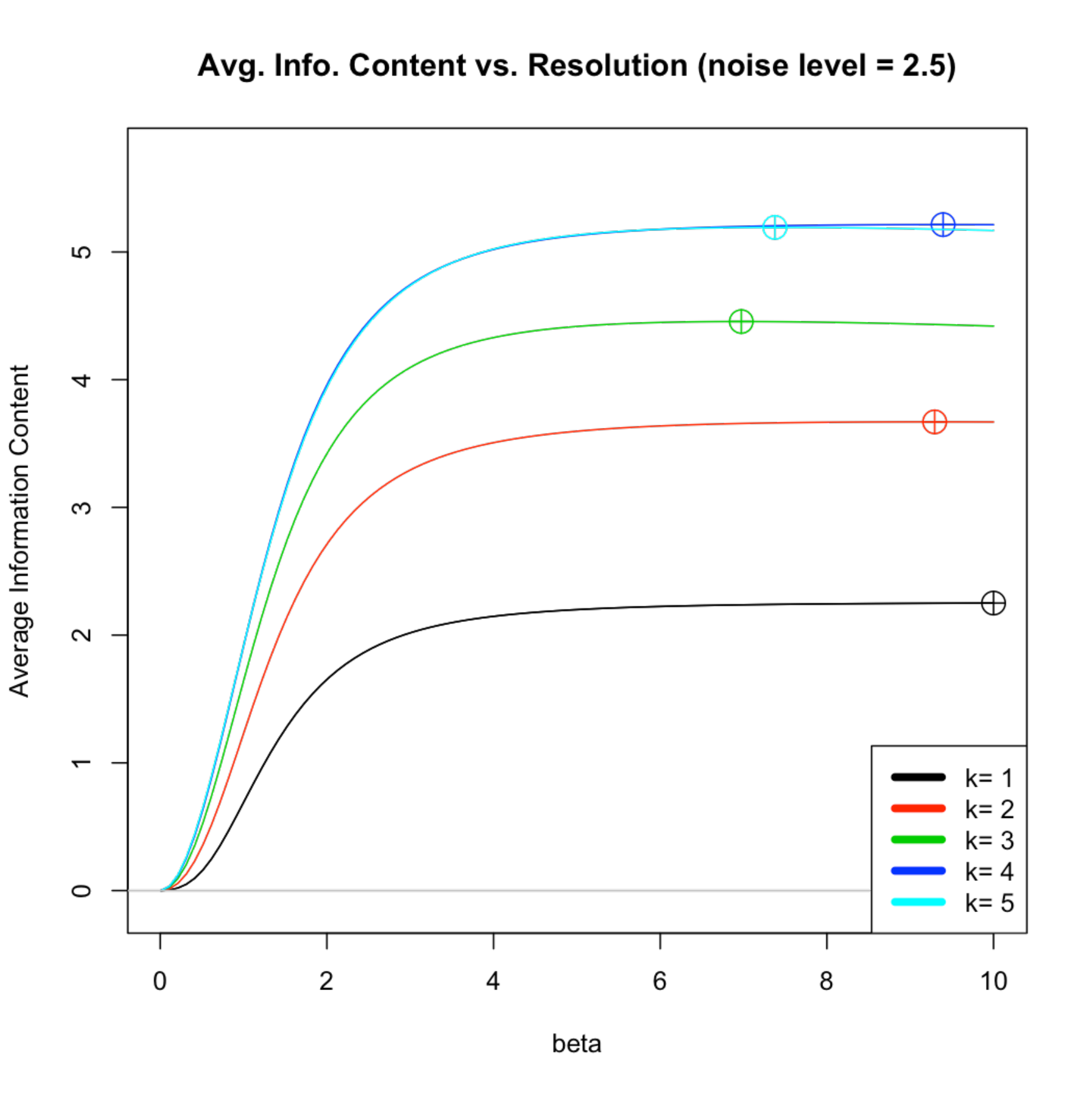} 
    \caption{$\sigma=2.5$}
    \label{fig:avgICsigma2.5}
  \end{subfigure}%
  \begin{subfigure}{.5\textwidth}
    \centering
    \includegraphics[width=1\textwidth]{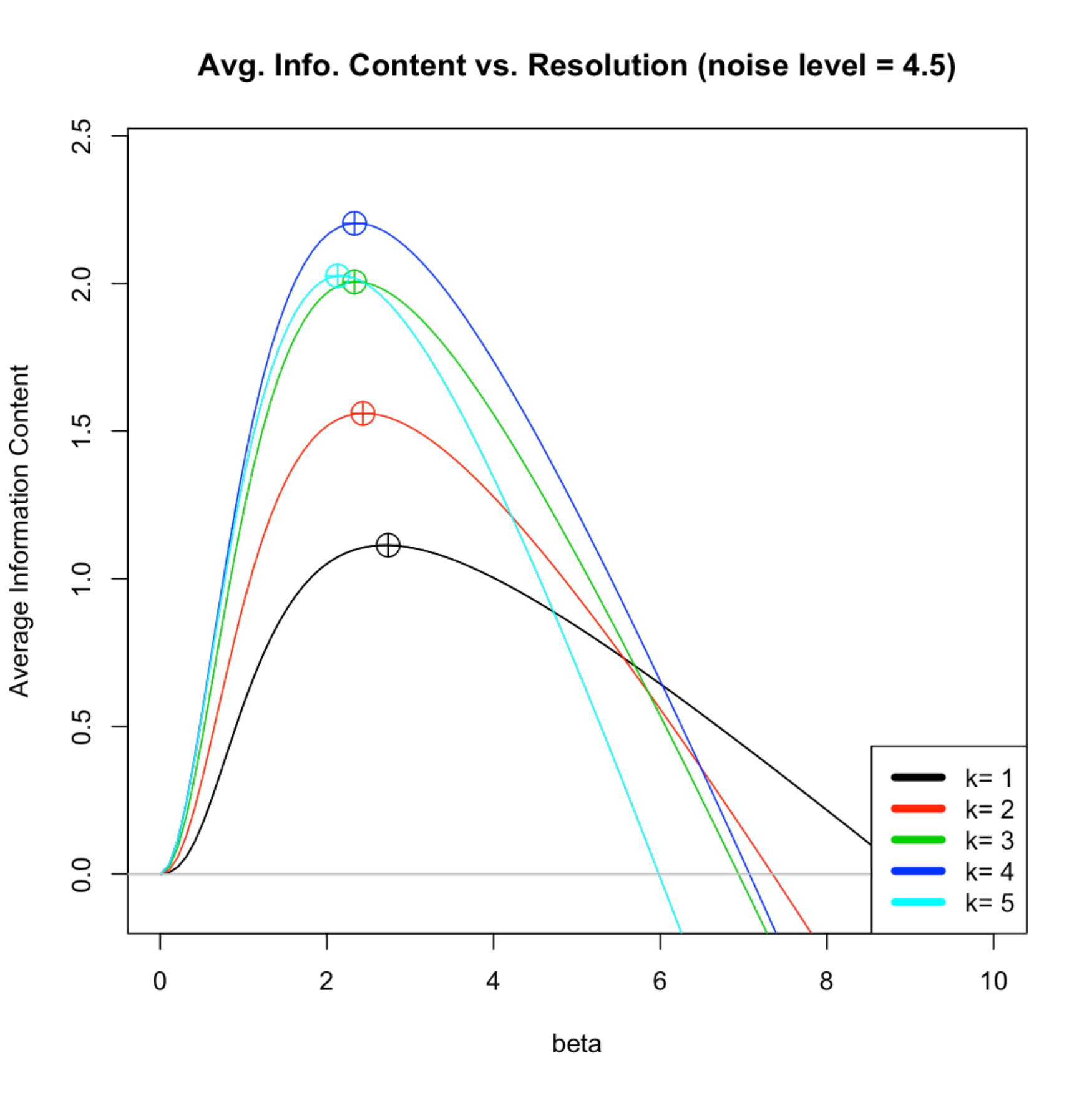}
    \caption{$\sigma=4.5$}
    \label{fig:avgICsigma4.5}
  \end{subfigure}
  %%         --- .5\textwidth stands for 50% of text width
  \caption[Average information content for different $\sigma$]%<<-- Legend for the list of figures at the beginning of you thesis
  {Average information content for different $\sigma$}% legend displayed below the graph.
  \label{fig:avgICsigma}
\end{figure}

	For a given $k$ we see that, as in the non-sparse case, for low noise levels, increases in resolution obtain diminishing gains in information content while for higher noise levels, once the optimum resolution $\beta^*$ is surpassed, the information content decreases. Also notice that as $k$ increases toward $\frac{d}{2}=5$ the information content also increases. This is because the size of the hypothesis space is increasing in $k$ from 0 to $\ceil{\frac{d}{2}}$. We now show the generalization capacity for different values of $k$.

	\begin{figure}[H]%--- Picture 'H'ere, 'B'ottom or 'T'op; '!' Try to
	                    %impose your will to LaTeX
	  \centering
	  \includegraphics[width=.5\textwidth]{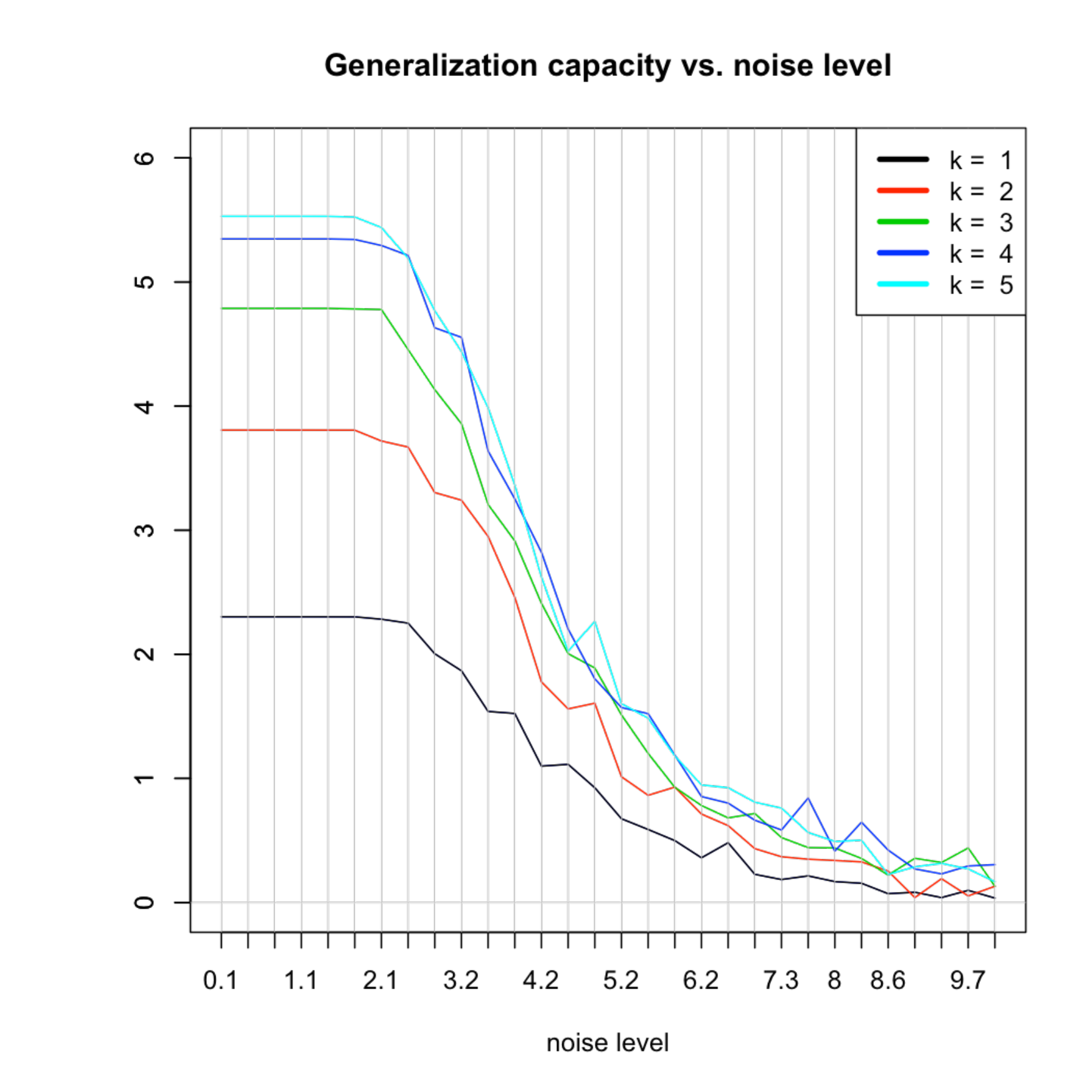} %<< no file extension
	  %%         --- .5\textwidth stands for 50% of text width
	  \caption[Generalization capacity for different $k$]%<<-- Legend for the list of figures at the beginning of you thesis
	  {Generalization capacity for different $k$}% legend displayed below the graph.
	  \label{fig:GCk}
	\end{figure}

Again, since values of $k$ closer to $\ceil{\frac{d}{2}}$ correspond to larger hypothesis spaces, the generalization capacity of the cost function $\hat{R}(\mu, \xi)$ is larger for these $k$. 

\section{Sampling algorithm} \label{SA}

As was mentioned in Section \ref{prbStmnt} we are ultimately interested in the sparse case where $k$ is kept constant and $d$ grows toward infinity. As $d$ grows, the size of the hypothesis space $\mathcal{C}^k$ grows exponentially fast which means computing the partition functions $Z_{\beta}(\xi)$ and $\dZbz$, which are sums over $\mathcal{C}^k$, quickly becomes unfeasible. In this section we use the sampling algorithm suggested in \cite{Buhm14} to estimate  $Z_{\beta}(\xi)$ and $\dZbz$ without summing over the entire hypothesis space. 

Recall that: 

\begin{align}
	\mathbb{E}_{(\xi_1, \xi_2)}[\IC] &= \log |\mathcal{C}^k|  + \mathbb{E}_{(\xi_1, \xi_2)}[\log \dZbz] \\
	                                 &- \mathbb{E}_{\xi_1}[\log \ZbOnez] - \mathbb{E}_{\xi_2}[ \log \ZbTwoz]\\
	&= \log |\mathcal{C}^k|  + \mathbb{E}_{(\xi_1, \xi_2)}\bigg[\log \sum_{\mu \in \mathcal{C}^k} e^{-\beta(\hat{R}(\mu, \xi_1)+\hat{R}(\mu, \xi_2))}\bigg]\\
	                                 &- \mathbb{E}_{\xi_1}\bigg[\log \sum_{\mu \in \mathcal{C}^k} e^{-\beta\hat{R}(\mu, \xi_1)}\bigg] - \mathbb{E}_{\xi_2}\bigg[ \log \sum_{\mu \in \mathcal{C}^k} e^{-\beta\hat{R}(\mu, \xi_2)}\bigg]
\end{align}	

If we let $p(\mu)=\frac{1}{|\mathcal{C}^k|}=\frac{1}{{d \choose k}}$ then, 

\begin{align}
	\mathbb{E}_{(\xi_1, \xi_2)}[\IC] &= \log |\mathcal{C}^k|  + \mathbb{E}_{(\xi_1, \xi_2)}\bigg[\log |\mathcal{C}^k|\sum_{\mu \in \mathcal{C}^k} \frac{1}{|\mathcal{C}^k|} e^{-\beta(\hat{R}(\mu, \xi_1)+\hat{R}(\mu, \xi_2))}\bigg]\\
	                                 &- \mathbb{E}_{\xi_1}\bigg[\log |\mathcal{C}^k| \sum_{\mu \in \mathcal{C}^k} \frac{1}{|\mathcal{C}^k|} e^{-\beta\hat{R}(\mu, \xi_1)}\bigg] \\
									 &- \mathbb{E}_{\xi_2}\bigg[ \log |\mathcal{C}^k| \sum_{\mu \in \mathcal{C}^k} \frac{1}{|\mathcal{C}^k|} e^{-\beta\hat{R}(\mu, \xi_2)}\bigg]\\
									 &= \log |\mathcal{C}^k| + \log |\mathcal{C}^k|  + \mathbb{E}_{(\xi_1, \xi_2)}\bigg[\log \sum_{\mu \in \mathcal{C}^k} \mathbb{E}_{\mu} \bigg[ e^{-\beta(\hat{R}(\mu, \xi_1)+\hat{R}(\mu, \xi_2))}\bigg]\bigg]\\
									 	                                 &- \log |\mathcal{C}^k|-\mathbb{E}_{\xi_1}\bigg[\log  \sum_{\mu \in \mathcal{C}^k} \mathbb{E}_{\mu} \bigg[ e^{-\beta\hat{R}(\mu, \xi_1)}\bigg]\bigg] \\
									 									 &- \log |\mathcal{C}^k| -\mathbb{E}_{\xi_2}\bigg[ \log  \sum_{\mu \in \mathcal{C}^k} \mathbb{E}_{\mu} \bigg[ e^{-\beta\hat{R}(\mu, \xi_2)}\bigg]\bigg]\\
								 &=  \mathbb{E}_{(\xi_1, \xi_2)}\bigg[\log \sum_{\mu \in \mathcal{C}^k} \mathbb{E}_{\mu} \bigg[ e^{-\beta(\hat{R}(\mu, \xi_1)+\hat{R}(\mu, \xi_2))}\bigg]\bigg]\\
								 	                                 &-\mathbb{E}_{\xi_1}\bigg[\log  \sum_{\mu \in \mathcal{C}^k} \mathbb{E}_{\mu} \bigg[ e^{-\beta\hat{R}(\mu, \xi_1)}\bigg]\bigg] \\
								 									 &-\mathbb{E}_{\xi_2}\bigg[ \log  \sum_{\mu \in \mathcal{C}^k} \mathbb{E}_{\mu} \bigg[ e^{-\beta\hat{R}(\mu, \xi_2)}\bigg]\bigg]
\end{align}	

The last expression suggests the following sampling algorithm to estimate the generalization capacity $I$: 

\begin{enumerate}[1.]
	\item Choose a grid of relevant $\beta$ values: $\underline{\beta}=(\beta_1,...,\beta_l)$
	\item For $i=1$ to $m$
		\begin{enumerate}[a.]
			\item Simulate $\xi_1^i,\xi_2^i \sim N(0,I_d)$ 
			\item Uniformly sample $r$ hypotheses $\mu_j \in \mathcal{C}^k$
			\item For $k=1$ to $l$
				\begin{itemize}
					\item Calculate \emph{quasi} information content:
						\begin{align}
							\tilde{I}^i_{\beta_k}&= \log \frac{1}{r} \sum_{j=1}^r e^{-\beta_k (\hat{R}(\mu_j, \xi_1^i)+\hat{R}(\mu_j, \xi_2^i))} \\
							&- \log \frac{1}{r} \sum_{j=1}^r e^{-\beta_k \hat{R}(\mu_j, \xi_1^i)} - \log \frac{1}{r} \sum_{j=1}^r e^{-\beta_k \hat{R}(\mu_j, \xi_2^i)}
						\end{align}	
				\end{itemize}	
		\end{enumerate}	
	\item For $k=1$ to $l$
			\begin{itemize}
				\item Estimate mean information content:
				\begin{align}
					\bar{I}_{\beta_k}= \frac{1}{m} \sum_{i=1}^m \tilde{I}^i_{\beta_k}  
				\end{align}
				\item Estimate generalization capacity:
				\begin{align}
					\hat{I}= \max_{k \in \{1,...,l\}} \bar{I}_{\beta_k}
				\end{align}
			\end{itemize}	
\end{enumerate}

\section{Simulation results: sampling algorithm} \label{SASR}

We used the following parameters for the simulation experiments:

\begin{itemize}
	\item $d=10$,
	\item $k=4$,
	\item $m=100$,
	\item $r=100,1000$,
	\item 20 different $\beta$ values from 0.01 to 10, and
	\item 20 different noise levels $\sigma$ from 0.1 to 15. 
\end{itemize}

We first compare the generalization capacity estimation using the exhaustive and sampling algorithms for two different choices of $r$.

\begin{figure}[H]%--- Picture 'H'ere, 'B'ottom or 'T'op; '!' Try to
                    %impose your will to LaTeX
 % \centering 
  
  \begin{subfigure}{.5\textwidth}
    \centering
	\includegraphics[width=1\textwidth]{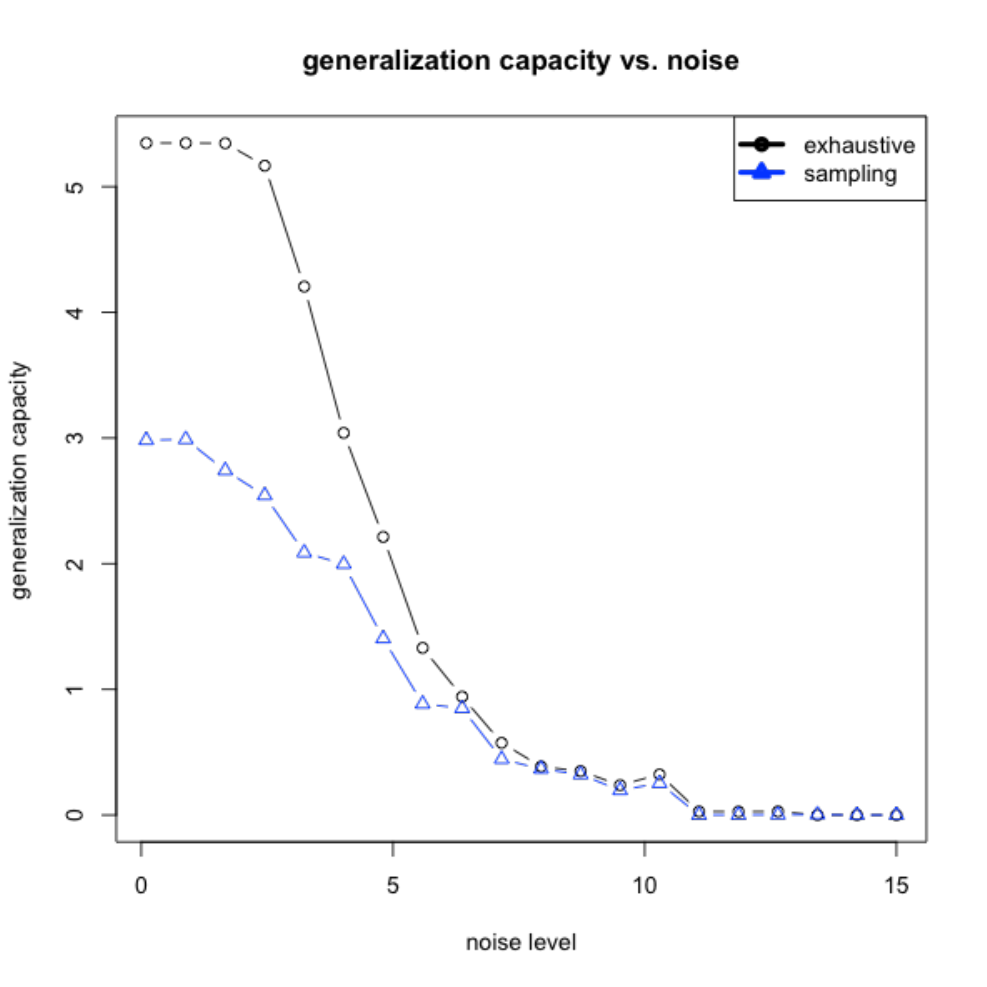} 
    \caption{$r=100$}
    \label{fig:GC100}
  \end{subfigure}%
  \begin{subfigure}{.5\textwidth}
    \centering
    \includegraphics[width=1\textwidth]{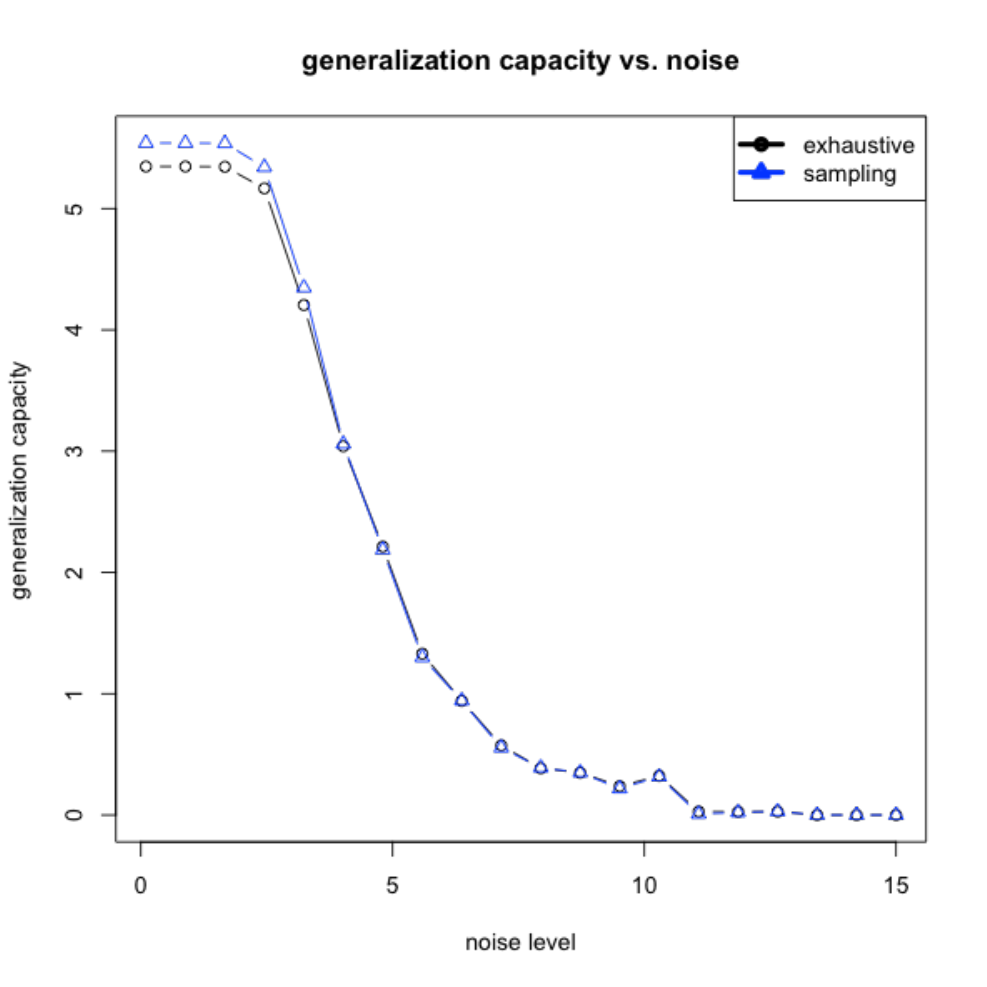}
    \caption{$r=1000$}
    \label{fig:GC1000}
  \end{subfigure}
  %%         --- .5\textwidth stands for 50% of text width
  \caption[Generalization capacity estimation for different $r$]%<<-- Legend for the list of figures at the beginning of you thesis
  {Generalization capacity estimation for different $r$}% legend displayed below the graph.
  \label{fig:GC}
\end{figure}

It is clear that using the sampling algorithm with $r=100$ we underestimate the generalization capacity since for very low noise levels we know the generalization capacity is equal to $\log |\mathcal{C}^k| = \log |{d \choose k}|= \log |{10 \choose 4}|\approx 5.35$. For $r=1000$ the estimation is much better, however since $|\mathcal{C}^k| = 210$ it is cheaper to use the exhaustive algorithm. Next we check the generalization capacity estimate for both algorithms as $d$ increases. The parameters used for the simulation experiments were the following:

\begin{itemize}
	\item $d=5,6,...,20$,
	\item $k=4$,
	\item $m=100$,
	\item $r=20,100,500,1000$,
	\item 20 different $\beta$ values from 0.01 to 10, and
	\item 2 different noise levels $\sigma=2,4$. 
\end{itemize}

\begin{figure}[H]%--- Picture 'H'ere, 'B'ottom or 'T'op; '!' Try to
                    %impose your will to LaTeX
 % \centering 
  
  \begin{subfigure}{.5\textwidth}
    \centering
	\includegraphics[width=1\textwidth]{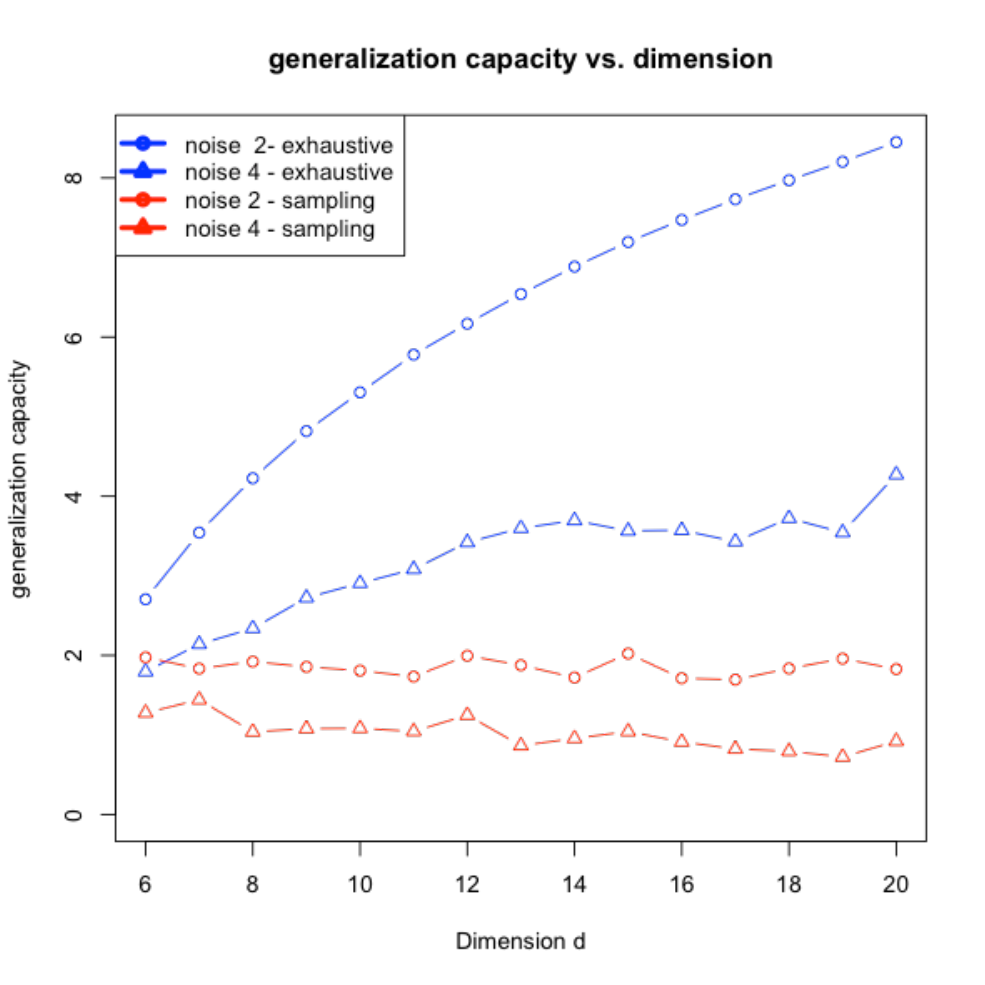} 
    \caption{$r=20$}
    \label{fig:GCd20}
  \end{subfigure}%
  \begin{subfigure}{.5\textwidth}
    \centering
    \includegraphics[width=1\textwidth]{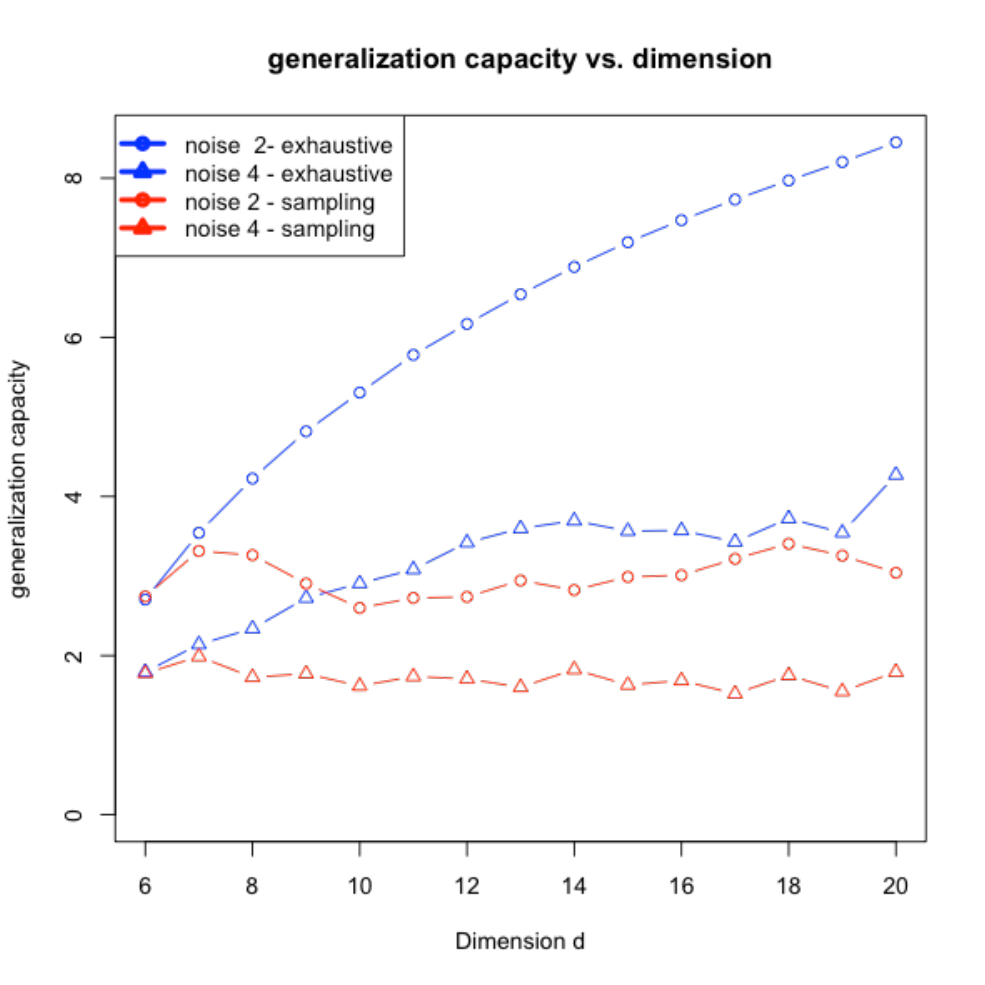}
    \caption{$r=100$}
    \label{fig:GCd100}
  \end{subfigure}
  \begin{subfigure}{.5\textwidth}
    \centering
    \includegraphics[width=1\textwidth]{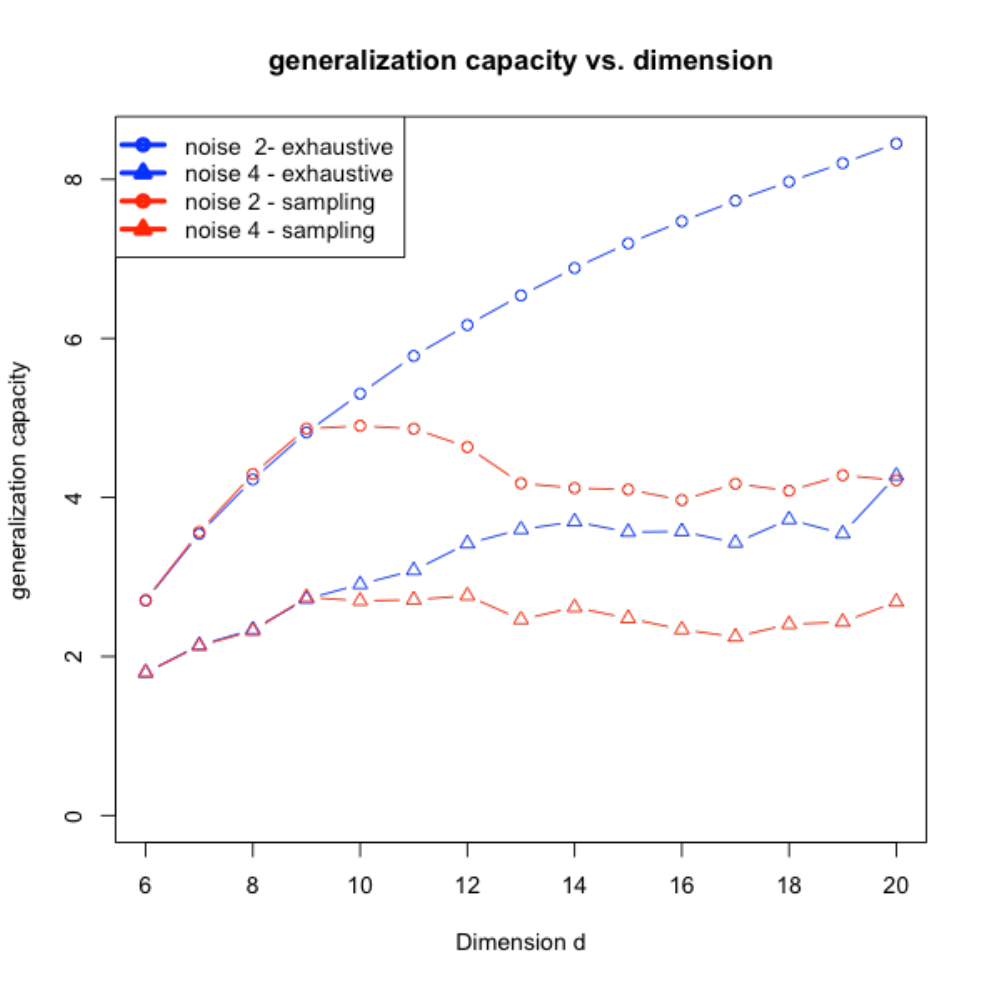}
    \caption{$r=500$}
    \label{fig:GCd500}
  \end{subfigure}
  \begin{subfigure}{.5\textwidth}
    \centering
    \includegraphics[width=1\textwidth]{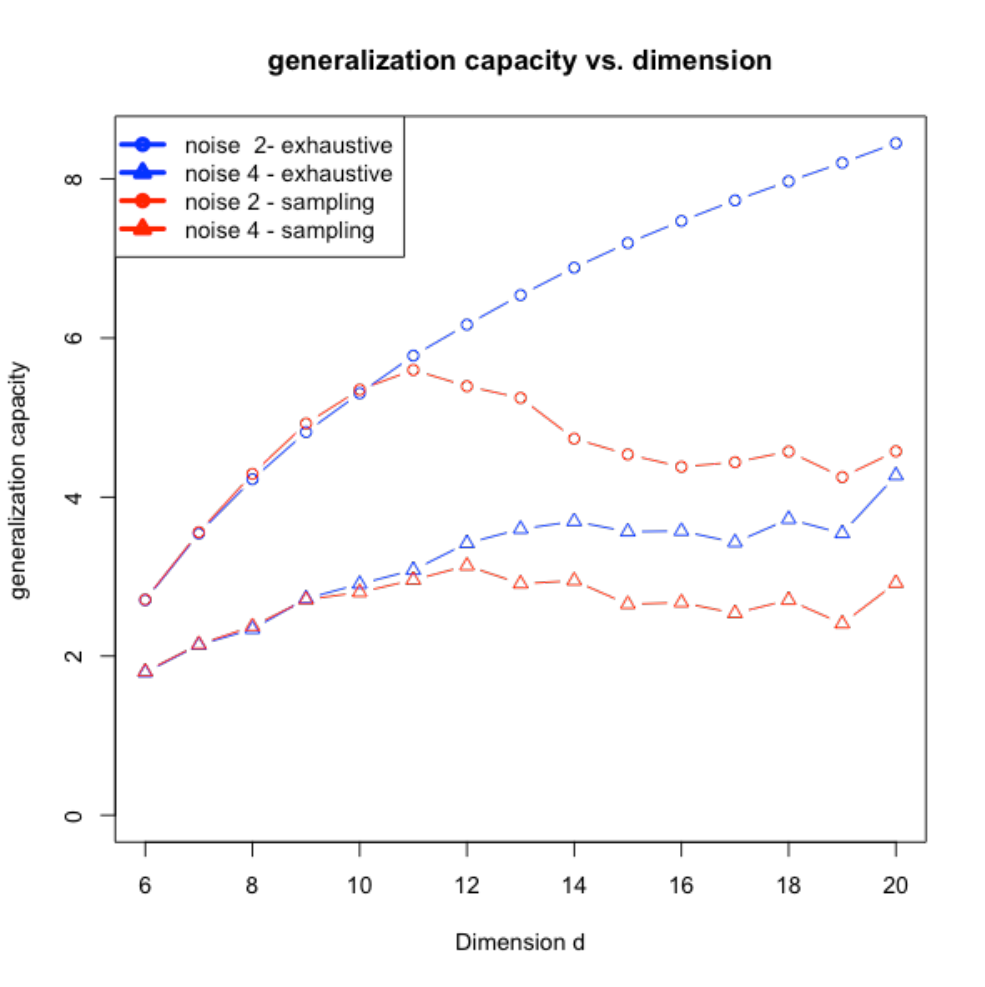}
    \caption{$r=1000$}
    \label{fig:GCd1000}
  \end{subfigure}
  %%         --- .5\textwidth stands for 50% of text width
  \caption[Generalization capacity estimation for different $d$ and $r$]%<<-- Legend for the list of figures at the beginning of you thesis
  {Generalization capacity estimation for different $d$ and $r$}% legend displayed below the graph.
  \label{fig:GC}
\end{figure}

The figure above confirms that the sampling algorithm proposed turns out to be more expensive than the exhaustive algorithm. For example, to estimate the generalization capacity accurately when $d=11$ we need $r=1000$ which is larger than the size of the hypothesis space ${11 \choose 4} = 330$. The sampling algorithm represents a way to estimate genearalization capacity by summing over a sample of the hypothesis class. The sample size required must be, at most, linear in $d$ if it is to be useful in estimating generalization capacity for sparse conditions when $k$ is fixed and $d$ is large. 

\section{Importance sampling algorithm} \label{ISA}

The aim of the sampling algorithm is to estimate $\mathbb{E}_{\mu}[e^{-\beta \hat{R}(\mu, \xi)}]$ with a sample from $p(\mu)=\frac{1}{|\mathcal{C}^k|}$ instead of exhaustively calculating it as:

\begin{align}
	\mathbb{E}_{\mu}[e^{-\beta \hat{R}(\mu, \xi)}]=\frac{1}{|\mathcal{C}^k|} \sum_{\mu \in \mathcal{C}^k} e^{-\beta\hat{R}(\mu,\xi)}
\end{align}	

This means sampling $\mu_j \in \mathcal{C}^k$, calculating the Boltzmann weights $w_{\beta}(\mu_j, \xi)$ and averaging them. In general, different hypothesis $\mu_j \in \mathcal{C}^k$ contribute differently to the average. The lower the cost of a hypothesis the larger the weight contributed. Those hypotheses which contribute most of the weight are relatively small in number. This means that with a small number of samples the proportion of \emph{important} and \emph{unimportant} hypotheses will not accurately reflect the population proportions and the result will be a biased estimate of $Z_{\beta}(\xi)$ and $\dZbz$. If we take a large sample, of size bigger than the size of $\mathcal{C}^k$, the simulation results seem to show that the estimates converge to those of the exhaustive algorithm, but this defeats the purpose of sampling hypotheses: to obtain an algorithm that is \emph{not} exponential in $d$ as the size of the hypothesis class is. 

We want to design an \emph{importance sampling} algorithm where:

\begin{enumerate}[1.]
	\item We sample according to a proposal distribution $q(\mu)$ which assigns more probability to more \emph{important} hypotheses.
	\item When estimating the expectation with the average we assign weights $w(\mu)$ to each sample to correct for the fact we sampled according to \emph{wrong} distribution:
\end{enumerate}	

\begin{align}
	\mathbb{E}_{p(\mu)}[e^{-\beta\hat{R}(\mu, \xi)}] &= \sum_{\mu \in \mathcal{C}^k} p(\mu)e^{-\beta\hat{R}(\mu, \xi)} &= \sum_{i = 1}^{{d \choose k}} p(\mu_i)e^{-\beta\hat{R}(\mu_i, \xi)}\\
	&=\sum_{i = 1}^{{d \choose k}} \frac{p(\mu_i)}{q(\mu_i)}q(\mu_i)e^{-\beta\hat{R}(\mu_i, \xi)} &= \mathbb{E}_{q(\mu)}[w(\mu)e^{-\beta\hat{R}(\mu, \xi)}]
\end{align}	
Where $w(\mu)=\frac{p(\mu)}{q(\mu)}$.

Recall that we may write:

\begin{align}
	\hat{R}(\mu, \xi) &= \mu^T(1-2\frac{\sigma}{\sqrt{n}}\xi - 2\mu^0)\\
	\hat{R}(\mu, \xi_1)+\hat{R}(\mu, \xi_2) &=2\mu^T(1-\frac{\sigma}{\sqrt{n}}(\xi_1+\xi_2) - 2\mu^0)
\end{align}	

Observe that $\mu^T1=k$ is constant for the case where the hypothesis space is $\mathcal{C}^k$ and not $\mathcal{C}=\binarynumbers^d$. Also note that $h:=\mu^T\mu^0 \in \{\max(0,2k-d),...,k\}$ is the number of \emph{hits} of hypothesis $\mu$: the number of components $j$ such that $\mu_j=\mu_j^0=1$. With this in mind we can obtain the following equivalent expression for $\hat{R}(\mu, \xi)$:

\begin{align}
	\hat{R}(\mu, \xi) &= -2[h + \frac{\sigma}{\sqrt{n}}\mu^T\xi] \\
	\hat{R}(\mu, \xi_1) + \hat{R}(\mu, \xi_2) &= -2[2h + \frac{\sigma}{\sqrt{n}}\mu^T(\xi_1 + \xi_2)] 
\end{align}	
Where $h:=h(\mu, \mu^0)=\mu^T\mu^0$. This means we may express the Boltzmann weights as:

\begin{align}
	w_{\beta}(\mu, \xi) = e^{-\beta\hat{R}(\mu, \xi)} = e^{2\beta[h+\frac{\sigma}{\sqrt{n}}\mu^T\xi]}= e^{2\beta h}e^{2\beta\frac{\sigma}{\sqrt{n}}\mu^T\xi} = f_{\beta}(h)g_{\beta}(\mu, \xi)
\end{align}	

The importance of a hypothesis $\mu$ is given by the Boltzmann weights $w_{\beta}(\mu, \xi)$ and these depend on the functions $f_{\beta}$ and $g_{\beta}$. We first analyse $g_{\beta}(\mu, \xi):= e^{2\beta\frac{\sigma}{\sqrt{n}}\mu^T\xi}$. We have that, for $\beta,k>0$:

\begin{align}
	\xi \sim N(0,I_d) \Rightarrow \mu^T\xi \sim N(0,k) &\Rightarrow \frac{2\beta\sigma}{\sqrt{n}}\mu^T\xi \sim N\bigg(0, \bigg(\frac{2\beta\sigma}{\sqrt{n}}\bigg)^2k\bigg)\\
	& \Rightarrow e^{\frac{2\beta\sigma}{\sqrt{n}}\mu^T\xi} \sim \log N\bigg(0, \bigg(\frac{2\beta\sigma}{\sqrt{n}}\bigg)^2k\bigg)\\
	& \Rightarrow \mathbb{E}_{\xi}[e^{\frac{2\beta\sigma}{\sqrt{n}}\mu^T\xi}] = e^{\frac{1}{2}(\frac{2\beta\sigma}{\sqrt{n}})^2k} > 1
\end{align}	

This means that we can expect $g_{\beta}(\mu, \xi)$ to increase $w_{\beta}(\mu, \xi)$ for half the samples $\xi$ and decrease it for the other half, although in general, we can expect $g_{\beta}(\mu, \xi)$ to increase $w_{\beta}(\mu, \xi)$. So we see that $g_{\beta}(\mu, \xi)$ may contribute to a hypothesis being more or less important depending on the $\xi$ sampled. If we want to take this effect into account in determining the proposal distribution $q(\mu)$ we would have to make it depend on the given $\xi$ sampled. For simplicity we only take into account $f_{\beta}(h)$ in determining $q(\mu)$. 

We now analyse $f_{\beta}(h)=e^{2\beta h}$. The more hits $h$ that a hypothesis $\mu$ has with respect to $\mu^0$ the lower the costs $\hat{R}(\mu, \xi)$ which means the larger he weight $w_{\beta}(\mu, \xi)$ contributed. Also note that there are ${k \choose h}{d-k \choose k-h}$ hypotheses $\mu \in \mathcal{C}^k$ such that $h(\mu, \mu^0)=\mu^T\mu^0=h$. The following figure illustrates the importance $f_{\beta}(h)$ of a hypothesis $\mu \in \mathcal{C}^k$ and the number of such hypotheses in $\mathcal{C}^k$ as a function of $h$ for $d=20$, $k=4$ and $\beta=1$. 

	\begin{figure}[H]%--- Picture 'H'ere, 'B'ottom or 'T'op; '!' Try to
	                    %impose your will to LaTeX
	  \centering
	  \includegraphics[width=.5\textwidth]{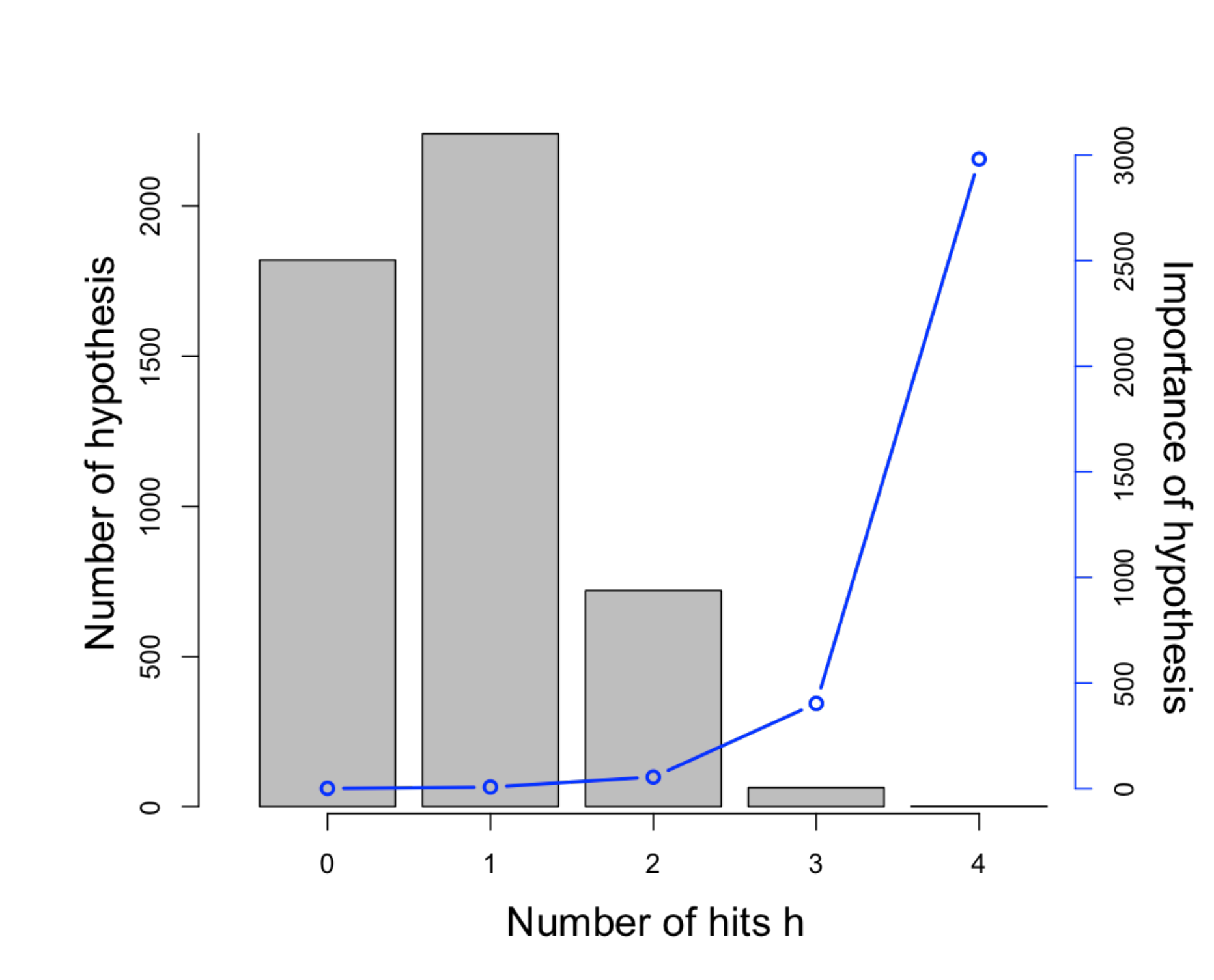} %<< no file extension
	  %%         --- .5\textwidth stands for 50% of text width
	  \caption[Importance and frequency of hypotheses]%<<-- Legend for the list of figures at the beginning of you thesis
	  {Importance and frequency of hypotheses}% legend displayed below the graph.
	  \label{fig:imp}
	\end{figure}
	
Hypothesis $\mu$ such that $h(\mu,\mu^0)$ is close to $k$ contribute moste of the weight but are relatively few in number. There are ${k \choose h}{d-k \choose k-h}$ hypotheses $\mu \in \mathcal{C}^k$ such that $h(\mu, \mu^0)=\mu^T\mu^0=h$ so there are $\sum_{h=\max(1,2k-d)}^k{k \choose h}{d-k \choose k-h}$ hypotheses such that $h>0$. As $d$ grows this becomes a low proportion of ${d \choose k}$, the total number of hypotheses. In other words, as $d$ grows, the most important hypotheses, those that contribute most of the weight, become a smaller proportion of all hypotheses. 
	
	\begin{figure}[H]%--- Picture 'H'ere, 'B'ottom or 'T'op; '!' Try to
	                    %impose your will to LaTeX
	  \centering
	  \includegraphics[width=.5\textwidth]{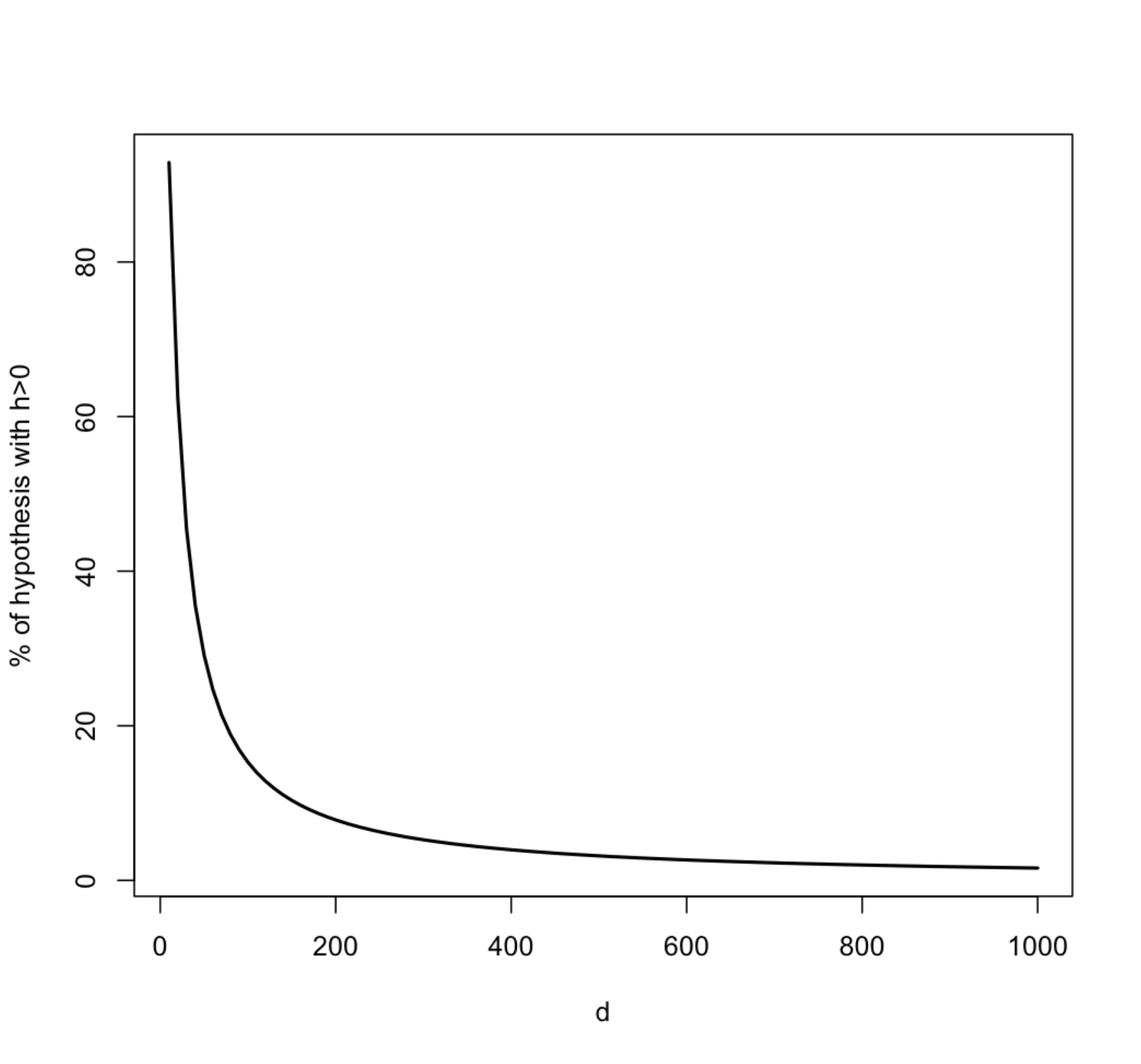} %<< no file extension
	  %%         --- .5\textwidth stands for 50% of text width
	  \caption[\% of hypotheses $\mu \in \mathcal{C}^k$ such that $h>0$ for $k=4$]%<<-- Legend for the list of figures at the beginning of you thesis
	  {\% of hypotheses $\mu \in \mathcal{C}^k$ such that $h>0$}% legend displayed below the graph.
	  \label{fig:imp2}
	\end{figure}

This means that to sample a representative proportion of important hypotheses we need a very large sample. To find a way around this we use a proposal distribution $q(\mu)$ such that the probability of sampling a hypothesis with $h$ hits is the same for all $h$. 	

Let, 

\begin{align}
	h \sim U\{\max(0,2k-d),...,k\} &\Rightarrow q(h) = \frac{1}{k-\max(0,2k-d)+1}\\
	\mu|h \sim U\bigg\{1,...,{k \choose h}{d-k \choose k-h}\bigg\} &\Rightarrow q(\mu|h) = \frac{\mathbbm{1}_{\{\mu^T\mu^0=h\}}}{{k \choose h}{d-k \choose k-h}} 
\end{align}	

This means that:

\begin{align}
	q(\mu) = \sum_{h=\max(0,2k-d)}^k q(\mu, h) &= \sum_{h=\max(0,2k-d)}^k q(\mu|h)q(h) \\
	&= \sum_{h=\max(0,2k-d)}^k \frac{\mathbbm{1}_{\{\mu^T\mu^0=h}\}}{(k-\max(0,2k-d)+1){k \choose h}{d-k \choose k-h}}\\
	&= \frac{1}{(k-\max(0,2k-d)+1){k \choose \mu^T\mu^0}{d-k \choose k-\mu^T\mu^0}}
\end{align}

\begin{align}
	w(\mu) = \frac{p(\mu)}{q(\mu)} = \frac{{k \choose \mu^T\mu^0}{d-k \choose k-\mu^T\mu^0}}{{d \choose k}} 
\end{align}

The above suggests the following importance sampling algorithm for estimating generalization capacity. 

\begin{enumerate}[I.]
	\item Choose a grid of relevant $\beta$ values: $\underline{\beta}=(\beta_1,...,\beta_l)$
	\item For $i=1$ to $m$
		\begin{enumerate}[1.]
			\item Simulate $\xi_1^i,\xi_2^i \sim N(0,I_d)$ 
			\item For $j=1$ to $r$
				\begin{enumerate}[a.]
					\item Sample from $h_j \sim q(h)=\frac{1}{k-\max(0,2k-d)+1}$
					\item Sample from $\mu_j|h_j \sim q(\mu|h)=\frac{\mathbbm{1}_{\{\mu^T\mu^0=h\}}}{{k \choose h}{d-k \choose k-h}}$ This can be done by: 
					\begin{enumerate}[i.]
						\item Calculate number of \emph{misses} $m_j=k-h_j$
						\item Identify $A_0 = \{a: \mu_a^0=0\}$ and $A_1 = \{a: \mu_a^0=1\}$
						\item Set $\mu_j \leftarrow \mu^0$
						\item Uniformly sample $m_j$ times $r \in A_0$ and $s \in A_1$ setting $\mu_{jr}=1$ and $\mu_{js}=0$ each time.
					\end{enumerate}	
				\end{enumerate}	
			\item For $k=1$ to $l$
				\begin{itemize}
					\item Calculate \emph{quasi} information content:
						\begin{align}
							\tilde{I}^i_{\beta_k}&= \log \frac{1}{r} \sum_{j=1}^r e^{-\beta_k (\hat{R}(\mu_j, \xi_1^i)+\hat{R}(\mu_j, \xi_2^i))} \\
							&- \log \frac{1}{r} \sum_{j=1}^r e^{-\beta_k \hat{R}(\mu_j, \xi_1^i)} - \log \frac{1}{r} \sum_{j=1}^r e^{-\beta_k \hat{R}(\mu_j, \xi_2^i)}
						\end{align}	
				\end{itemize}	
		\end{enumerate}	
	\item For $k=1$ to $l$
			\begin{itemize}
				\item Estimate mean information content:
				\begin{align}
					\bar{I}_{\beta_k}= \sum_{i=1}^m w_i \tilde{I}^i_{\beta_k}
				\end{align}
				Where $w_i=w(\mu_i)=\frac{{k \choose \mu_i^T\mu^0}{d-k \choose k-\mu_i^T\mu^0}}{{d \choose k}}$
				\item Estimate generalization capacity:
				\begin{align}
					\hat{I}= \max_{k \in \{1,...,l\}} \bar{I}_{\beta_k}
				\end{align}
			\end{itemize}	
\end{enumerate}

\section{Simulation results: importance sampling algorithm} \label{impSampRes}

We used the following parameters for the simulation experiments:

\begin{itemize}
	\item $d=10$,
	\item $k=4$,
	\item $m=100$,
	\item $r=100,1000$,
	\item 20 different $\beta$ values from 0.01 to 10, and
	\item 20 different noise levels $\sigma$ from 0.1 to 15. 
\end{itemize}

We first compare the generalization capacity estimation using the exhaustive, sampling and importance sampling algorithms for two different choices of $r$.

\begin{figure}[H]%--- Picture 'H'ere, 'B'ottom or 'T'op; '!' Try to
                    %impose your will to LaTeX
 % \centering 
  
  \begin{subfigure}{.5\textwidth}
    \centering
	\includegraphics[width=1\textwidth]{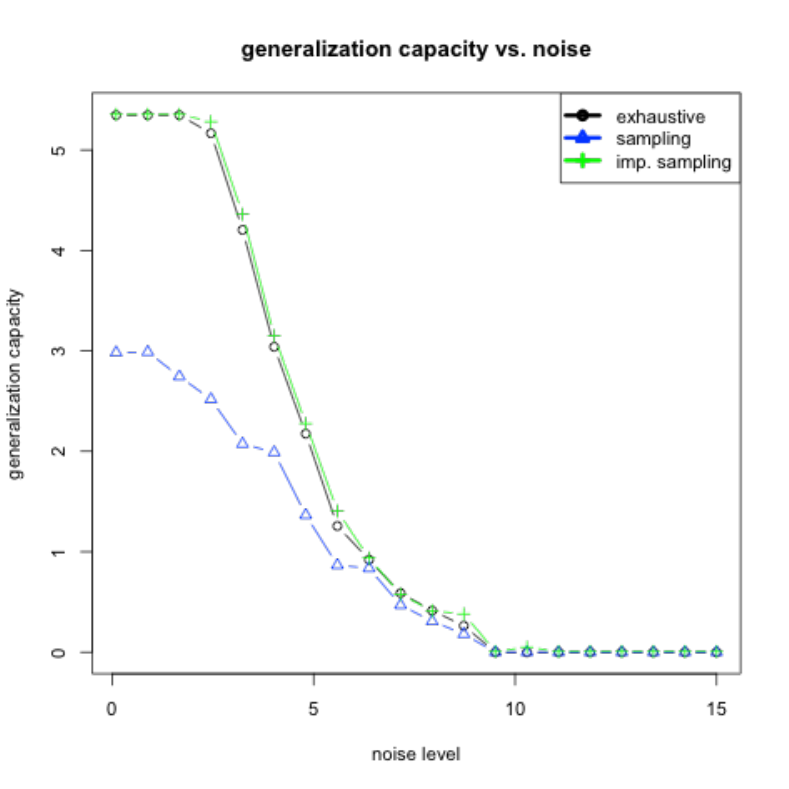} 
    \caption{$r=100$}
    \label{fig:GC100}
  \end{subfigure}%
  \begin{subfigure}{.5\textwidth}
    \centering
    \includegraphics[width=1\textwidth]{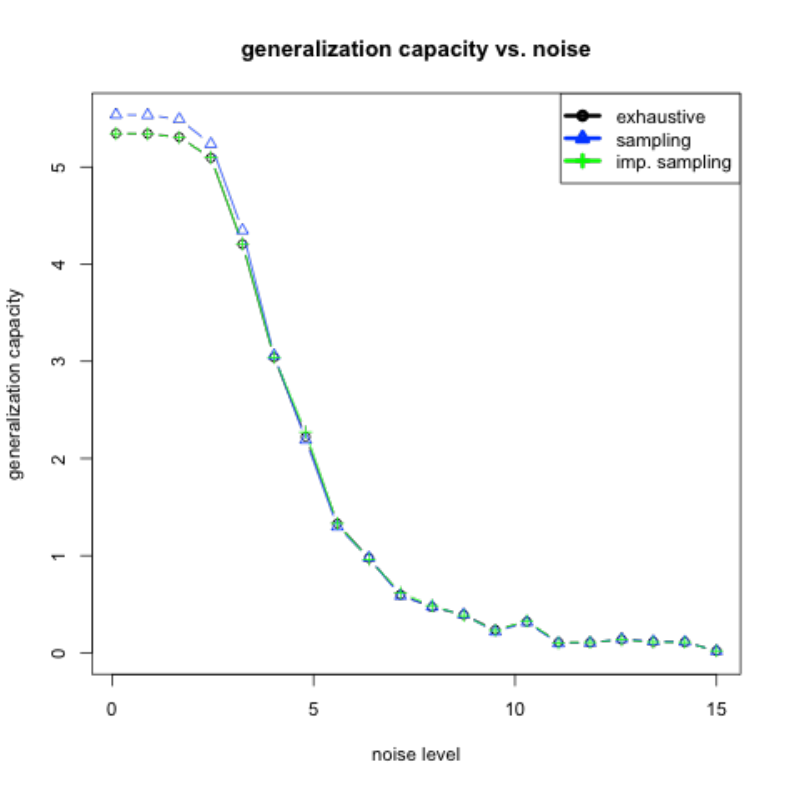}
    \caption{$r=1000$}
    \label{fig:GC1000}
  \end{subfigure}
  %%         --- .5\textwidth stands for 50% of text width
  \caption[Generalization capacity estimation for different $r$]%<<-- Legend for the list of figures at the beginning of you thesis
  {Generalization capacity estimation for different $r$}% legend displayed below the graph.
  \label{fig:GC}
\end{figure}

The figures suggest that the generalization capacity estimated with the importance sampling algorithm converges to the exhaustive algorithm results for a sample size much smaller than is required for the sampling algorithm. Next we check the generalization capacity estimate for all three algorithms as $d$ increases. The parameters used for the simulation experiments were the following:

\begin{itemize}
	\item $d=5,6,...,20$,
	\item $k=4$,
	\item $m=100$,
	\item $r=20,100,500,1000$,
	\item 20 different $\beta$ values from 0.01 to 10, and
	\item 2 different noise levels $\sigma=2,4$. 
\end{itemize}

\begin{figure}[H] %--- Picture 'H'ere, 'B'ottom or 'T'op; '!' Try to
                    %impose your will to LaTeX
 % \centering 
  
  \begin{subfigure}{.5\textwidth}
    \centering
	\includegraphics[width=1\textwidth]{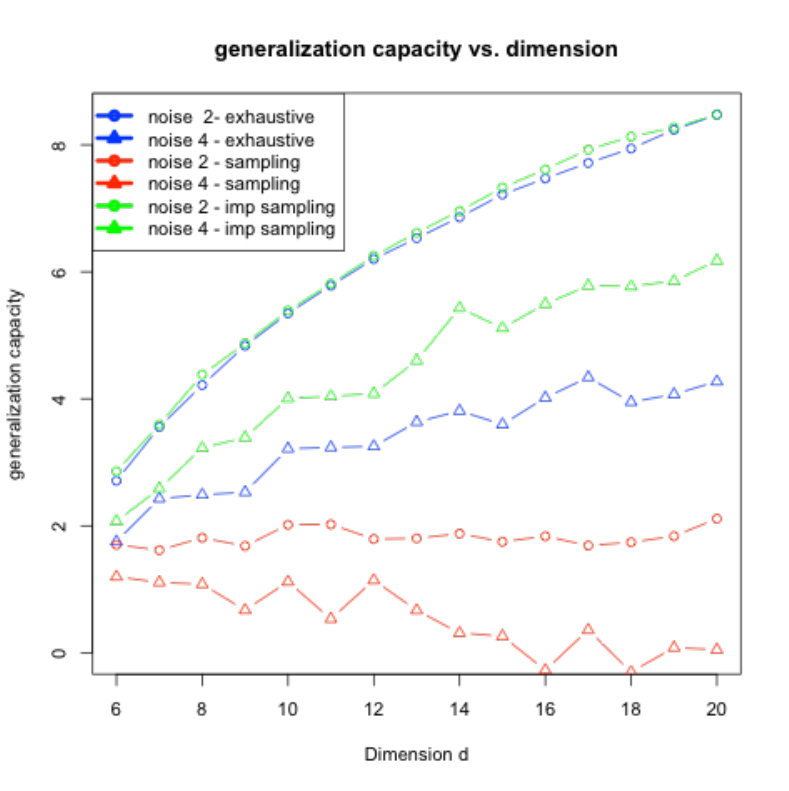} 
    \caption{$r=20$}
    \label{fig:GCd20}
  \end{subfigure}%
  \begin{subfigure}{.5\textwidth}
    \centering
    \includegraphics[width=1\textwidth]{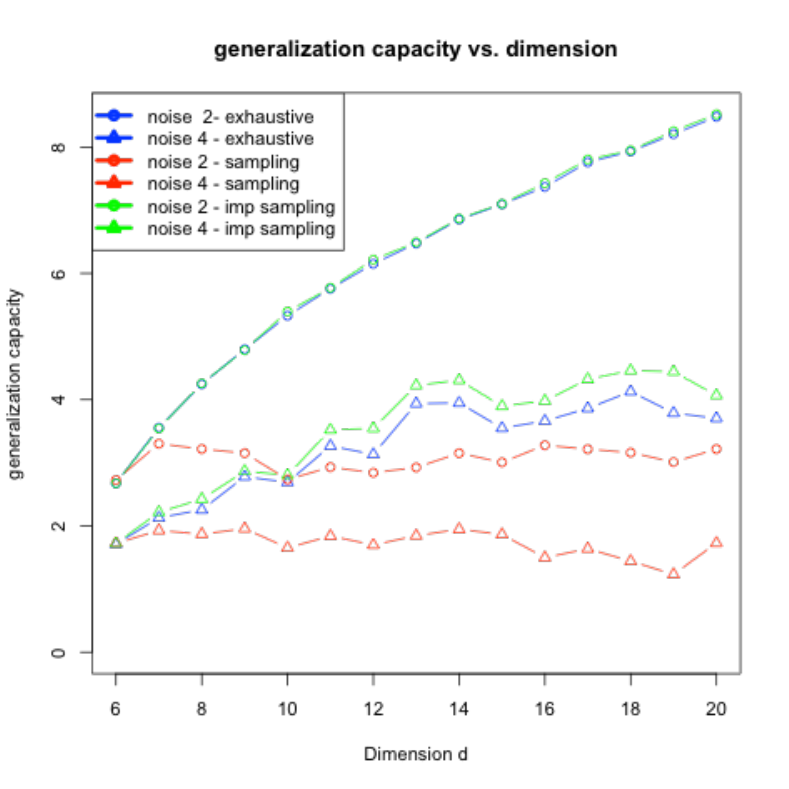}
    \caption{$r=100$}
    \label{fig:GCd100}
  \end{subfigure}
  \begin{subfigure}{.5\textwidth}
    \centering
    \includegraphics[width=1\textwidth]{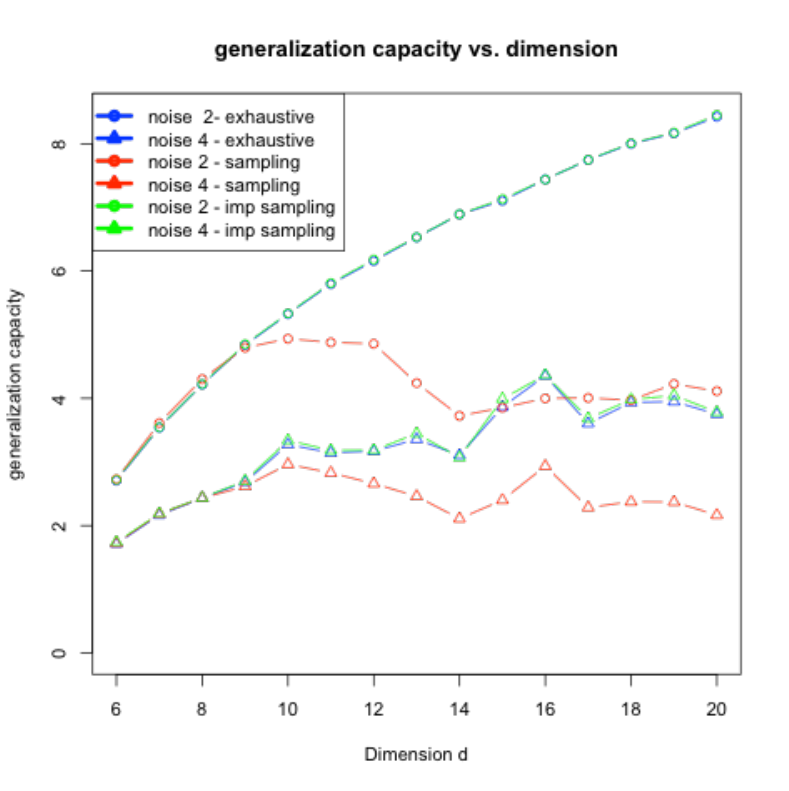}
    \caption{$r=500$}
    \label{fig:GCd500}
  \end{subfigure}
  \begin{subfigure}{.5\textwidth}
    \centering
    \includegraphics[width=1\textwidth]{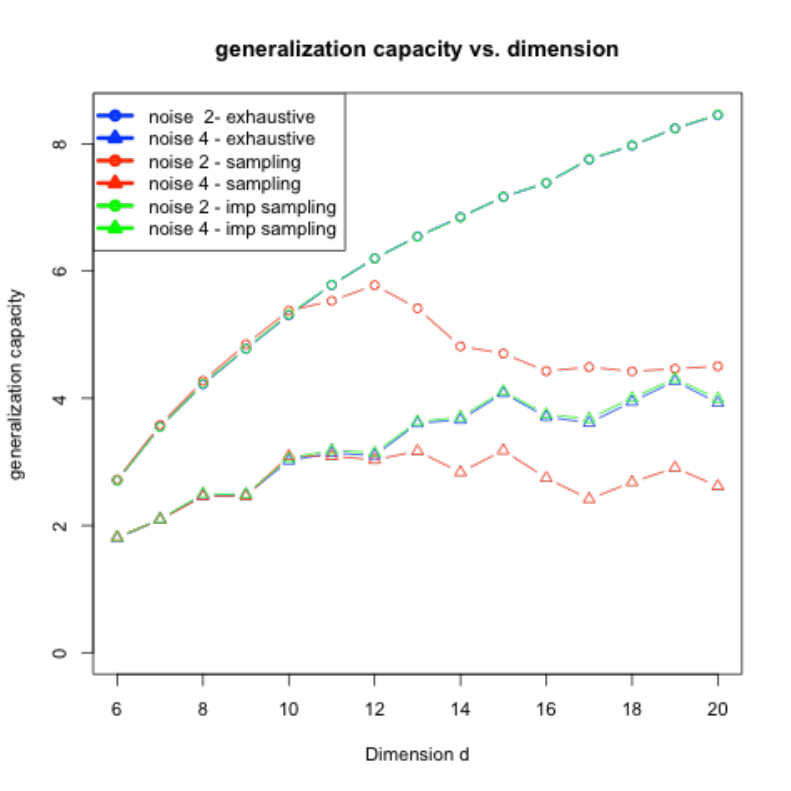}
    \caption{$r=1000$}
    \label{fig:GCd1000}
  \end{subfigure}
  %%         --- .5\textwidth stands for 50% of text width
  \caption[Generalization capacity estimation for different $d$ and $r$]%<<-- Legend for the list of figures at the beginning of you thesis
  {Generalization capacity estimation for different $d$ and $r$}% legend displayed below the graph.
  \label{fig:GC}
\end{figure}

The above figures seem to confirm that the importance sampling algorithm need a much smaller sample size to converge than does the sampling algorithm suggesting it will be useful in estimating generalization capacity for large values of $d$ when we can no longer \emph{exhaustively} evaluate the partition functions. In the following simulation experiments we estimate the generalization capacity of the cost function $\hat{R}(\mu, \xi)$ in sparse settings where $k$ is constant and $d$ is large. We used the following parameters for these simulation experiments: 

\begin{itemize}
	\item $d=10,...,10000$,
	\item $k=4$,
	\item $m=100$,
	\item $r=100$,
	\item 100 different $\beta$ values from 0.01 to 30, and
	\item 4 different noise levels $\sigma=3,4,5,6$. 
\end{itemize}

Recall we ended Section \ref{Shannon} by mentioning that we can use generalization capacity to choose between alternative cost functions. The following simulation experiments, carried out with the above parameters, were performed for two different cost functions, the squared-loss based risk function and an absolute-loss based risk function. Recall expression \ref{sqrLoss} for the squared-loss based risk function:

\begin{align} 
	\hat{R}(\mu, X) = ||\mu - \bar{X}||^2_2
\end{align}	

If we replace the L2 norm with the L1 norm we get te absolute-loss based risk function: 

\begin{align} 
	\hat{R}(\mu, X) = ||\mu - \bar{X}||^2_1
\end{align}	

The following figure shows the estimated generalization capacity of both cost functions for different number of components $d$ and noise levels $\sigma$. Each scatter point represents the estimation, by simulation, of the generalization capacity for a given $d$ and cost function $\hat{R}(\mu,\xi)$. Blue points represent the generalization capacity of the L1 norm cost function and black points that of the L2 norm cost function. The blue and black lines are smoothing splines applied to the blue and black points respectively. 

\begin{figure}[H]%--- Picture 'H'ere, 'B'ottom or 'T'op; '!' Try to
                    %impose your will to LaTeX
 % \centering 
  
  \begin{subfigure}{.5\textwidth}
    \centering
	\includegraphics[width=1\textwidth]{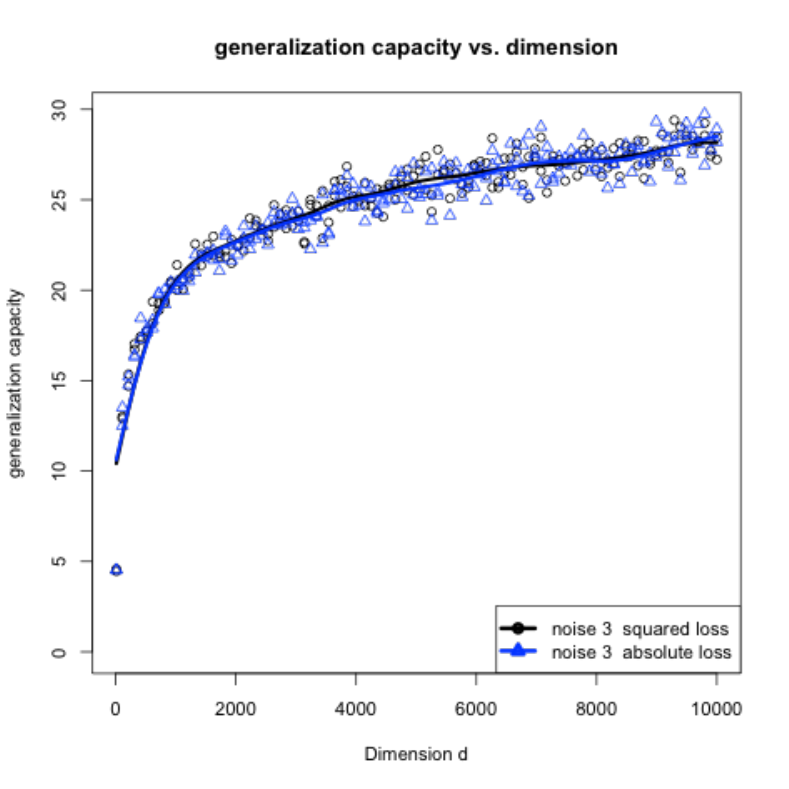} 
    \caption{$r=20$}
    \label{fig:GCdsigma3}
  \end{subfigure}%
  \begin{subfigure}{.5\textwidth}
    \centering
    \includegraphics[width=1\textwidth]{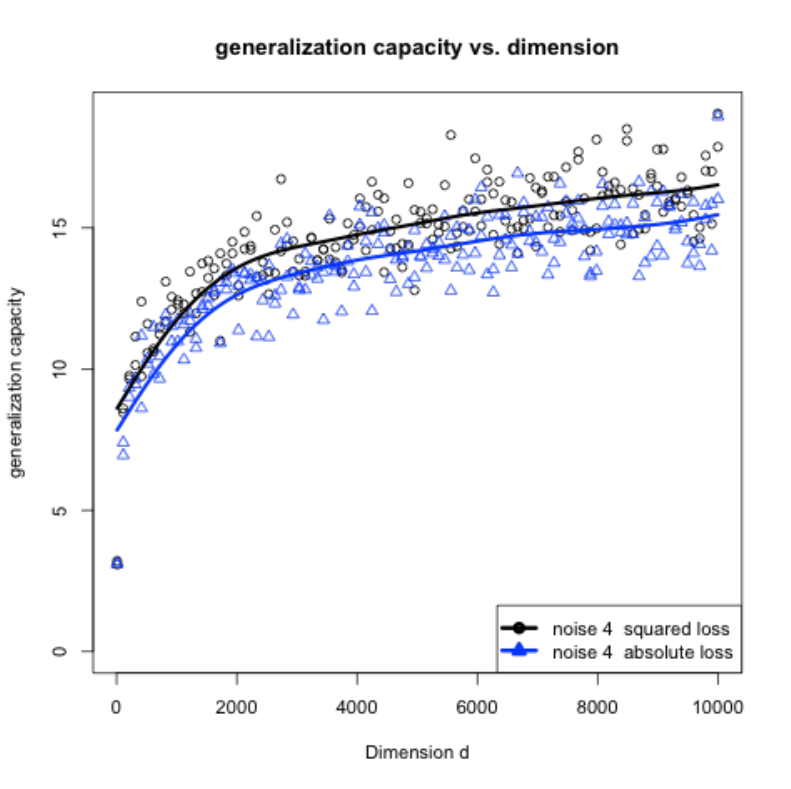}
    \caption{$r=100$}
    \label{fig:GCdsigma4}
  \end{subfigure}
  \begin{subfigure}{.5\textwidth}
    \centering
    \includegraphics[width=1\textwidth]{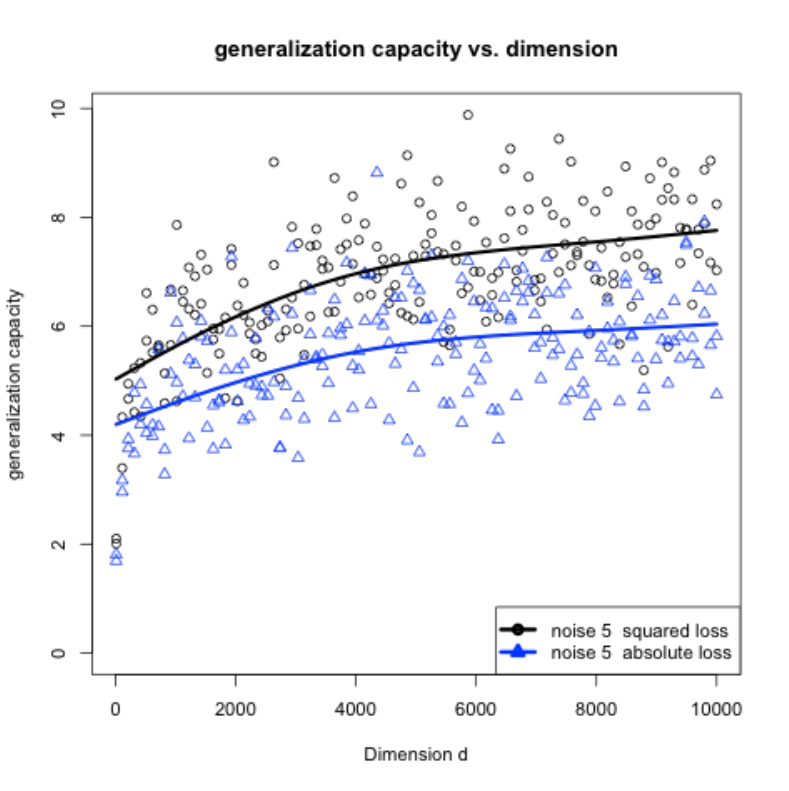}
    \caption{$r=500$}
    \label{fig:GCdsigma5}
  \end{subfigure}
  \begin{subfigure}{.5\textwidth}
    \centering
    \includegraphics[width=1\textwidth]{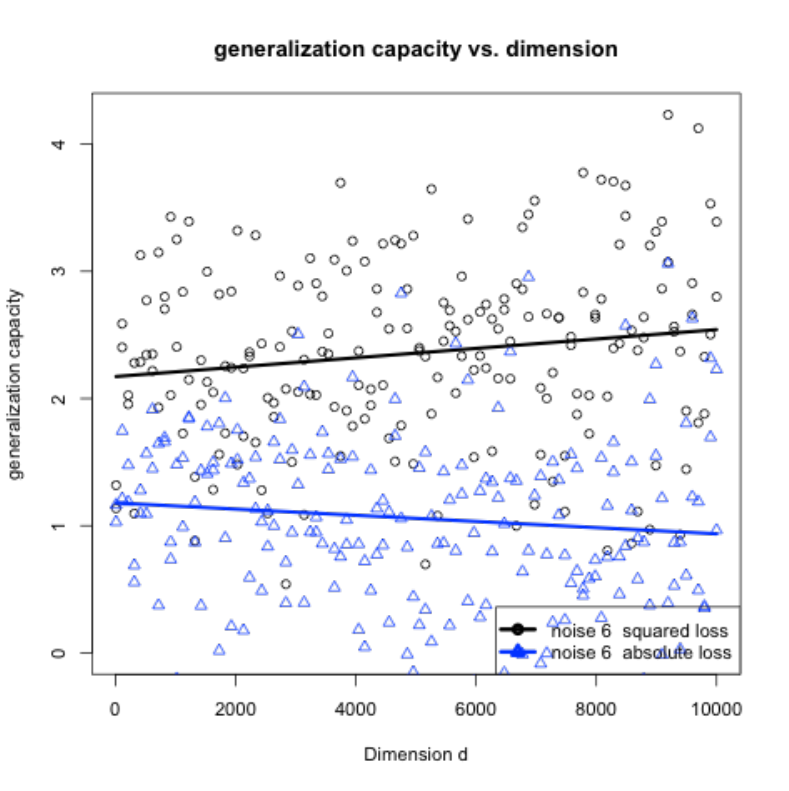}
    \caption{$r=1000$}
    \label{fig:GCdsigma6}
  \end{subfigure}
  %%         --- .5\textwidth stands for 50% of text width
  \caption[Generalization capacity estimation for different $d$ and $\sigma$]%<<-- Legend for the list of figures at the beginning of you thesis
  {Generalization capacity estimation for different $d$ and $\sigma$}% legend displayed below the graph.
  \label{fig:GChighD}
\end{figure}

We can see that the generalization capacity for both cost functions, as a function of the number of components $d$,  displays concave behavior. Initially, as the hypothesis set size grows we see rapid increase in the generalization capacity however after a certain threshold, the large number of hypotheses makes detection of the \emph{relevant} components difficult so that additional gains in generalization capacity are marginal. The value of this threshold depends on the noise level of the data: the higher the noise level the lower the threshold. Additionally, observe that, as we would expect given that the data has a gaussian distribution, the L2 norm cost function has a greater generalization capacity than that of the L1 norm cost function. 
\chapter{Towards sparse feature selection for correlated features} 
\label{ch:TowardsSparseFeatureSelection}

As with the model described in Section \ref{prbStmnt} we deal with the statistical model:

$$
		X_i = \mu^0 + \epsilon_i
$$

Except in this case we have that: 

\begin{itemize}
		\item $||\mu^0||_1=k$,
		\item $\mu^0 \in \binarynumbers^d_k=\{\mu \in \binarynumbers^d: ||\mu||_1=k\}$,
		\item $|\binarynumbers^d_k|={d \choose k} $,
		\item It is assumed that $k$ is known,
		\item $\epsilon \sim N(0,\Gamma)$ where $\Gamma_{ii}=\sigma^2$ for all $i \in \{1,...,d\}$, and
		\item observations $X_i$ with $i=1,...,n$ are i.i.d. 
\end{itemize}

Notice that in contrast to the problem statement in section \ref{prbStmnt}, the features are now correlated. This is because each entry represents a variable or feature and the normal situation is that features are correlated. We now change the sparsity condition from a small, fixed $k$ and large $d$ to $k\approx\frac{d}{\log(d)}$. This reflects the assumption that the number of relevant features grows as the number of total features grows, albeit at a slower rate. 

We are ultimately interested in estimating, by simulation, the generalization capacity of the quadratic loss based, empirical risk function. We could try to use the importance sampling algorithm of Section \ref{ISA} although we haven't tested for the case where $k \approx \frac{d}{\log d}$ and $\Sigma \neq I_d$. In this Chapter we explore an alternate way to approximate the expected value of the log partition function $\mathbb{E}_X [\log Z_{\beta}(X)]$, which, as we have seen, is a component of the generalization capacity. The following derivations and approximations are based on \cite{BuhmNotes}.

In this case we have that

\begin{align}
	X \sim N \bigg(\mu^0,\Gamma \bigg) \Rightarrow \bar{X} \sim N \bigg(\mu^0, \frac{\Gamma}{n} \bigg)
\end{align}	

So we let $\xi = \frac{\bar{X}-\mu^0}{\sfrac{\sigma}{\sqrt{n}}} \sim N(0,\Sigma)$, where $\Sigma = \frac{\Gamma}{\sigma^2}$ so that $\bar{X}= \mu^0 + \frac{\sigma}{\sqrt{n}}\xi$. Recall from \ref{risk2} that we may write the square based empirical risk function as: 

\begin{align}
	&\hat{R}(\mu, \xi) = \mu^T(1-2\frac{\sigma}{\sqrt{n}}\xi-2\mu^0)\\
\end{align}	

Let $\hat{r}(\xi) = 1-2\frac{\sigma}{\sqrt{n}}\xi-2\mu^0$ and $\mathcal{C}^k_h = \{\mu \in \mathcal{C}^k: \mu^T\mu^0=h\}$. Then:

\begin{align}
	\hat{R}(\mu, \xi) &= \mu^T\hat{r}(\xi)\\
	 Z_{\beta}(\xi) &= \sum_{\mu \in \mathcal{C}^k} w_{\beta}(\mu, \xi) = \sum_{\mu \in \mathcal{C}^k} e^{-\beta \hat{R}(\mu, \xi)}\\ 
	 				&= \sum_{\mu \in \mathcal{C}^k} e^{-\beta \mu^T\hat{r}(\xi)} = \sum_{h=max(0,2k-d)}^k \sum_{\mu \in \mathcal{C}^k_h} e^{-\beta \mu^T\hat{r}(\xi)} 
\end{align}	

with which

\begin{align}
	\mathbb{E}_\xi\bigg[\log Z_\beta(\xi) \bigg] &= \mathbb{E}_\xi\bigg[\log  \sum_{h=max(0,2k-d)}^k \sum_{\mu \in \mathcal{C}^k_h} e^{-\beta \mu^T\hat{r}(\xi)}  \bigg]\\
								  &= \mathbb{E}_\xi\bigg[\log \bigg\{ \sum_{\mu \in \mathcal{C}^k_k} e^{-\beta \mu^T\hat{r}(\xi)} +  \sum_{h=max(0,2k-d)}^{k-1} \sum_{\mu \in \mathcal{C}^k_h} e^{-\beta \mu^T\hat{r}(\xi)}\bigg\}\bigg]
\end{align}	

but since $\mathcal{C}^k_k = \{\mu^0\}$ we have that

\begin{align}
	\mathbb{E}_\xi\bigg[\log Z_\beta(\xi)\bigg] &= \mathbb{E}_\xi\bigg[\log \bigg\{  e^{-\beta \mu^{0T}\hat{r}(\xi)} + \sum_{h=max(0,2k-d)}^{k-1} {k \choose h} {d-k \choose k-h} \sum_{\mu \in \mathcal{C}^k_h} \frac{e^{-\beta \mu^T\hat{r}(\xi)}}{{k \choose h} {d-k \choose k-h}} \bigg\} \bigg] \\
											&= \mathbb{E}_\xi\bigg[\log \bigg\{  e^{-\beta \mu^{0T}\hat{r}(\xi)} \bigg( 1 + e^{\beta \mu^{0T}\hat{r}(\xi)} \sum_{h=max(0,2k-d)}^{k-1} {k \choose h} {d-k \choose k-h} \sum_{\mu \in \mathcal{C}^k_h} \frac{e^{-\beta \mu^T\hat{r}(\xi)}}{{k \choose h} {d-k \choose k-h}} \bigg) \bigg\} \bigg] \\
											&= \mathbb{E}_\xi\bigg[-\beta \mu^{0T}\hat{r}(\xi) \bigg]  +  \mathbb{E}_\xi\bigg[\log \bigg\{1 +  \sum_{h=max(0,2k-d)}^{k-1} {k \choose h} {d-k \choose k-h} \sum_{\mu \in \mathcal{C}^k_h} \frac{e^{-\beta (\mu-\mu^0)^T\hat{r}(\xi)}}{{k \choose h} {d-k \choose k-h}} \bigg\} \bigg] 
\end{align}	

Let
\begin{align}
 \bar{B}_h := 	\sum_{\mu \in \mathcal{C}^k_h} \frac{e^{-\beta (\mu-\mu^0)^T\hat{r}(\xi)}}{{k \choose h} {d-k \choose k-h}} 
\end{align}	

Notice that $\bar{B}_k = 1$ so that 

\begin{align}
	\mathbb{E}_\xi\bigg[\log Z_\beta(\xi)\bigg] &= -\beta \mu^{0T}\mathbb{E}_\xi\bigg[\hat{r}(\xi) \bigg]  +  \mathbb{E}_\xi\bigg[\log \sum_{h=max(0,2k-d)}^{k} {k \choose h} {d-k \choose k-h}  \bar{B}_h  \bigg] 
\end{align}	

If we take a hypothesis $\mu$ at random the number of hits $h:=h(\mu)$ is a hypergeometric random variable:

\begin{align}
	H &\sim Hypergeometric(S=k,N=d,n=k)
\end{align}	

Where
\begin{itemize}
	\item $H$ measures the number of succesesses from $n$ draws without replacement,
	\item $S$ is the number of success states and
	\item $n$ is the number of draws.
\end{itemize}	

In this case we pick a $\mu$ at random which represents picking $k$ components to be equal to one (without replacement), from a population of $d$ where there are exactly $k$ components equal to one.  This random variable has the following probability distribution function:

\begin{align}
	\mathbb{P}(H=h)&=\frac{{k \choose h}{d-k \choose k-h}}{{d \choose k}}
\end{align}

Where $h \in \{ \max(0,n+S-N),...,\min(n,S)\}= \{\max(0,2k-d),...,k \}$ which means that 

\begin{align}
	\mathbb{E}_\xi\bigg[\log Z_\beta(\xi)\bigg] &= -\beta \mu^{0T}\mathbb{E}_\xi\bigg[\hat{r}(\xi) \bigg]  +  \mathbb{E}_\xi\bigg[\log {d \choose k} \sum_{h=\max(0,2k-d)}^{k} \frac{{k \choose h} {d-k \choose k-h}}{{d \choose k}}  \bar{B}_h  \bigg] \\
	&= -\beta \mu^{0T}\mathbb{E}_\xi\bigg[\hat{r}(\xi) \bigg]  +  \mathbb{E}_\xi\bigg[\log {d \choose k} \sum_{h=\max(0,2k-d)}^{k}    \mathbb{P}(H=h) \bar{B}_h  \bigg] \\
	&= -\beta \mu^{0T}\mathbb{E}_\xi\bigg[\hat{r}(\xi) \bigg]  +  \mathbb{E}_\xi\bigg[\log {d \choose k} \mathbb{E}_H \bigg[ \bar{B}_H \bigg] \bigg] 
\end{align}

Now $|\mathcal{C}_h^k|$ is exponential in $d$  and $\bar{B}_h$ is a sum over a very large hypothesis space so we will need a way to approximate it. We have that:

\begin{align}
	\mu^T \hat{r}(\xi) &= \mu^T(1-2\mu^0-2\frac{\sigma}{\sqrt{n}}) = \mu^T 1 -2\mu^T\mu^0 - 2\frac{\sigma}{\sqrt{n}}\mu^T\xi \\
					  &= k - 2h - \eta(\mu)
\end{align}	

where $\eta(\mu):=2\frac{\sigma}{\sqrt{n}}\mu^T\xi$. First notice that if $\mu=\mu^0$ then $h=k$, $\mu^{T}\hat{r}(\xi)=-k-\eta(\mu)$ and since $\xi  \sim N(0,\Sigma)$ we have that $\mu^{T}\mathbb{E}_\xi\bigg[\hat{r}(\xi) \bigg]=-k$ so that: 

\begin{align}
	\mathbb{E}_\xi\bigg[\log Z_\beta(\xi)\bigg]	&= \beta k  +  \mathbb{E}_\xi\bigg[\log {d \choose k} \mathbb{E}_H \bigg[ \bar{B}_H \bigg] \bigg]\\ 
		&= \beta k  +  \mathbb{E}_\xi\bigg[\log {d \choose k} \mathbb{E}_H \bigg[ \sum_{\mu \in \mathcal{C}^k_H} \frac{e^{-\beta (\mu-\mu^0)^T\hat{r}(\xi)}}{{k \choose H} {d-k \choose k-H}}  \bigg] \bigg]\\ 
		&= \beta k  +  \mathbb{E}_\xi\bigg[\log {d \choose k} \mathbb{E}_H \bigg[ \sum_{\mu \in \mathcal{C}^k_H} \frac{e^{-\beta \{(k - 2H - \eta(\mu))-(-k-\eta(\mu^0))\}}}{{k \choose H} {d-k \choose k-H}}  \bigg] \bigg]\\ 
		&= \beta k  +  \mathbb{E}_\xi\bigg[\log {d \choose k} e^{-2\beta k}e^{-\beta \eta(\mu^0)} \mathbb{E}_H \bigg[ \frac{e^{2\beta H}}{{k \choose H} {d-k \choose k-H}} \sum_{\mu \in \mathcal{C}^k_H} e^{\beta \eta(\mu)}  \bigg] \bigg]\\ 
		&= \beta k  +  \mathbb{E}_\xi\bigg[\log {d \choose k} -2\beta k -\beta \eta(\mu^0) + \log \mathbb{E}_H \bigg[ \frac{e^{2\beta H}}{{k \choose H} {d-k \choose k-H}} Y_H(\xi)  \bigg] \bigg]\\ 
		&= -\beta k  + \log {d \choose k} + \mathbb{E}_\xi\bigg[ -\beta \eta(\mu^0) \bigg] + \mathbb{E}_\xi\bigg[ \log \mathbb{E}_H \bigg[ \frac{e^{2\beta H}}{{k \choose H} {d-k \choose k-H}} Y_H(\xi)  \bigg] \bigg]
\end{align}	

Where $Y_h :=\sum_{\mu \in \mathcal{C}^k_h} e^{\beta \eta(\mu)}$ . Since 

\begin{align}
	\mathbb{E}_\xi\bigg[ -\beta \eta(\mu^0) \bigg]=\mathbb{E}_\xi\bigg[ -2\beta \frac{\sigma}{\sqrt{n}}\mu^{0T}\xi  \bigg]=-2\beta \frac{\sigma}{\sqrt{n}}\mu^{0T}\mathbb{E}_\xi[\xi]=0
\end{align}	

so that 

\begin{align}
	\mathbb{E}_\xi\bigg[\log Z_\beta(\xi)\bigg]	&= -\beta k  + \log {d \choose k} +  \mathbb{E}_\xi\bigg[ \log \mathbb{E}_H \bigg[ \frac{e^{2\beta H}}{{k \choose H} {d-k \choose k-H}} Y_H(\xi)  \bigg] \bigg]
\end{align}	

The problem with evaluating $Y_h$ is that it is a sum over $\mathcal{C}^k_h$ which is very large if $d$ is large and $k \approx \frac{d}{\log d}$. If we can come up with a good approximation $\widetilde{Y}_h$ of $Y_h$ then the above formula suggests the following algorithm for estimating $\mathbb{E}_\xi\bigg[\log Z_\beta(\xi)\bigg]$, for a given $\beta$, $d$ and $k$:

\begin{enumerate}[1.]
	\item For $i=1$ to $m$
		\begin{enumerate}[a.]
			\item Simulate $\xi^i \sim N(0,\Sigma)$ 
			\item For $j=1$ to $p$
				\begin{itemize}
					\item Simulate $H_j \sim Hypergeometric(k,d,k)$
					\item Calculate $a_j$
						\begin{align}
							a_j := \frac{e^{2\beta H_j}}{{k \choose H_j} {d-k \choose k-H_j}} \widetilde{Y}_{H_j}(\xi^i)
						\end{align}	
				\end{itemize}	
			\item Calculate $\bar{a}^i = \frac{1}{p} \sum_{j=1}^p a_j$
		\end{enumerate}	
	\item Estimate $\mathbb{E}_\xi\bigg[\log Z_\beta(\xi)\bigg]$
		\begin{align}
			\hat{\mathbb{E}}_\xi\bigg[\log Z_\beta(\xi)\bigg]= -\beta k + \log {d \choose k} + \frac{1}{m} \sum_{i=1}^m \log \bar{a}^i
		\end{align}	
\end{enumerate}

In order to explore possible ways to approximate $Y_h$ we take a look at its distribution. 

\begin{align}
	Y_h =\sum_{\mu \in \mathcal{C}^k_h} e^{\beta \eta(\mu)} = \sum_{\mu \in \mathcal{C}^k_h} e^{2\beta \frac{\sigma}{\sqrt{n}}\mu^T\xi}
\end{align}	

Since

\begin{align}
	\xi \sim N(0,\Sigma) &\Rightarrow \mu^T \xi \sim N(0, \mu^T\Sigma \mu) \\
						&\Rightarrow \eta(\mu) = \frac{2\sigma}{\sqrt{n}}\mu^T \xi \sim N\bigg(0,\frac{4 \sigma^2}{n} \mu^T\Sigma \mu \bigg)\\
						&\Rightarrow e^{\beta \eta(\mu)}  \sim \log N\bigg(0,\frac{4 \beta^2 \sigma^2}{n} \mu^T\Sigma \mu \bigg)
\end{align}	

and

\begin{align}
	Cov(\mu_1^T \xi, \mu_2^T \xi) = \mu_1^T \Sigma \mu_2 \Rightarrow Cov(\beta\eta(\mu_1),\beta\eta(\mu_2)) = \frac{4 \beta^2 \sigma^2}{n} \mu_1^T \Sigma \mu_2
\end{align}	

we have that even in the case that $\Sigma=I_d$,  $Y_h(\xi)$ is a sum of $|\mathcal{C}^k_h|$ correlated log-normals with correlation matrix $\Lambda$ such that

\begin{align}
	\Lambda_{ij} = \frac{4 \beta^2 \sigma^2}{n} \mu_i^T \Sigma \mu_j
\end{align}	

If $\Sigma = I_d$ we have that 

\begin{align}
	\beta \eta(\mu) &\sim N\bigg(0,\frac{4 \beta^2 \sigma^2 k}{n}  \bigg)\\
	\Rightarrow  e^{\beta \eta(\mu)} &\sim \log N\bigg(0,\frac{4 \beta^2 \sigma^2 k}{n}  \bigg)
\end{align}	

and

\begin{align}
	Cov(\beta \eta(\mu_i), \beta \eta(\mu_j)) & = \Lambda_{ij} = \frac{4 \beta^2 \sigma^2 c_{ij}}{n} \\
	Cor(\beta \eta(\mu_i), \beta \eta(\mu_j)) & = \frac{\Lambda_{ij}}{\sqrt{\Lambda_{ii}}\sqrt{\Lambda_{jj}}} = \frac{c_{ij}}{k}
\end{align}	

Where $c_{ij}:= \mu_i^T\mu_j$ is the number of overlapping ones in $\mu_i$ and $\mu_j$. If we can use the approximation proposed in \cite{BuhmNotes} where we assume $\eta(\mu) \approx \eta_h$ we have that:

\begin{align}
	\mathbb{E}_\xi\bigg[\log Z_\beta(\xi)\bigg]	&= -\beta k  + \log {d \choose k} +  \mathbb{E}_\xi\bigg[ \log \mathbb{E}_H \bigg[ \frac{e^{2\beta H}}{{k \choose H} {d-k \choose k-H}} Y_H(\xi)  \bigg] \bigg]\\
	&= -\beta k  + \log {d \choose k} +  \mathbb{E}_\xi\bigg[ \log \mathbb{E}_H \bigg[ \frac{e^{2\beta H}}{{k \choose H} {d-k \choose k-H}} \sum_{\mu \in \mathcal{C}^k_H} e^{\beta \eta(\mu)}   \bigg] \bigg]\\
	&\approx -\beta k  + \log {d \choose k} +  \mathbb{E}_\xi\bigg[ \log \mathbb{E}_H \bigg[ \frac{e^{2\beta H}}{{k \choose H} {d-k \choose k-H}} \sum_{\mu \in \mathcal{C}^k_H} e^{\beta \eta_H}   \bigg] \bigg]\\
	&= -\beta k  + \log {d \choose k} +  \mathbb{E}_\xi\bigg[ \log \mathbb{E}_H \bigg[ \frac{e^{2\beta H}}{{k \choose H} {d-k \choose k-H}} |\mathcal{C}^k_H| e^{\beta \eta_H}   \bigg] \bigg]\\
		&= -\beta k  + \log {d \choose k} +  \mathbb{E}_\xi\bigg[ \log \mathbb{E}_H \bigg[ e^{2\beta H}  e^{\beta \eta_H}   \bigg] \bigg]\\
		&= -\beta k  + \log {d \choose k} +  \mathbb{E}_\xi\bigg[ \log \mathbb{E}_H \bigg[ e^{\beta (2H + \eta_H)}   \bigg] \bigg]
\end{align}	

However it is not clear that this approximation will be helpful since it corresponds to assuming that if $\mu_i,\mu_j \in \mathcal{C}^k_{h1}$ and $\mu_p \in \mathcal{C}^k_{h2}$  then $Cor(\beta \eta(\mu_i), \beta \eta(\mu_j))=1$ and $Cor(\beta \eta(\mu_i), \beta \eta(\mu_p))=0$  for all $i,j \in \{1,...,{k \choose h1}{d-k \choose k-h1}\}$ and $p \in \{1,...,{k \choose h2}{d-k \choose k-h2}\}$. 

%We explore the sparse feature selection problem no further. 

\chapter{Summary}
\label{s:Summary}

The work presented in this report falls into the following three categories:

\begin{enumerate}
	\item Exposition of the theory related to approximation set coding and generalization capacity: Sections \ref{PA}-\ref{Shannon}, 
	\item Implementation of the generalization capacity concept to the sparse mean localization: Section \ref{prbStmnt} and Chapters \ref{MeanLoc}-\ref{chp3}, and 
	\item Exploration of how generalization capacity can be implemented for a more realistic version of the sparse feature selection problem: Chapter \ref{ch:TowardsSparseFeatureSelection}.
\end{enumerate}	

In the first case we introduce pattern analysis which is a broad framework with which to deal with learning problems. We describe the approximation set approach to learning where we look for a set of hypotheses with \emph{similar} performance instead of looking for just one. With the definitions from pattern analysis and approximation sets we then introduce information theoretic concepts, originally developed within the field of statistical physics, which we need to define GC. To motivate the meaning of GC we study Shannon's noisy channel coding theorem and related theory drawing an extensive analogy between channel capacity and generalization capacity. We describe in detail the differences and similarities between the communication scenario from which Shannon's concept of channel capacity arises and the learning scenario from which \citeauthor{Buhm13}'s concept of generalization capacity can be derived.  

% We consider this exposition a useful contribution since the channel capacity is crucial in understanding generalization capacity but the relationship between the two, although obviously known to those who have developed or worked with GC, has not been presented in such detail in the available literature to the best of our knowledge. 

The second type of contribution presented here involves the implementation of generalization capacity to the problem of sparse mean localization. In \cite{Buhm14} an expression for GC for the sparse mean localization problem (with respect to the squared loss based empirical risk function) is derived that suggests a simulating algorithm for its estimation.  The simulating algorithm is briefly described and results are presented. Based on this work we designed and detailed successive simulating algorithms for GC estimation for non-sparse and sparse mean localization problems. We first designed an \emph{exhaustive} sampling algorithm for non-sparse mean localization in low dimensional cases where the dimension $d$ of the mean vector $\mu^0$ is in the order of 10. We implemented various versions of \emph{common random numbers} to reduce variance and chose the best one. We then applied it to the sparse mean localization problem and tried to scale it up to deal with high dimensional cases, but found that since the size of the hypothesis class is exponential in the dimension $d$ the exhaustive algorithm is unfeasible since it involves summing over the entire hypothesis space. To deal with this we designed a uniform sampling algorithm, also suggested in \cite{Buhm14}, but found that for this algorithm to converge the number of sampled hypotheses necessary was actually larger than the hypothesis class size. We solved this difficulty by designing an importance sampling alogorithm which seems to be adequate for high dimensions in the order of 10,000 (asymptotic confidence intrevals need to be derived to verify this). In \cite{Buhm14} results showing the estimation of GC for high dimensional sparse mean localization are shown however the simulation method is not detailed. The contribution of this work is to detail and justify numerical estimation of GC for the sparse mean localization problem. It is also worth mentioning that the results of the estimatation of GC are qualitatively different to those found in \cite{Buhm14}:compare for example figures 1 and 2 in \cite{Buhm14} to figure \ref{fig:GChighD}. 

Finally, based on \cite{BuhmNotes},  we explored estimating GC in a more realistic version of the sparse feature selection problem and sketched a general simulation algorithm, describing certain difficulties that need to be resolved.

\section{Future work}
\label{ss:FutureWork}

Possible ways to extend the work presented here are: 

\begin{itemize}
	\item Derive asymptotic confidence intervals for the different GC estimators (exhaustive, sampling, importance sampling) so as to verify when these are fullfilled and so determine the number of simulations necessary. This is especially important for the sparse high dimensional cases ($d$ large) so as to establish the reliability of the estimates produced and the feasability of the algorithms.
	\item Explore the sparse feature selection problem further: does the $\eta(\mu) \approx \eta_h$ approximation work? What other approximations can we use to make the estimation of $Y_h(\xi)$ feasible for large $d$ and $k \approx \frac{d}{\log d}$.
	
	\item Implement the importance sampling algorithm to the more general sparse feature selection problem and compare with simulation methods based on approximations of the type discussed in Chapter \ref{ch:TowardsSparseFeatureSelection}.
	
	\item Extend the study to include estimation of sparse histograms. In this case a histogram can be represented as a vector $X_i$ composed of a mean $\mu^0$ and noise $\epsilon$: i.e. $X_i = \mu^0 + \epsilon_i$ as in the sparse mean localization problem. However in this case $\mu^0 \in [0,1]^d$ instead of $\mu^0 \in \{0,1\}^d$ and $|\mu^0|_1=1$ instead of $|\mu^0|_1=k$. The number of non-zero entries is still assumed to be $k$ to enforce a sparsity condition. 
\end{itemize}	

%%% Local Variables: 
%%% mode: latex
%%% TeX-master: "MasterThesisSfS"
%%% End: 

%%%%%%%%%%%%%%%%%%%%%%%%%%%%%%%%%%%%%%%%%%%%%%%%%
%%% Bibliography                              %%%
%%%%%%%%%%%%%%%%%%%%%%%%%%%%%%%%%%%%%%%%%%%%%%%%%
\addtocontents{toc}{\vspace{.5\baselineskip}}
\cleardoublepage
\phantomsection
\addcontentsline{toc}{chapter}{\protect\numberline{}{Bibliography}}
\bibliography{myReferences}
%% All books from our library (SfS) are already in a BiBTeX file
%% (Assbib). You can use Assbib combined with your personal BiBTeX file:
%% \bibliography{Myreferences,Assbib}. Of course, this will only work on
%% the computers at SfS, unless you copy the Assbib file 
%%  --> /u/sfs/bib/Assbib.bib

%%%%%%%%%%%%%%%%%%%%%%%%%%%%%%%%%%%%%%%%%%%%%%%%% 
%%% Appendices (if needed)                    %%%
%%%%%%%%%%%%%%%%%%%%%%%%%%%%%%%%%%%%%%%%%%%%%%%%%

\begin{appendices}
\chapter{\Rp Code}
\label{app:rcode}

We include the code in \Rp that was used to implement the importance sampling algorithm. Other algorithms described to estimate generalization capacity are very similar so we omit them. 

\section{Functions}
\begin{lstlisting}
#############################################################################
# Aproximation Set Coding
# Sparse Mean localization
# Importance Sampling Algorithm
#############################################################################

#############################################################################
# Functions
#############################################################################

#Generate any of the 1,...,choose(d,k) hypothesis
get.hyp.i <- function(i,d,k){
	
	one.pos <- as.numeric()
	m <- i-1 

	for(l in k:1){
		candidates <- d:(l-1)
		try.ck <- choose(candidates,l)
		indx.ck <- which(try.ck <= m)[1]	
		ck <- candidates[indx.ck]
		one.pos <- c(one.pos,ck)
		m <- m - choose(ck,l)
		
	}
	res <- rep(0,d)
	res[one.pos+1] <-1
	return(res)
}

#Generate unsrestricted hypothesis space
get.hyp.sp <- function(d){
	hyp.size <- 2^d
	as.numeric(strsplit(substr(paste(as.integer(intToBits(2^d-1)), collapse=""),1,d),"")[[1]])
	C <- sapply(0:(2^d-1), function(i) as.numeric(strsplit(substr(paste(as.integer(intToBits(i)), collapse=""),1,d),"")[[1]]))
	dim(C)
	dimnames(C) <- list(component=seq(d), hypothesis=seq(hyp.size))
	return(C)
}

#Generate a hypothesis with a certain number of errors out of the k 1-components
get.hyp.hit <- function(mu0, num.hit){
	d <- length(mu0)
	k <- sum(mu0)
	num.miss <- k-num.hit
	indx.1 <- which(mu0==1)
	indx.0 <- which(mu0==0)
	
	
	mu <- apply(num.miss, c("hypothesis","repetition"), function(miss){
			indx.one <- as.numeric(sample(as.character(indx.1), size=miss, replace=F))
			indx.zero <- as.numeric(sample(as.character(indx.0), size=miss, replace=F))
			res <- mu0
			res[indx.one] <- 0
			res[indx.zero] <- 1
			return(res)
	})
	names(dimnames(mu))[1] <- "component"
	dimnames(mu)[[1]] <- 1:d
	
	return(mu)
}

#Generate restricted hypothesis spaces
get.res.hyp.sp <- function(d, k){
	
	Cs <- lapply(k, function(ak) sapply(1:choose(d,ak), function(i) get.hyp.i(i, d, ak)))
	names(Cs) <- k
	return(Cs)
}

#Generate restricted hypothesis spaces
get.res.hyp.sp2 <- function(C, k){
	Cs <- lapply(k, function(num1) C[,which(apply(C,2,sum)==num1)])
	names(Cs) <- k
	return(Cs)
}

#Simlate Data
sim.data <- function(n, noise.level, dim.data, dimnames.data){
	d <- dim.data["component"]
	num.noise <- dim.data["noise"]
	reps <- dim.data["repetition"]
	num.spars <- dim.data["sparsity"]
	
	if("Zs" %in% names(dim.data)){
		num.Zs <- dim.data["Zs"]
	} else{
		num.Zs <- 1
	}

	#simulate normal data
	data.norm <- rnorm(num.spars*d*reps*num.Zs*num.noise,0,rep(noise.level/sqrt(n), rep(num.spars*d*reps*num.Zs, num.noise)))
	data.arr <- array(data.norm,dim=dim.data, dimnames=dimnames.data)
	return(data.arr)
}

#Create cost array
cost.mean.n2.old <- function(mu, mu0, X){	

	a <- 1 
	b <- 2 
	res1 <- apply(X, c("repetition","Zs","noise.level"), function(vec) a*t(mu)%*%(1-b*vec-2*mu0))
	a <- 2
	b <- 1
	Y <- (X[,,1,]+X[,,2,])
	res2 <- apply(Y, c("repetition","noise.level"), function(vec) a*t(mu)%*%(1-b*vec-2*mu0))
	dimnames <- dimnames(res1)
	dimnames$Zs <- c(dimnames$Zs, "nZ12")
	res <- abind(res1[,1,],res1[,2,],res2, along=3)
	res <- aperm(res, c(1,3,2))
	dimnames(res) <- dimnames
	return(res)

}

cost.mean.n2 <- function(mu, mu0, X){	

	res <- apply(X, c("repetition","Zs","noise.level"), function(vec) t(mu- (vec+mu0))%*%(mu- (vec+mu0)))
    dimnames <- dimnames(res)
    dimnames$Zs <- c(dimnames$Zs, "nZ12")	
	res <- abind(res[,1,],res[,2,],res[,1,]+res[,2,], along=3)
	res <- aperm(res, c(1,3,2))
	dimnames(res) <- dimnames
	return(res)
}

cost.mean.n1 <- function(mu, mu0, X){	

	d <- length(mu0)
	res <- apply(X, c("repetition","Zs","noise.level"), function(vec) t(abs(mu- (vec+mu0)))%*%rep(1, d))
    dimnames <- dimnames(res)
    dimnames$Zs <- c(dimnames$Zs, "nZ12")	
	res <- abind(res[,1,],res[,2,],res[,1,]+res[,2,], along=3)
	res <- aperm(res, c(1,3,2))
	dimnames(res) <- dimnames
	return(res)
}


#Create cost array sample case
# squared loss
cost.mean.smpl.n2 <- function(C, mu0, X){	


	rep.Z <- length(dimnames(X)$repetition)
	num.noise <- length(dimnames(X)$noise.level)
	
	res <- sapply(1:rep.Z, function(i) apply(X[,i,,], c("Zs","noise.level"), function(vec) t(C[,i]- (vec+mu0))%*%(C[,i]- (vec+mu0))), simplify="array")
	res <- aperm(res, c(3,1,2))
	names(dimnames(res))[1] <- "repetition"
	dimnames(res)[[1]] <- 1:rep.Z


    dimnames <- dimnames(res)
    dimnames$Zs <- c(dimnames$Zs, "nZ12")	
	res <- abind(res[,1,],res[,2,],res[,1,]+res[,2,], along=3)
	res <- aperm(res, c(1,3,2))
	dimnames(res) <- dimnames
	return(res)

}
# absolute loss
cost.mean.smpl.n1 <- function(C, mu0, X){	


	rep.Z <- length(dimnames(X)$repetition)
	num.noise <- length(dimnames(X)$noise.level)
	d <- length(dimnames(X)$component)
	
	res <- sapply(1:rep.Z, function(i) apply(X[,i,,], c("Zs","noise.level"), function(vec) t(abs(C[,i]- (vec+mu0)))%*%rep(1,d)), simplify="array")
	res <- aperm(res, c(3,1,2))
	names(dimnames(res))[1] <- "repetition"
	dimnames(res)[[1]] <- 1:rep.Z


    dimnames <- dimnames(res)
    dimnames$Zs <- c(dimnames$Zs, "nZ12")	
	res <- abind(res[,1,],res[,2,],res[,1,]+res[,2,], along=3)
	res <- aperm(res, c(1,3,2))
	dimnames(res) <- dimnames
	return(res)

}


#Create cost array importance sampling case
# squared loss
cost.mean.imp.smpl.n2 <- function(C, mu0, X){	


	rep.Z <- length(dimnames(X)$repetition)
	num.noise <- length(dimnames(X)$noise.level)
	rep.mu <- length(dimnames(C)$hypothesis)
	num.Zs <- length(dimnames(X)$Zs)
	

	res <- sapply(1:rep.Z, function(i) sapply(1:num.Zs, function(z) 
															sapply(1:num.noise, function(n) 
																					sapply(1:rep.mu, function(hyp){
																						t(C[,hyp,i]- (X[,i,z,n]+mu0))%*%(C[,hyp,i]- (X[,i,z,n]+mu0))
	 }, simplify="array"), simplify="array"), simplify="array"),simplify="array")
	
	
	
	dimnames(res) <- list(hypothesis=1:rep.mu, noise.level=seq(num.noise), Zs=c("Z1","Z2"), repetition=seq(rep.Z))
    dimnames <- dimnames(res)
    dimnames$Zs <- c(dimnames$Zs, "nZ12")	
	res <- abind(res[,,1,],res[,,2,],res[,,1,]+res[,,2,], along=4)
	
	
	res <- aperm(res, c(1,2,4,3))
	dimnames(res) <- dimnames
	return(res)

}
# absolute loss
cost.mean.imp.smpl.n1 <- function(C, mu0, X){	


	rep.Z <- length(dimnames(X)$repetition)
	num.noise <- length(dimnames(X)$noise.level)
	rep.mu <- length(dimnames(C)$hypothesis)
	num.Zs <- length(dimnames(X)$Zs)
	d <- length(dimnames(X)$component)
	

	res <- sapply(1:rep.Z, function(i) sapply(1:num.Zs, function(z) 
															sapply(1:num.noise, function(n) 
																					sapply(1:rep.mu, function(hyp){
																						t(abs(C[,hyp,i]- (X[,i,z,n]+mu0)))%*%rep(1,d)
	 }, simplify="array"), simplify="array"), simplify="array"),simplify="array")
	
	
	
	dimnames(res) <- list(hypothesis=1:rep.mu, noise.level=seq(num.noise), Zs=c("Z1","Z2"), repetition=seq(rep.Z))
    dimnames <- dimnames(res)
    dimnames$Zs <- c(dimnames$Zs, "nZ12")	
	res <- abind(res[,,1,],res[,,2,],res[,,1,]+res[,,2,], along=4)
	
	
	res <- aperm(res, c(1,2,4,3))
	dimnames(res) <- dimnames
	return(res)

}


#Calculate boltzman weight, probabiity and partition arrays
# without log-sum-exp trick
get.arrays <- function(Cs, data.arr, mu0, beta, cost.fun){

	
	d <- length(dimnames(data.arr)$component)
	num.betas <- length(beta)
	k <- as.numeric(dimnames(data.arr)$sparsity)
	num.spars <- length(k)
	hyp.size <- choose(d,k)
	max.hyp.size <- 2^d
	
	

	Rs <- lapply(1:num.spars, function(sp){
			R <- sapply(as.data.frame(Cs[[sp]]),function(hyp) cost.fun(hyp,mu0[,sp],data.arr[sp,,,,]), simplify="array")
			R <- aperm(R,c(4,1,2,3))
			names(dimnames(R))[[1]] <- "hypothesis"
			return(R)
	})	
	

	#Calculate boltzman weights
	ws <- lapply(Rs, function(el) {
		w <- outer(el, beta, FUN=function(x,y) exp(-y*x)) 
		names(dimnames(w))[length(names(dimnames(w)))] <- "beta"
		dimnames(w)[[length(names(dimnames(w)))]] <- seq(num.betas)
		return(w)
	})
	
	names(ws) <- 1:num.spars
	
	#Calculate partition functions
	Z <- sapply(ws, function(el) apply(el, c("repetition","Zs","noise.level","beta"),sum), simplify="array")
	Z <- aperm(Z, c(5,1,2,3,4))
	names(dimnames(Z))[1] <- "sparsity"
	dimnames(Z)[[1]] <- k
	

	#Calculate gibbs probability distribution over hypothesis space
	#prs <- lapply(1:num.spars, function(i){
	#
	#	pr <- sapply(1:hyp.size[i], function(j) ws[[i]][j,,1:2,,]/Z[i,,1:2,,], simplify="array")
	#	pr <- aperm(pr, c(5,1,2,3,4))
	#	dimnames <- dimnames(ws[[i]])
	#	dimnames$Zs <- dimnames$Zs[1:2]
	#	dimnames(pr) <- dimnames
	#	return(pr)
	#})
	
	#names(prs) <- seq(num.spars)
	
	
	res <- list(Rs=Rs, ws=ws, Z=Z) #, prs=prs
}
# with log-sum-exp trick
get.uf.arrays <- function(Cs, data.arr, mu0, beta, cost.fun){

	#cost.fun <- cost.mean.n2
	d <- length(dimnames(data.arr)$component)
	num.betas <- length(beta)
	k <- as.numeric(dimnames(data.arr)$sparsity)
	num.spars <- length(k)
	hyp.size <- choose(d,k)
	max.hyp.size <- 2^d
	
	

	Rs <- lapply(1:num.spars, function(sp){
			R <- sapply(as.data.frame(Cs[[sp]]),function(hyp) cost.fun(hyp,mu0[,sp],data.arr[sp,,,,]), simplify="array")
			R <- aperm(R,c(4,1,2,3))
			names(dimnames(R))[[1]] <- "hypothesis"
			return(R)
	})	
	

	
	#Calculate -beta*R so that we may apply log-sum-exp trick to deal with underflow

	#Calculate boltzman weights
	as <- lapply(Rs, function(el) {
		a <- outer(el, beta, FUN=function(x,y) -y*x) 
		names(dimnames(a))[length(names(dimnames(a)))] <- "beta"
		dimnames(a)[[length(names(dimnames(a)))]] <- seq(num.betas)
		return(a)
	})
	
	names(as) <- 1:num.spars
	
	
	#Calculate boltzman weights
	logZ <- sapply(as, function(el){ 
		res <- apply(el, c("noise.level","Zs","repetition","beta"), function(vec){
			b <- max(vec)
			return(log(sum(exp(vec-b)))+b)
		 	})
		return(res)
	}, simplify="array")
	
	logZ <- aperm(logZ, c(5,1,2,3,4))
	names(dimnames(logZ))[1] <- "sparsity"
	dimnames(logZ)[[1]] <- k
	
	res <- list(Rs=Rs, logZ=logZ) #, prs=prs
}


#Calculate boltzman weight, probabiity and partition arrays - sample version
# without log-sum-exp trick
get.smpl.arrays <- function(reps.mu, data.arr, mu0, beta, cost.fun){

	
	d <- length(dimnames(data.arr)$component)
	reps.Z <- length(dimnames(data.arr)$repetition)
	num.betas <- length(beta)
	k <- as.numeric(dimnames(data.arr)$sparsity)
	num.spars <- length(k)
	hyp.size <- choose(d,k)
	max.hyp.size <- 2^d
	
	# generate a list with the possible hypothesis numbers that can be sampled
	hyp.no.list <- sapply(hyp.size, function(size) 1:size)
	
	
	smpls <- sapply(1:num.spars, function(sp) sapply(1:reps.Z, function(i) sample(hyp.no.list[[sp]],size=reps.mu, replace=T) ), simplify="array")
	dimnames(smpls) <- list(hypothesis=1:reps.mu, repetition=1:reps.Z, sparsity=1:num.spars)
	
	
	R <- sapply(1:num.spars, function(sp){
			R.aux <- sapply(1:reps.mu, function(j){
				C <- sapply(smpls[j,,sp], function(i) get.hyp.i(i, d, k[sp])) 
				dimnames(C) <- list(component=1:d, repetition=1:reps.Z)
				return(cost.fun(C,mu0[,sp],data.arr[sp,,,,]))
			}, simplify="array")
			R.aux <- aperm(R.aux,c(4,1,2,3))
			names(dimnames(R.aux))[1] <- "hypothesis"
			dimnames(R.aux)[[1]] <- 1:reps.mu
			return(R.aux)
	}, simplify="array")
	
	R <- aperm(R, c(5,1,2,3,4))
	names(dimnames(R))[1] <- "sparsity"
	dimnames(R)[[1]] <- k

	
	
	#Calculate boltzman weights
	w <- outer(R, beta, FUN=function(x,y) exp(-y*x)) 
	names(dimnames(w))[length(names(dimnames(w)))] <- "beta"
	dimnames(w)[[length(names(dimnames(w)))]] <- seq(num.betas)

	#Calculate partition functions
	Z <- apply(w, c("sparsity","repetition","Zs","noise.level","beta"),mean)#*hyp.size

	#Calculate gibbs probability distribution over hypothesis space
	#pr <- sapply(1:reps.mu, function(i) w[,i,,1:2,,]/(Z[,,1:2,,]*reps.Z), simplify="array")
	#pr <- aperm(pr, c(1,6,2,3,4,5))
	#dimnames <- dimnames(w)
	#dimnames$Zs <- dimnames$Zs[1:2]
	#dimnames(pr) <- dimnames

	res <- list(R=R, w=w, Z=Z) #, pr=pr
}
# with log-sum-exp trick
get.smpl.uf.arrays <- function(reps.mu, data.arr, mu0, beta, cost.fun){

	#cost.fun <- cost.mean.smpl.n2
	d <- length(dimnames(data.arr)$component)
	reps.Z <- length(dimnames(data.arr)$repetition)
	num.betas <- length(beta)
	k <- as.numeric(dimnames(data.arr)$sparsity)
	num.spars <- length(k)
	hyp.size <- choose(d,k)
	max.hyp.size <- 2^d
	
	# generate a list with the possible hypothesis numbers that can be sampled
	hyp.no.list <- sapply(hyp.size, function(size) 1:size)
	
	
	smpls <- sapply(1:num.spars, function(sp) sapply(1:reps.Z, function(i) sample(hyp.no.list[[sp]],size=reps.mu, replace=T) ), simplify="array")
	dimnames(smpls) <- list(hypothesis=1:reps.mu, repetition=1:reps.Z, sparsity=1:num.spars)
	
	
	R <- sapply(1:num.spars, function(sp){
			R.aux <- sapply(1:reps.mu, function(j){
				C <- sapply(smpls[j,,sp], function(i) get.hyp.i(i, d, k[sp])) 
				dimnames(C) <- list(component=1:d, repetition=1:reps.Z)
				return(cost.fun(C,mu0[,sp],data.arr[sp,,,,]))
			}, simplify="array")
			R.aux <- aperm(R.aux,c(4,1,2,3))
			names(dimnames(R.aux))[1] <- "hypothesis"
			dimnames(R.aux)[[1]] <- 1:reps.mu
			return(R.aux)
	}, simplify="array")
	
	R <- aperm(R, c(5,1,2,3,4))
	names(dimnames(R))[1] <- "sparsity"
	dimnames(R)[[1]] <- k

	#Calculate boltzman weights
	a <- outer(R, beta, FUN=function(x,y) -y*x) 
	names(dimnames(a))[length(names(dimnames(a)))] <- "beta"
	dimnames(a)[[length(names(dimnames(a)))]] <- seq(num.betas)
	
	
	#Calculate boltzman weights
	logZ <- apply(a, c("sparsity","noise.level","Zs","repetition","beta"), function(vec){
			b <- max(vec)
			res <- log(sum(exp(vec-b)))+b
		 	return(res)
		})
		
	logZ <- logZ - log(reps.mu)

	res <- list(R=R, logZ=logZ) 
}


#Calculate boltzman weight, probabiity and partition arrays - importance sampling version
# without log-sum-exp trick
get.imp.smpl.arrays <- function(reps.mu, data.arr, mu0, beta, cost.fun){

	#cost.fun <- cost.mean.imp.smpl.n2
	d <- length(dimnames(data.arr)$component)
	reps.Z <- length(dimnames(data.arr)$repetition)
	num.betas <- length(beta)
	k <- as.numeric(dimnames(data.arr)$sparsity)
	num.spars <- length(k)
	hyp.size <- choose(d,k)
	max.hyp.size <- 2^d
	
	# generate a list with the possible hypothesis numbers that can be sampled
	#hyp.no.list <- sapply(hyp.size, function(size) 1:size)
	
	names(beta) <- 1:length(beta)
	
	weights <- lapply(k, function(i){
		hits.rng <- max(0,2*i-d):i
		res <- rep(1/length(hits.rng), length(hits.rng))
		names(res) <- hits.rng
		return(res)
	})
	
	num.type <- lapply(k, function(i){
		hits.rng <- max(0,2*i-d):i
		res <- choose(i,hits.rng)*choose(d-i, i-(hits.rng))
		names(res) <- hits.rng
		return(res)
	})
	
	
	#lapply(weights, function(el) apply(el,"beta", sum))
	
	smpls <- sapply(1:num.spars, function(sp)  sapply(1:reps.Z, function(i) 
			as.numeric(sample(names(weights[[sp]]),size=reps.mu, replace=T, prob=weights[[sp]])), simplify="array"), simplify="array")
	
	dimnames(smpls) <- list(hypothesis=1:reps.mu, repetition=1:reps.Z, sparsity=k)
	
	#get the weights of each sampled hypothesis for when we calculate the weighted mean Z
	
	reweights <- sapply(1:reps.mu, function(hyp) 
										sapply(1:reps.Z, function(i) sapply(1:num.spars,
																		function(sp){
																				indx <- match(smpls[hyp,i,sp],names(weights[[sp]]))
																				(1/reps.mu)*(1/hyp.size[sp])*(num.type[[sp]][indx]/weights[[sp]][indx])
																		 }, simplify="array"), simplify="array"), simplify="array")
	
	dimnames(reweights) <- list(sparsity=k, repetition=1:reps.Z, hypothesis=1:reps.mu)
	
	reweights <- aperm(reweights, c(1,3,2))
	
	C <- sapply(1:num.spars, function(sp){
		 #print(sp)
		 get.hyp.hit(mu0=mu0[,sp], num.hit=smpls[,,sp])
	 }, simplify="array")
	C <- aperm(C, c(4,1,2,3))
	names(dimnames(C))[1] <- "sparsity"
	dimnames(C)[[1]] <- k
	
	
	#apply(C,c("sparsity","hypothesis","repetition"), function(vec) sum(vec))
	#apply(C[2,,,],c("hypothesis","repetition"), function(vec) t(vec)%*%mu0[,2])==smpls[,,2]
	
	
	
	R <- sapply(1:num.spars, function(sp) cost.fun(C=C[sp,,,], mu0=mu0[,sp], X=data.arr[sp,,,,]), simplify="array")


	
	R <- aperm(R, c(5,1,2,3,4))
	names(dimnames(R))[1] <- "sparsity"
	dimnames(R)[[1]] <- k

	#apply(R, c("sparsity","noise.level"), summary)
	
	#Calculate boltzman weights
	w <- outer(R, beta, FUN=function(x,y) exp(-y*x)) 
	names(dimnames(w))[length(names(dimnames(w)))] <- "beta"
	dimnames(w)[[length(names(dimnames(w)))]] <- seq(num.betas)
	
	
	
	#Calculate partition functions
	Z <- sapply(1:num.noise, function(noi) 
								sapply(1:3, function(z) 
												sapply(1:num.betas, function(bet)
																		w[,,noi,z,,bet]*reweights
																			,simplify="array"), simplify="array"), simplify="array")
		 
	names(dimnames(Z))[(length(names(dimnames(Z)))-2):length(names(dimnames(Z)))] <- c("beta","Zs","noise.level")
	dimnames(Z)[[(length(names(dimnames(Z)))-2)]] <- seq(num.betas)
	dimnames(Z)[[(length(names(dimnames(Z)))-1)]] <- c("Z1","Z2","nZ12")
	dimnames(Z)[[(length(names(dimnames(Z))))]] <- seq(num.noise)
	
	Z <- apply(Z, c("sparsity","repetition","Zs","noise.level","beta"),sum)#*hyp.size
	Z <- aperm(Z, c(1,4,3,2,5))

	#Calculate gibbs probability distribution over hypothesis space
	pr <- sapply(1:reps.mu, function(i) w[,i,,1:2,,]/(Z[,,1:2,,]*reps.Z), simplify="array")
	pr <- aperm(pr, c(1,6,2,3,4,5))
	dimnames <- dimnames(w)
	dimnames$Zs <- dimnames$Zs[1:2]
	dimnames(pr) <- dimnames

	res <- list(R=R, w=w, pr=pr, Z=Z)
}
# with log-sum-exp trick
get.imp.smpl.uf.arrays <- function(reps.mu, data.arr, mu0, beta, cost.fun){

	#cost.fun <- cost.mean.imp.smpl.n2
	d <- length(dimnames(data.arr)$component)
	reps.Z <- length(dimnames(data.arr)$repetition)
	num.betas <- length(beta)
	k <- as.numeric(dimnames(data.arr)$sparsity)
	num.spars <- length(k)
	hyp.size <- choose(d,k)
	max.hyp.size <- 2^d
	
	# generate a list with the possible hypothesis numbers that can be sampled
	#hyp.no.list <- sapply(hyp.size, function(size) 1:size)
	
	names(beta) <- 1:length(beta)
	
	weights <- lapply(k, function(i){
		hits.rng <- max(0,2*i-d):i
		res <- rep(1/length(hits.rng), length(hits.rng))
		names(res) <- hits.rng
		return(res)
	})
	
	num.type <- lapply(k, function(i){
		hits.rng <- max(0,2*i-d):i
		res <- choose(i,hits.rng)*choose(d-i, i-(hits.rng))
		names(res) <- hits.rng
		return(res)
	})
	
	
	#lapply(weights, function(el) apply(el,"beta", sum))
	
	smpls <- sapply(1:num.spars, function(sp)  sapply(1:reps.Z, function(i) 
			as.numeric(sample(names(weights[[sp]]),size=reps.mu, replace=T, prob=weights[[sp]])), simplify="array"), simplify="array")
	
	dimnames(smpls) <- list(hypothesis=1:reps.mu, repetition=1:reps.Z, sparsity=k)
	
	#get the weights of each sampled hypothesis for when we calculate the weighted mean Z
	
	reweights <- sapply(1:reps.mu, function(hyp) 
										sapply(1:reps.Z, function(i) sapply(1:num.spars,
																		function(sp){
																				indx <- match(smpls[hyp,i,sp],names(weights[[sp]]))
																				return((1/reps.mu)*(1/hyp.size[sp])*(num.type[[sp]][indx]/weights[[sp]][indx]))
																		 }, simplify="array"), simplify="array"), simplify="array")
	
	dimnames(reweights) <- list(sparsity=k, repetition=1:reps.Z, hypothesis=1:reps.mu)
	
	reweights <- aperm(reweights, c(1,3,2))
	
	#lets check that the reweights add up to one, or 1/reps.mu in this case since we already added that 
#	apply(reweights, c("sparsity", "repetition"), sum); 1/reps.mu
	
	
	C <- sapply(1:num.spars, function(sp){
		 #print(sp)
		 get.hyp.hit(mu0=mu0[,sp], num.hit=smpls[,,sp])
	 }, simplify="array")
	C <- aperm(C, c(4,1,2,3))
	names(dimnames(C))[1] <- "sparsity"
	dimnames(C)[[1]] <- k
	
	
	#apply(C,c("sparsity","hypothesis","repetition"), function(vec) sum(vec))
	#apply(C[2,,,],c("hypothesis","repetition"), function(vec) t(vec)%*%mu0[,2])==smpls[,,2]
	
	
	
	R <- sapply(1:num.spars, function(sp) cost.fun(C=C[sp,,,], mu0=mu0[,sp], X=data.arr[sp,,,,]), simplify="array")
	
	R <- aperm(R, c(5,1,2,3,4))
	names(dimnames(R))[1] <- "sparsity"
	dimnames(R)[[1]] <- k
	#apply(R, c("sparsity","noise.level"), summary)
	
	
	
	#Calculate -beta*R so that we may apply log-sum-exp trick to deal with underflow
	
	a <- outer(R, beta, FUN=function(x,y) -y*x)
	names(dimnames(a))[length(names(dimnames(a)))] <- "beta"
	dimnames(a)[[length(names(dimnames(a)))]] <- seq(num.betas)
	b <- apply(a, c("sparsity","noise.level","Zs","repetition","beta"),max)
	
	
	#Calculate boltzman weights
	w <- apply(a, c("sparsity","noise.level","Zs","repetition","beta"), function(vec) exp(vec-max(vec)))
	w <- aperm(w, c(2,1,3,4,5,6))
	
	#Calculate partition functions
	Z <- sapply(1:num.noise, function(noi) 
								sapply(1:3, function(z) 
												sapply(1:num.betas, function(bet)
																		w[,,noi,z,,bet]*reweights
																			,simplify="array"), simplify="array"), simplify="array")
		 
	names(dimnames(Z))[(length(names(dimnames(Z)))-2):length(names(dimnames(Z)))] <- c("beta","Zs","noise.level")
	dimnames(Z)[[(length(names(dimnames(Z)))-2)]] <- seq(num.betas)
	dimnames(Z)[[(length(names(dimnames(Z)))-1)]] <- c("Z1","Z2","nZ12")
	dimnames(Z)[[(length(names(dimnames(Z))))]] <- seq(num.noise)
	
	Z <- apply(Z, c("sparsity","repetition","Zs","noise.level","beta"),sum)#*hyp.size
	Z <- aperm(Z, c(1,4,3,2,5))
	
	logZ <- log(Z) + b


	res <- list(R=R, w=w, Z=Z, logZ=logZ) 
}

#calculate information content, average info. content and generalization capacity
# exhaustive algo.
get.info <- function(hyp.size, logZ){

	cardinality <-  array(hyp.size, dim=dim(logZ[,,1,,]), dimnames=dimnames(logZ[,,1,,])) 
	
	num.noise <- length(dimnames(logZ)$noise.level)
	#Calculate Information Content
	info.content <- log(cardinality) + logZ[,,3,,] - logZ[,,1,,] -logZ[,,2,,] 

	#Calculate average information content
	avg.info.content <- apply(info.content, c("sparsity","noise.level","beta"), mean)

	#Calculate standard deviation information content
	sd.info.content <- apply(info.content, c("sparsity","noise.level","beta"), sd)

	#Obtain generalization capacity and corresponding beta
	indx.beta.opt <- apply(avg.info.content,c("sparsity","noise.level"),which.max)
	gen.capacity <- apply(avg.info.content,c("sparsity","noise.level"),max)
	
	indx.arr <- melt(indx.beta.opt)
	names(indx.arr)[3] <- "beta"
	indx.arr$sparsity <- match(indx.arr$sparsity, k)
	indx.arr$sd <- sd.info.content[as.matrix(indx.arr)]
	sd.gen.capacity <- cast(indx.arr, sparsity~noise.level, value="sd")
	sd.gen.capacity <- as.matrix(sd.gen.capacity[,2:dim(sd.gen.capacity)[2]])
	dimnames(sd.gen.capacity) <- dimnames(gen.capacity)
	
	return(list(info.content=info.content, avg.info.content=avg.info.content, sd.info.content=sd.info.content, indx.beta.opt=indx.beta.opt ,gen.capacity=gen.capacity, sd.gen.capacity=sd.gen.capacity))
}

#calculate information content, average info. content and generalization capacity
# sampling algo.
get.smpl.info <- function(hyp.size, logZ){

	k <- as.numeric(dimnames(logZ)$sparsity)
	num.noise <- length(dimnames(logZ)$noise.level)
	#Calculate Information Content
	info.content <-   logZ[,,3,,] - logZ[,,1,,] -logZ[,,2,,] 

	#Calculate average information content
	avg.info.content <- apply(info.content, c("sparsity","noise.level","beta"), mean)

	#Calculate standard deviation information content
	sd.info.content <- apply(info.content, c("sparsity","noise.level","beta"), sd)

	#Obtain generalization capacity and corresponding beta
	indx.beta.opt <- apply(avg.info.content,c("sparsity","noise.level"),which.max)
	gen.capacity <- apply(avg.info.content,c("sparsity","noise.level"),max)
	
	indx.arr <- melt(indx.beta.opt)
	names(indx.arr)[3] <- "beta"
	indx.arr$sparsity <- match(indx.arr$sparsity, k)
	indx.arr$sd <- sd.info.content[as.matrix(indx.arr)]
	sd.gen.capacity <- cast(indx.arr, sparsity~noise.level, value="sd")
	sd.gen.capacity <- as.matrix(sd.gen.capacity[,2:dim(sd.gen.capacity)[2]])
	dimnames(sd.gen.capacity) <- dimnames(gen.capacity)
	
	return(list(info.content=info.content, avg.info.content=avg.info.content, sd.info.content=sd.info.content, indx.beta.opt=indx.beta.opt ,gen.capacity=gen.capacity, sd.gen.capacity=sd.gen.capacity))
}

# importance sampling algo. 
get.imp.smpl.info <- function(hyp.size, logZ){

	k <- as.numeric(dimnames(logZ)$sparsity)
	num.noise <- length(dimnames(logZ)$noise.level)
	#Calculate Information Content
	info.content <-   logZ[,,3,,] - logZ[,,1,,] -logZ[,,2,,] #+ log(hyp.size) 

	#Calculate average information content
	avg.info.content <- apply(info.content, c("sparsity","noise.level","beta"), mean)

	#Calculate standard deviation information content
	sd.info.content <- apply(info.content, c("sparsity","noise.level","beta"), sd)

	#Obtain generalization capacity and corresponding beta
	indx.beta.opt <- apply(avg.info.content,c("sparsity","noise.level"),which.max)
	gen.capacity <- apply(avg.info.content,c("sparsity","noise.level"),max)
	
	indx.arr <- melt(indx.beta.opt)
	names(indx.arr)[3] <- "beta"
	indx.arr$sparsity <- match(indx.arr$sparsity, k)
	indx.arr$sd <- sd.info.content[as.matrix(indx.arr)]
	sd.gen.capacity <- cast(indx.arr, sparsity~noise.level, value="sd")
	sd.gen.capacity <- as.matrix(sd.gen.capacity[,2:dim(sd.gen.capacity)[2]])
	dimnames(sd.gen.capacity) <- dimnames(gen.capacity)
	
	return(list(info.content=info.content, avg.info.content=avg.info.content, sd.info.content=sd.info.content, indx.beta.opt=indx.beta.opt ,gen.capacity=gen.capacity, sd.gen.capacity=sd.gen.capacity))
}

#plot avg info. content vs beta for different noise levels
plot.avg.info.content <- function(info, beta, noise.level, choose.noise, choose.sparsity){
	
	choose.num.noise <- length(choose.noise)
	choose.num.sparsity <- length(choose.sparsity)
	k <- as.numeric(dimnames(info$info.content)$sparsity)
	
	
	
	for(j in 1:choose.num.sparsity){	
	plot(beta, info$avg.info.content[choose.sparsity[j],choose.noise[1],], type="l", ylim=c(-0.5,max(info$avg.info.content[choose.sparsity[j],choose.noise,])*1.1), xlab="beta",ylab="Average Information Content", main=paste("sparsity = ",k[choose.sparsity[j]]))		
		for(i in 1:choose.num.noise){
			lines(beta, info$avg.info.content[choose.sparsity[j],choose.noise[i],], col=i)
			lines(x=beta[info$indx.beta.opt[choose.sparsity[j],choose.noise[i]]], info$gen.capacity[choose.sparsity[j],choose.noise[i]], type="p", col=i)	
			text(x=beta[info$indx.beta.opt[choose.sparsity[j],choose.noise[i]]],y=info$gen.capacity[choose.sparsity[j], choose.noise[i]]+0.1,labels=round(noise.level[choose.noise[i]],1),cex=0.8)
		}
	}
	abline(h=0, col="grey")
	
	for(i in 1:choose.num.noise){
	plot(beta, info$avg.info.content[choose.sparsity[1],choose.noise[i],], type="l", ylim=c(-0.5,max(info$avg.info.content[choose.sparsity,choose.noise[i],])*1.1), xlab="beta",ylab="Average Information Content", main=paste("noise level = ",round(noise.level[choose.noise[i]],1)))
	for(j in 1:choose.num.sparsity){
			lines(beta, info$avg.info.content[choose.sparsity[j],choose.noise[i],], col=j)
			lines(x=beta[info$indx.beta.opt[choose.sparsity[j],choose.noise[i]]], info$gen.capacity[choose.sparsity[j],choose.noise[i]], type="p", col=j)
			text(x=beta[info$indx.beta.opt[choose.sparsity[j],choose.noise[i]]],y=info$gen.capacity[choose.sparsity[j], choose.noise[i]]+0.1,labels=k[choose.sparsity[j]],cex=0.8)
		}
	}
	abline(h=0,col="grey")
	
}	

#plot generalization capacity vs noise level
plot.gen.capacity <- function(n, noise.level, prs, info, Cs, mu0, choose.sparsity){
	
	d <- dim(Cs[[1]])[1]
	
	reps <- length(dimnames(prs[[1]])$repetition)
	num.noise <- length(noise.level)
	bayes.hit   <- (pnorm(sqrt(n)/(2*noise.level)))^d
	bayes.error <- 1-bayes.hit
	num.spars <- length(prs)

	indx.mat <- melt(info$indx.beta.opt)
	colnames(indx.mat)[3] <- "beta"
	pr.opt <- lapply(1:num.spars, function(i) apply(prs[[i]], c("hypothesis","repetition","Zs"), function(mat) mat[as.matrix(indx.mat[which(indx.mat$sparsity==i),c("noise.level","beta")])]))
	names(pr.opt) <- 1:num.spars
	pr.opt <- lapply(pr.opt, function(el){
		 names(dimnames(el))[1] <- "noise.level"
		 dimnames(el)[[1]] <- 1:num.noise
		 el <- aperm(el, c(2,3,4,1))
		 return(el)
	 })
	 indx.true.hyp <- sapply(1:num.spars, function(i) which(apply(Cs[[i]]==mu0[,i], "hypothesis", all)))
	 pr.gibbs.true <- sapply(1:num.spars, function(i) apply(pr.opt[[i]][indx.true.hyp[i],,1,], c("noise.level"), mean))
	names(dimnames(pr.gibbs.true)) <- c("noise.level","sparsity")
	dimnames(pr.gibbs.true)$noise.level <- 1:num.noise
	dimnames(pr.gibbs.true)$sparsity <- 1:num.spars
	


	plot(noise.level, pr.opt[[choose.sparsity]][indx.true.hyp[choose.sparsity],1,1,], type="l",ylim=range(pr.opt[[choose.sparsity]][indx.true.hyp[choose.sparsity],,,]))
	for(i in 1:reps) for(j in 1:2) lines(noise.level, pr.opt[[choose.sparsity]][indx.true.hyp[choose.sparsity],i,j,])
	lines(noise.level, pr.gibbs.true[,choose.sparsity], col="blue")
	
	y.gen <- pretty(c(0,info$gen.capacity[choose.sparsity,]))
	 # add extra room to the left of the plot 
	 par(oma=c(0,2,0,0))
	 #generalization capacity
	 plot(x=noise.level, y=info$gen.capacity[choose.sparsity,], col='blue', type='l', ylim=range(y.gen), main=paste("Generalization capacity by noise level, sparsity= ", k[choose.sparsity]), xlab='noise level', ylab='',xaxt='n', yaxt='n', lwd=0.75) 
	 abline(h=0, col="grey")
	 axis(2, col='blue', at=y.gen, labels=y.gen)	 
	 #gibbs probability
	 par(new=T)
	 plot(x=noise.level, y=pr.gibbs.true[,choose.sparsity], ylim=c(0,1),col='red', type='l', lwd=0.75,xaxt='n', axes=F, ylab='')
	 axis(side=2, at=y.gen/max(y.gen), labels=round(y.gen/max(y.gen),1), col='red', line=2)
	 #1-bayes error
	 par(new=T)
	 plot(x=noise.level, y=bayes.hit, ylim=c(0,1),col='green', type='l', lwd=0.75,xaxt='n', axes=F, ylab='')
	 axis(side=2, at=y.gen/max(y.gen), labels=rep("",length(y.gen)), col='green', line=2.1)
	 #legend
	 legend(x=0, y=.8, legend=c('gen. cap', 'sd. info cont','gibbs true', '1-bayes err.'), col=c("blue","black","red","green"), lwd=c(3.5,  3.5, 3.5))
	 #standar deviation information content
	 par(new=T)
	 plot(x=noise.level, y=info$sd.gen.capacity[choose.sparsity,], ylim=c(0,max(info$sd.gen.capacity[choose.sparsity,])),col='black', type='l', lwd=0.75,xaxt='n', axes=F, ylab='')
	 axis(side=2, at=y.gen/max(y.gen)*max(info$sd.gen.capacity[choose.sparsity,]), labels=round(y.gen/max(y.gen)*max(info$sd.gen.capacity)), col='black', line=4)
	 #x-axis
	 axis(side=1, at=noise.level,   labels=round(noise.level,1))
	 abline(v=noise.level, col='grey', lwd=0.5)
	 
	 #All sparsities
	  y.gen <- pretty(c(0,gen.capacity))
	  # add extra room to the left of the plot 
	  par(oma=c(0,2,0,0))
	  #generalization capacity
	  plot(x=noise.level, y=info$gen.capacity[1, ], col='blue', type='l', ylim=range(y.gen), main="Generalization capacity by noise level and sparsity", xlab='noise level', ylab='', lwd=0.75)
	 for(i in 1:num.spars) lines(x=noise.level, y=info$gen.capacity[i, ], col=i)
	 #x-axis
	  axis(side=1, at=noise.level,   labels=round(noise.level,1))
	  abline(v=noise.level, col='grey', lwd=0.5)
 
	  legend(x=10, y=5, legend=paste("sparsity = ", 1:num.spars), col=1:num.spars, lwd=c(3.5,  3.5, 3.5))
	 
	
}

#plot generalization capacity vs noise level
plot.smpl.gen.capacity <- function(d, n, noise.level, info){
	

	reps.Z <- length(dimnames(info$info.content)$repetition)
	num.noise <- length(noise.level)
	bayes.hit   <- (pnorm(sqrt(n)/(2*noise.level)))^d
	bayes.error <- 1-bayes.hit

	
	y.gen <- pretty(c(0,info$gen.capacity))
	 # add extra room to the left of the plot 
	 par(oma=c(0,2,0,0))
	 #generalization capacity
	 plot(x=noise.level, y=info$gen.capacity, col='blue', type='l', ylim=range(y.gen), main="Generalization capacity by noise level", xlab='noise level', ylab='',xaxt='n', yaxt='n', lwd=0.75)
	 #segments(x0=noise.level, y0=info$gen.capacity-info$sd.gen.capacity, x1 = noise.level, y1 = info$gen.capacity+info$sd.gen.capacity, col="grey", lwd=3)
	 
	 abline(h=0, col="grey")
	 axis(2, col='blue', at=y.gen, labels=y.gen)	 
	 #1-bayes error
	 par(new=T)
	 plot(x=noise.level, y=bayes.hit, ylim=c(0,1),col='green', type='l', lwd=0.75,xaxt='n', axes=F, ylab='')
	 axis(side=2, at=y.gen/max(y.gen), labels=rep("",length(y.gen)), col='green', line=2.1)
	 
	 legend(x=0, y=.8, legend=c('gen. cap', 'sd. info cont', '1-bayes err.'), col=c("blue","black","green"), lwd=c(3.5,  3.5, 3.5))
	 #standar deviation information content
	 par(new=T)
	 plot(x=noise.level, y=info$sd.gen.capacity, ylim=c(0,max(info$sd.gen.capacity)),col='black', type='l', lwd=0.75,xaxt='n', axes=F, ylab='')
	 axis(side=2, at=y.gen/max(y.gen)*max(info$sd.gen.capacity), labels=round(y.gen/max(y.gen)*max(info$sd.gen.capacity)), col='black', line=4)
	 
	 #x-axis
	 axis(side=1, at=noise.level,   labels=round(noise.level,1))
	 abline(v=noise.level, col='grey', lwd=0.5)
	
}

#obtain generalization capacity and standard deviation of information content 
get.gen.capacity <- function(data.arr, Cs, mu0, beta, cost.fun){
	
	
	hyp.size <- sapply(Cs, function(x) dim(x)[2])
	
	
	arrays <- get.uf.arrays(Cs, data.arr, mu0, beta, cost.fun)
	info <- get.info(hyp.size, arrays$logZ)
	return(list(arrays=arrays, info=info))
}

#obtain generalization capacity and standard deviation of information content -sample version 
get.smpl.gen.capacity <- function(reps.mu, data.arr, mu0, beta, cost.fun){
	
	
	d <- dim(data.arr)["component"]
	k <- as.numeric(dimnames(data.arr)$sparsity)
	hyp.size <- choose(d,k)
	
	arrays <- get.smpl.uf.arrays(reps.mu, data.arr, mu0, beta, cost.fun)
	info <- get.smpl.info(hyp.size, logZ=arrays$logZ)
	return(list(arrays=arrays, info=info))
}

#obtain generalization capacity and standard deviation of information content -importance sample version 
get.imp.smpl.gen.capacity <- function(reps.mu, data.arr, mu0, beta, cost.fun){
	
	
	d <- dim(data.arr)["component"]
	k <- as.numeric(dimnames(data.arr)$sparsity)
	hyp.size <- choose(d,k)
	arrays <- get.imp.smpl.uf.arrays(reps.mu, data.arr, mu0, beta, cost.fun)
	info <- get.imp.smpl.info(hyp.size, logZ=arrays$logZ)
	return(list(arrays=arrays, info=info))
}
\end{lstlisting}
\section{Script}
\lstinputlisting{script.R}

\end{appendices}

%\addtocontents{toc}{\vspace{.5\baselineskip}}
%\appendix
%\include{AppendixSimResults}
%\include{AppendixRCode}

%% Epilogue (optional)
%\addtocontents{toc}{\vspace{.5\baselineskip}}
%\cleardoublepage
%\phantomsection
%\addcontentsline{toc}{chapter}{\protect\numberline{}{Epilogue}}
%\markboth{Epilogue}{Epilogue}
%\include{Epilogue}

%%%%%%%%%%%%%%%%%%%%%%%%%%%%%%%%%%%%%%%%%%%%%%%%%% 
%%% Declaration of originality (Do not remove!)%%%
%%%%%%%%%%%%%%%%%%%%%%%%%%%%%%%%%%%%%%%%%%%%%%%%%%
%% Instructions:
%% -------------
%% fill in the empty document confirmation-originality.pdf electronically
%% print it out and sign it
%% scan it in again and save the scan in this directory with name
%% confirmation-originality-scan.pdf 
%%
%% General info on plagiarism:
%% https://www.ethz.ch/students/en/studies/performance-assessments/plagiarism.html 
%\cleardoublepage
%\includepdf[pages={-}, frame=true,scale=1]{confirmation-originality-scan.pdf}
\end{document}